\documentclass[aps,prx,longbibliography,twocolumn,notitlepage,citeautoscript,superscriptaddress]{revtex4-2}
\usepackage{amsmath,amssymb,bm,graphicx,color,gensymb,bbold,appendix,hyperref}

\usepackage{cancel}

\usepackage[normalem]{ulem} 

\begin{document}
\title{Roller-Coaster in a Flatland:  Magnetoresistivity in Eu-intercalated Graphite}

\author{A. L. Chernyshev}
\affiliation{Department of Physics and Astronomy, University of California, Irvine, California 92697, USA}
\author{O. A. Starykh}
\affiliation{Department of Physics and Astronomy, University of Utah, Salt Lake City, Utah 84112, USA}
\date{\today}
\begin{abstract}
Novel phenomena in magnetically-intercalated graphite has been 
a subject of much research, pioneered and promoted by M.~S. and G.~Dresselhaus and many others  in the 1980s.
Among the most enigmatic findings of that era was a dramatic, roller-coaster-like behavior
of the magnetoresistivity in EuC$_6$ compound, in which magnetic Eu$^{2+}$ ions form a
triangular lattice that is commensurate to graphite honeycomb planes. 
In this study, we provide a long-awaited {\it microscopic} explanation of this behavior, demonstrating 
that the resistivity of EuC$_6$ is dominated by  spin excitations in Eu-planes 
and their highly nontrivial evolution with the magnetic field. Together with 
our theoretical analysis, the present study showcases the 
power of the synthetic 2D materials 
as a source of potentially significant  insights into the nature of exotic spin excitations.
\end{abstract}
\maketitle
\tableofcontents

\section{Introduction}
\label{Sec_intro}

The two-dimensional (2D) world of Flatland, a mathematical abstraction and a cultural reference 
 \cite{Flatland}, has, arguably, received its ultimate physical realization in the form of 
graphene \cite{NovoselovNobel,GeimMacDonald}, whose unique properties \cite{Neto} have ushered 
in a new era of making  artificial heterostructures via a Lego-like \cite{GeimGrigorieva} 
assembly of  layered materials. 
Together with the research in the twisted bi- and $n$-layer graphene \cite{macdonald2019}, 
the fledging field of van der Waals magnets holds a lot of promise in opening new horizons for  the
fundamental studies and applications along the path of using this technology 
\cite{Mandrus,Park,Herrero18,Gong19,zhang2021metamagnetic}.

Historically, a more traditional, if not ancient \cite{ancient}, way of achieving similar goals 
of synthesizing materials with novel properties from a stack of carbon layers and various elements 
and compounds has relied on the process  of intercalation, 
suggesting another cultural metaphor \cite{Between}. 
The research in graphite intercalation compounds (GICs) has attracted significant attention in the past, 
with the evolution of such studies and understanding of these materials outlined in several books and, 
specifically, in the reviews by M.~S. and G.~Dresselhaus, whose efforts contributed to much of the progress 
in this area,  see Refs.~\cite{TwoDresselhaus1981,Dresselhaus88,Dresselhaus_review,Endo}. 

\begin{figure*}[t]
\includegraphics[width=\linewidth]{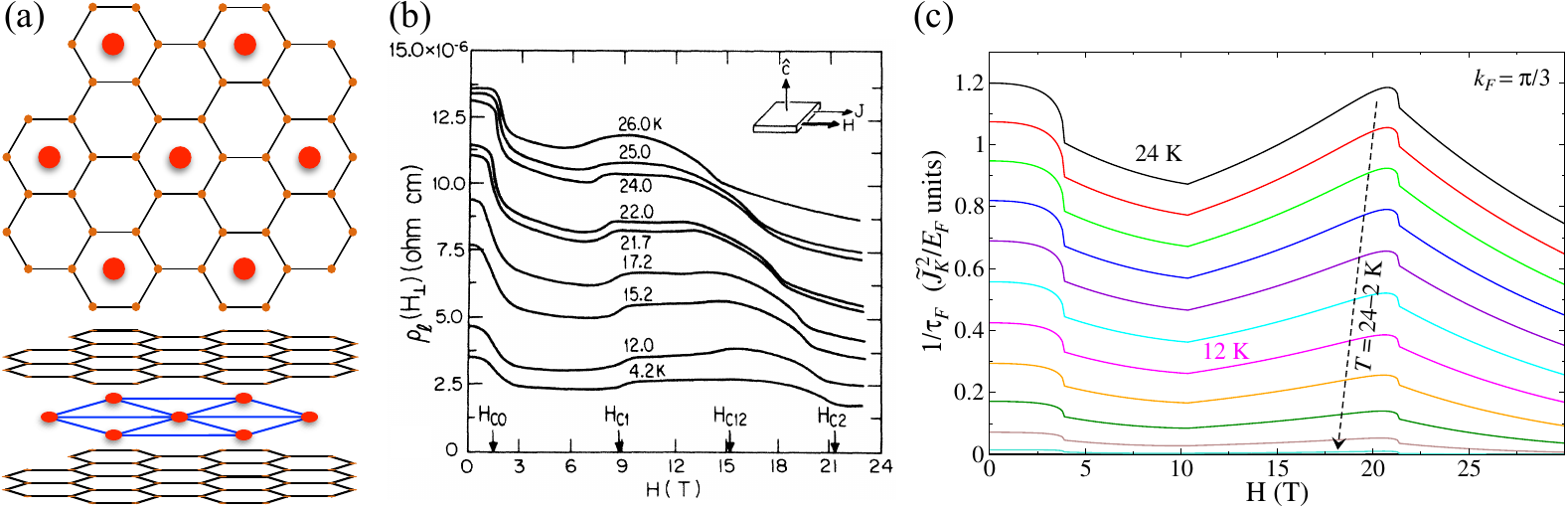}
\caption{(a) The schematics of  EuC$_6$. Small and large dots are C and Eu atoms, respectively. 
(b) The EuC$_6$ in-plane resistivity  data vs magnetic field, $\rho(H)$, for various temperatures, see Ref.~\cite{Chen}.  
Arrows are the fields of the anomalies in $\rho(H)$ that correspond to transitions between magnetic states, 
extrapolated to $T\!=\!0$.
[Reprinted with permission from S.~T.~Chen, M. S. Dresselhaus, G. Dresselhaus, H. Suematsu, H. Minemoto, 
K. Ohmatsu, and Y. Yosida, Phys. Rev. B {\bf 34}, 423 (1986), Ref.~\cite{Chen}. 
Copyright (1986) by the American Physical Society.]
(c) Our representative results for the transport scattering rate.}
\label{Fig_EuC6}
\vskip -0.2cm
\end{figure*}

Of the fundamental footprint of this research, it is the {\it magnetically}-intercalated compounds that 
have produced the most  intriguing phenomena \cite{Dresselhaus_review}. The case of 
EuC$_6$, made of alternating  honeycomb layers of carbon and triangular-lattice layers of Eu ions, 
shown schematically in Fig.~\ref{Fig_EuC6}(a), particularly stands out. A highly dramatic, 
roller-coaster-like dependence of the in-plane resistivity on a magnetic field, 
reproduced  from Ref.~\cite{Chen} in Fig.~\ref{Fig_EuC6}(b), 
is clearly indicative of an intricately intertwined magnetic and electronic degrees of freedom of this material.
Incidentally, EuC$_6$ is also the first magnet to exhibit the fabled $1/3$-magnetization plateau \cite{date1981} 
and was inspirational for an understanding of this state \cite{Chubukov1991}.

The pioneering studies of EuC$_6$ \cite{Sakikabara1,Sakikabara2,Sakikabara3,Chen}  have analyzed and successfully 
identified key exchange terms of the  triangular-lattice spin-$7/2$ model Hamiltonian of the localized 
$4f$ orbitals of Eu$^{2+}$ ions that are necessary to understand the field-induced phases and  
the concomitant  magnetization data \cite{date1981}. 
However,  while yielding a reasonable estimate of the Kondo coupling, 
the sole attempt to explain magnetoresistivity itself \cite{Sugihara1985}
has provided a largely unsuccessful  modeling of it via a crude consideration of the spin-scattering of electrons
and suggested a rather relic backflow mechanism to explain the $T$-dependence of the resistivity. 

Thus, it is fair to say that, by and large and to the best of our knowledge, 
there exists no proper explanation of the key resistivity results 
observed in EuC$_6$, shown in Fig.~\ref{Fig_EuC6}(b). 
Furthermore,  the refocusing of the research 
of the 1980s and 1990s 
on  correlated systems and high-temperature superconductivity
has left these striking results in their enigmatic state.

In this work, we provide a microscopic theory  of the magnetoresistivity of EuC$_6$
and demonstrate that its highly nontrivial evolution with the magnetic field
can be fully accounted for by the scattering off the spin excitations in Eu-planes.
Figure~\ref{Fig_EuC6}(c) demonstrates representative results of our theory, 
which capture most of the qualitative and  quantitative features of the experimental data in 
Fig.~\ref{Fig_EuC6}(b), with the details of the theory provided below.
Our effort  brings  together research in the magnetically-intercalated graphite 
compounds with that  in the novel graphite-derived artificial magnetic materials \cite{Park,Endo,Mak2019}.

More broadly, we would also like to highlight that there is a number of conducting magnetic materials 
that exhibit a highly non-monotonic magnetoresistivity 
\cite{Canfield,Maeno,Mydosh,Xia},  showing that such measurements can serve as 
a very sensitive probe of the field-induced phase transitions. However, most theoretical
explanations, if any, are limited to an associative construction of  phenomenological spin models 
to match the number of phase transitions and broad trends in magnetization 
\cite{Xia}, without any attempt to explicate scattering mechanisms and calculate resistivity. 
In that respect, our present study is also the one that accomplishes  precisely this goal: a fully 
microscopic calculation of the resistivity throughout all the phases in the 
phase diagram of the underlying spin model. We anticipate  our results  not only  be inspirational 
for the broader research in  metallic magnets, but  also to provide the technical guidance for similar studies. 

We outline, in broad strokes, our approach and results. We build on the achievements of the prior 
work on  EuC$_6$ \cite{Sakikabara1,Sakikabara2,Sakikabara3,Chen} and reanalyze phenomenological 
constraints on the  triangular-lattice spin-$7/2$ model of  Eu$^{2+}$ layers. In this analysis, 
we also use more recent experimental insights into the magnetic ground state of EuC$_6$ \cite{Lamura2012}
and density-functional theory of its electronic structure   \cite{Molodtsov1996}.

Thus, we establish bounds on the exchange parameters as related to the phenomenology of
different magnetic phases of EuC$_6$, examine ranges of parameters that make transitions between the phases 
first-order, and formulate a minimal model to describe EuC$_6$. We proceed by constructing the spin-wave
theory for all the field-induced phases of that model. 
Although a numerical procedure is generally needed to obtain magnon eigenenergies,
the approach leading to it, as well as the results for some of the phases, are fully analytical. 

While the Kondo coupling between conducting electronic states and Eu$^{2+}$ spins is fully local,
the matrix elements of electron scattering on magnons have a nontrivial form,
owing to the  internal structure of quasiparticle eigenstates in different phases. This structure
leads, among  other things, to the non-spin-flip scattering processes in the non-collinear phases.
We articulate that these matrix elements are essential for a consistent calculation of the transport scattering 
rate. The expression for the latter, given in a concise form, 
is derived using Boltzmann formalism, which we revisit for both spin-flip and non-spin-flip
channels, providing a thorough derivation of the relaxation-time approximation in the process. 

The temperature-dependence of the resistivity anticipated from our theory is discussed for all  field-induced phases. 
Significantly, the zero-field results of our theory demonstrate an analogue of the phonon-dominated 
resistivity behavior, but due to scattering off the acoustic magnons of the 120$\degree$ state, 
with a 2D  equivalent of the Bloch-Gr\"{u}neisen low-temperature asymptote of $\rho\!\propto\!T^4$ and the 
high-temperature  Ohm's  law, $\rho\!\propto\!T$. Given the extent of the magnon bandwidth, 
the  nearly linear trend of $\rho(T)$ observed in Ref.~\cite{Sugihara1985} above 8~K is shown to be well 
within the onset of the Ohm's regime.

The resistivity calculations   are performed at experimentally relevant temperatures  
for various parameters of the minimal model to demonstrate qualitative trends and for a 
specific set of parameters that best describes EuC$_6$. 
We also investigate the dependence of our results on the filling fraction of electronic bands, encoded in the 
Fermi momentum $k_F$, and conclude that the relatively smaller values of $k_F\!\alt\!\pi/3$ provide a better 
correspondence to the EuC$_6$ phenomenology, inviting more research into a  verification of its
electronic properties. Other intriguing features of the resistivity  for the larger values of 
$k_F$, potentially controllable by doping, are also discussed.

Altogether, the results of our model for the transport relaxation rate, offered in Fig.~\ref{Fig_EuC6}(c) 
for a representative $k_F \!=\!\pi/3$, show a striking similarity to the experimental 
data in Fig.~\ref{Fig_EuC6}(b), with the possible origin of the discrepancies at higher field discussed below. 
Our theory implicitly contains the field-dependence via that of the magnon spectra and scattering matrix elements,
which, in turn, depend on the spin arrangement in each of the field-induced phases. It
also properly accounts for the effect of the thermal population of magnetic scatterers on the resistivity. 
One of the qualitative messages of our study is the importance of the non-spin-flip channel of the scattering,
which is present in the  phases with the non-collinear spin configurations, but is absent for the collinear ones.
This effect explains the weaker scattering and lower resistivity   in the $1/3$-magnetization plateau 
and fully polarized phases.

The general picture that emerges from our analysis is that of the resistivity as a very informative probe of 
not only field-induced phase transitions, but also of the elementary spin excitations in these phases.
The provided thorough theoretical analysis of the iconic  two-dimensional
triangular-lattice antiferromagnet coupled to conduction electrons showcases a largely untapped 
power of the synthetic 2D materials as a source of potentially significant  insights into the nature of 
exotic spin excitations. Our approach and findings can be  applied, 
for example, to the electron scattering by the fractionalized spinons of the Kitaev spin liquid 
\cite{Mashhadi2019,Knolle} and to the other magnetically-intercalated systems such as 
chalcogenides \cite{Analytis,Musfeldt17,MacDonald_Kondo21}. 

The paper is organized as follows.
Section~\ref{sec_FS} discusses electronic structure of EuC$_6$ and the approximate values of the 
Fermi momenta. Section~\ref{Sec_model} gives an overview of the phenomenologically-motivated
spin model of EuC$_6$, its classical ground states and critical fields, and parameters of the 
minimal model. Details on the first-order transitions are delegated to Appendix~\ref{app:1st}.
Spin excitations of the model for all field-induced phases are discussed in Section~\ref{Sec_SWT},
which provides details of the spin-wave formalism and results for representative 
magnon eigenenergies and the eigenfunctions. The fully analytical results
for the polarized, 120$\degree$, and plateau phases are given in Appendix~\ref{Sec_particular}.

The Kondo coupling and its estimate, as well as  resistivity and some qualitative insight into it, are discussed
in Section~\ref{Sec_Kondo_res}. This consideration relies, not in a small way, on a detailed derivation of the 
relaxation rates for the spin-flip and non-spin-flip channels from the Boltzmann formalism, 
provided in Appendix~\ref{App_transport}, which also discusses possible limitations of this approach 
and potential new phenomena at large values of $2k_F$.
The temperature-dependence of magnetoresistivity,  results for various values of the key model parameters 
and Fermi momentum,  and an outlook on the possible future extensions are given in Section~\ref{Sec_results}.
We provide a summary in Section~\ref{Sec_summary}. 

\begin{figure*}[t]
\includegraphics[width=\linewidth]{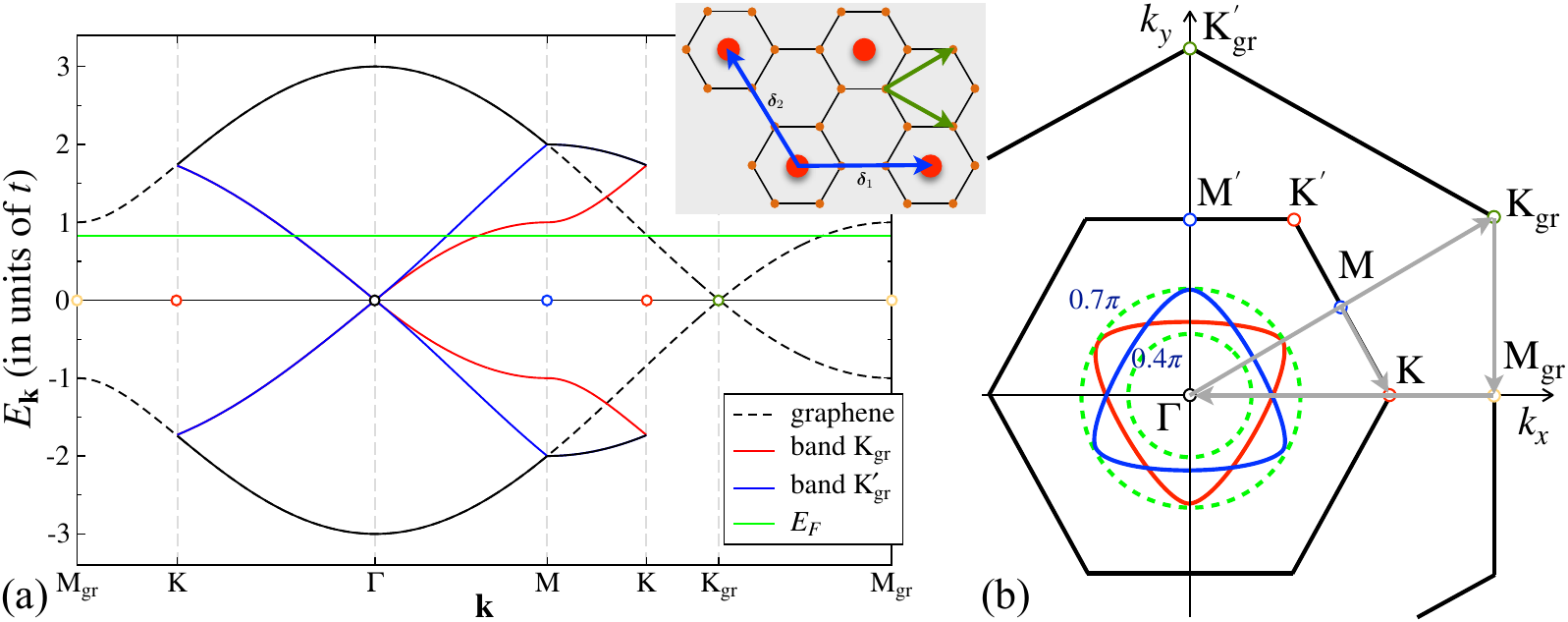}
\caption{(a) Energy bands of graphene in its full BZ (dashed lines), and folded onto Eu-lattice BZ to represent
rigid-band structure of Eu-intercalated graphite (solid lines) along the paths 
M$_{\rm gr}\Gamma$M(K)K$_{\rm gr}$M$_{\rm gr}$ shown in (b); high-symmetry points are highlighted. 
Two Dirac bands are color-coded to indicate their origin before folding. 
Energies are in units of the graphene tight-binding hopping parameter $t\!=\!3.16$~eV \cite{Neto}. 
Horizontal line is the Fermi energy $E_F\!=\!0.83t\!\approx\!2.62$~eV. It matches the Fermi momenta 
$k_{F, {\rm min}}\!\approx\!0.48\pi/a$ and $k_{F, {\rm min}}\!\approx\!0.7\pi/a$ from Ref.~\cite{Molodtsov1996}
of the trigonally warped Fermi surfaces shown in (b) that approximately correspond to the $0.5e$  filling of 
the bands.
Insert: crystal structure of Eu-GIC from Fig.~\ref{Fig_EuC6} with primitive translational vectors of 
Eu and C lattices. (b) Brillouin zones of the graphene and of the Eu-based triangular lattice. 
Fermi surfaces at $E_F$ are color-coded according to the bands in (a). Two representative circular Fermi 
surfaces, with $k_F \!=\! 0.4 \pi/a$ and $0.7\pi/a$, (dashed lines). 
High-symmetry paths are indicated by the arrows.}
\vskip -0.2cm
\label{Fig_bands}
\end{figure*}

\section{Phenomenology and modeling}
\label{Sec_phenom}
\subsection{Electronic structure of EuC$_6$}
\label{sec_FS}

The electronic structure of Eu-intercalated graphite EuC$_6$ has been investigated experimentally and 
theoretically in the mid-90s of the last century  \cite{Molodtsov1996}, with the 
summary of these efforts given in Ref.~\cite{Endo}. 

Structurally, EuC$_6$ is the so-called stage-I intercalated compound, meaning that the Eu layers 
alternate with that of carbon.  Viewed from a graphite layer, the rare-earth atoms are located on top of the
centers of the graphite hexagons and form a  $\sqrt{3}\times\!\sqrt{3}$ superstructure 
as is illustrated in Fig.~\ref{Fig_EuC6}(a). 
The material is characterized by the so-called $A\alpha A\beta$ stacking (space group $P6_3/mmc$), 
in which Eu atoms form a hexagonal closed packed structure with alternating positions $\alpha$ and 
$\beta$ between consecutive layers, 
while carbons follow the $AA$ stacking \cite{ElMakrini1980,Rida2010,Lamura2012}. 
This arrangement of carbon sheets is different from the $AB$, or Bernal, stacking of the  graphite. 

As a result, the principal unit  associated with the Eu-based triangular lattice can be seen as 
containing one Eu and six carbon atoms, while the structural unit cell contains two 
Eu atoms and twelve carbons. As is shown in Fig.~\ref{Fig_bands}(b), the two-dimensional (2D) 
Brillouin zone  (BZ) of the triangular Eu lattice is three times 
smaller than that of the  graphene. The lattice constants of the 
triangular Eu lattice and that of the honeycomb graphene lattice are related as $a\!=\!\sqrt{3} a_{\rm gr}$,
see Fig.~\ref{Fig_bands}.

The key features of the electronic band structure of EuC$_6$ can be understood within the ``rigid-band'' 
approximation, see Ch. 5 of Ref.~\cite{Endo}. One assumes that the band structure of the graphene layer 
is not changed by the Eu intercalation, with the latter resulting only  in a partial filling of the graphene bands 
up to a Fermi energy $E_F$, illustrated in Fig.~\ref{Fig_bands}(a) by a horizontal line. 

Upon folding onto the Eu-based Brillouin zone, the Dirac bands are mapped from the proximities of the 
K$_{\rm gr}$ and K$^\prime_{\rm gr}$ points of the graphene Brillouin zone  onto the neighborhood of the 
$\Gamma$ point, see Fig.~\ref{Fig_bands}.
These bands are equivalent up to a $\pi/3$ rotation, with a representative constant-energy cut 
demonstrating characteristic ``flower-petal'' Fermi surfaces originating from the trigonal 
$C_3$ symmetry of the graphene lattice, see Fig.~\ref{Fig_bands}(b). 
These two bands from the two valleys at K$_{\rm gr}$ and K$^\prime_{\rm gr}$ are the ones 
being filled away from the charge-neutrality point by the doping provided by the intercalated Eu.

To estimate the size of the Fermi surfaces produced by doping, one can 
approximate them as circles with a  radius $k_F$, neglecting their trigonal warping.
Naturally, the Fermi momentum $k_F$  is determined by the 2D density of electrons donated 
to a graphene sheet by the Eu layer.  Taking into account  band (valley)  and spin degeneracy factors
yields $n_e^{2D}\!= \!k_F^2/\pi$ \cite{Endo}. The nominal valence state of Eu is Eu$^{2+}$. 
Assuming that all 2$e$/Eu go into the conduction bands and using the 2D volume of the Eu unit cell 
$V_c\!=\!a^2\sqrt{3}/2$, one obtains 
$k_{F, 2e} \!= \!(4\pi/\sqrt{3}a^2)^{1/2}\!\approx \!0.86 \pi/a$.
The same result can be obtained by matching the area (2D volume) of the fully occupied, doubly-degenerate 
triangular-lattice Brillouin zone of the Eu lattice, $V_{BZ}^\triangle\!=\!8\pi^2/\sqrt{3}a^2$, 
with the four-fold degenerate (valley$\times$spin) Fermi circle of radius $k_{F, 2e}$. 
Altogether,  the Fermi surface in EuC$_6$, estimated within this approach, is expected to be large. 

The detailed calculations of   electronic structure of EuC$_6$ in Ref.~\cite{Molodtsov1996} 
feature the band-structure that is not unlike the rigid-band picture in Fig.~\ref{Fig_bands}(a), with the 
bands that are crossing the Fermi level  clearly reminiscent of the folded graphene bands.
However,  two key differences are a significantly lower doping of the carbon $\pi$-orbitals, 
which accounts  for  about $0.5e$ per Eu$^{2+}$, and the rest of 
electrons filling up  the Eu-derived $spd$-hybrid band, with the latter absent in the rigid-band 
description \cite{Molodtsov1996,Endo}. These findings are also supported by the angle-resolved 
photoemission studies of stage-I EuC$_6$ and stage-II EuC$_{12}$ materials, reported in 
Ref.~\cite{Molodtsov1996}. 

The most direct implication of the first result for our analysis of the Fermi surfaces is the four times smaller 
density of donated electrons, which straightforwardly translates into the two times smaller Fermi momentum 
in the graphene conduction bands,  $k_{F, e/2}\! \approx\! 0.43 \pi/a$.  
We also estimate the Fermi momenta of the ``true,'' trigonally warped Fermi surfaces 
from the band structure in Ref.~\cite{Molodtsov1996}  
as $k_{F, {\rm max}}\!\approx\!0.7 \pi/a$  and $k_{F, {\rm min}}\!\approx\!0.45\pi/a$, in a qualitative 
agreement with the estimate of $k_{F, e/2}$  above. 
Our choice of the representative $E_F\!=\!0.83t$ (in units of $t\!=\!3.16$~eV \cite{Neto}) 
in Fig.~\ref{Fig_bands}(a) and of  the resultant Fermi surfaces in Fig.~\ref{Fig_bands}(b)
is made to match the Fermi momenta from Ref.~\cite{Molodtsov1996},
which, in turn, should   approximately correspond to the $0.5e$  filling of the bands.

The other shortcoming of the rigid-band approximation is the omission of the 
Eu-derived partially filled $sd$-hybrid band \cite{Molodtsov1996,Endo}. It was also argued that the 
hybridization of the Eu $sd$-orbitals and graphene $\pi$-orbitals is responsible for the mediation 
of the strong Kondo interaction between the localized $4f$-orbital spins of Eu and conduction 
$\pi$-orbital electrons of graphite, estimated at $J_K\!\approx\! 0.15$~eV \cite{Sugihara1985}.  
This key element of our study is described in Sec. \ref{Sec_Kondo}.

In our analytical treatment of the  scattering rate in Sec.~\ref{Sec_resistivity} and Appendix~\ref{App_transport},
we are motivated by the analysis and discussion provided in this Section and approximate the 
relevant electronic degrees of freedom of EuC$_6$ 
by the two degenerate bands with the circular Fermi surfaces of radius $k_F$  centered around $\Gamma$ point.
We treat $k_F$ as a parameter and show how the key  features of the calculated
magnetoresistivity evolve with it, see Sec.~\ref{Sec_results}. 
We expect the renewed interest in the problem to result in a convergence of the   
band-structure calculations with the  experimental data regarding the relevant electronic structure 
and parameters of EuC$_6$ and other GICs.

\subsection{Spin model and parameters}
\label{Sec_model}

It has been proposed in Refs.~\cite{Chen,Sakikabara1,Sakikabara2,Sakikabara3} that the minimal model that  
describes  phenomenology of the magnetism in EuC$_6$  is the triangular-lattice $S\!=\!7/2$ model
\begin{align}
\label{H_1}
\mathcal{H} = &\sum_{\langle ij \rangle_n}J_n \, \mathbf{S}_i \cdot \mathbf{S}_j 
-B \sum_{\langle ij \rangle_1}\left(\mathbf{S}_i \cdot \mathbf{S}_j\right)^2
-\mathbf{h} \cdot\sum_{i}   \mathbf{S}_i ,
\end{align}
where  $\langle ij \rangle_{1(2)}$ denote the (next-)nearest-neighbor bonds with 
the corresponding exchanges $J_{1(2)}$  and ${\bf h}\!=\!g\mu_B\mathbf{H}$ in the Zeeman term.
A crucial ingredient of this model is the biquadratic term. While 
$B$ may be small compared to the exchanges, it is important 
because of the  $S^2$ amplification factor. 

It was argued in Refs.~\cite{Chen,Sakikabara1,Sakikabara2,Sakikabara3} that 
this minimal model would not be complete without 
the ring-exchange term, which is discussed below in some more detail.
While  the biquadratic and ring-exchange terms play  similar role in stabilizing the 
up-up-down (UUD or plateau) state in a wide range of fields, our  
analysis of the EuC$_6$ phenomenology provided below 
points to the values of the ring exchange that are secondary to $B$, differing from the values advocated in 
Refs.~\cite{Chen,Sakikabara1}. However, given a close similarity of their effects, this variation 
is likely inconsequential and amounts to a different parametrization of such effects within an effective model. 
In the spin-wave consideration that follows, we will ignore the ring-exchange term
entirely, citing cumbersomeness  of its treatment. 

Another difference of our model from the  consideration of 
Refs.~\cite{Chen,Sakikabara1,Sakikabara2,Sakikabara3} is that the exchange terms in (\ref{H_1}) 
are taken as Heisenberg, not $XY$. This makes no difference for the classical phase 
diagram in the in-plane field, which was simulated using classical Monte-Carlo 
in Ref.~\cite{Chen} in the $XY$ limit. However, the actual  
anisotropy in EuC$_6$ is unlikely to exceed 10\%, as is evidenced by the very similar saturation 
fields in the in-plane and out-of-plane magnetization and by the nearly isotropic $g$-factors \cite{Chen},
justifying our choice of the isotropic limit of the model.

\begin{figure}[t]
\includegraphics[width=\linewidth]{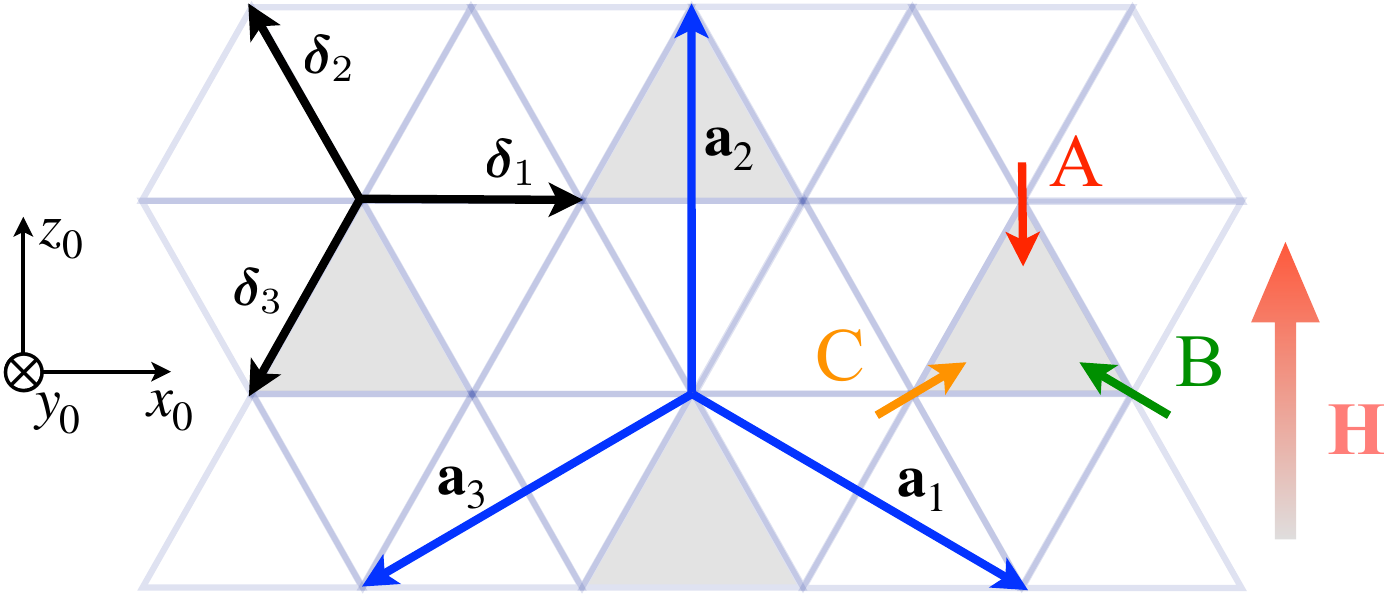}
\caption{Triangular lattice, its elementary translation vectors ${\bm \delta}_\alpha$, 
primitive unit cell for the three-sublattice spin structures (shaded) with its basis vectors ${\bf a}_\alpha$,
and an example of such a structure with $\{{\rm A},{\rm B},{\rm C}\}$ sublattices. 
Laboratory reference frame $\{x_0,y_0,z_0\}$ and the field direction are indicated.}
\label{Fig_lattice}
\end{figure}

\subsubsection{Classical ground states}
\label{Sec_classical}

In this work, we focus exclusively on the field orientation that is in the plane of Eu$^{2+}$ ions,
see Fig.~\ref{Fig_lattice}. While for the isotropic approximation that we choose in  model (\ref{H_1}) 
the direction of the field is irrelevant, the phenomenology that follows identifies with that of 
the in-plane field data for EuC$_6$, which exhibits a weak easy-plane ($XXZ$) anisotropy \cite{Sakikabara1}.  
The out-of-plane field direction in this latter case yields a different, and much simpler, 
magnetization and ground state evolution \cite{Yamamoto}. 

Triangular-lattice antiferromagnets host a rather astonishing variety of the 
unconventional field-induced phases, see  Refs.~\cite{Alicea2009,Starykh2015,Ye-Chubukov}.
As we have noted above, EuC$_6$ was the first  material in which 
the best-known of such unconventional phases, the UUD magnetization plateau state, 
has been identified \cite{date1981}. 

\begin{figure}[t]
\includegraphics[width=\linewidth]{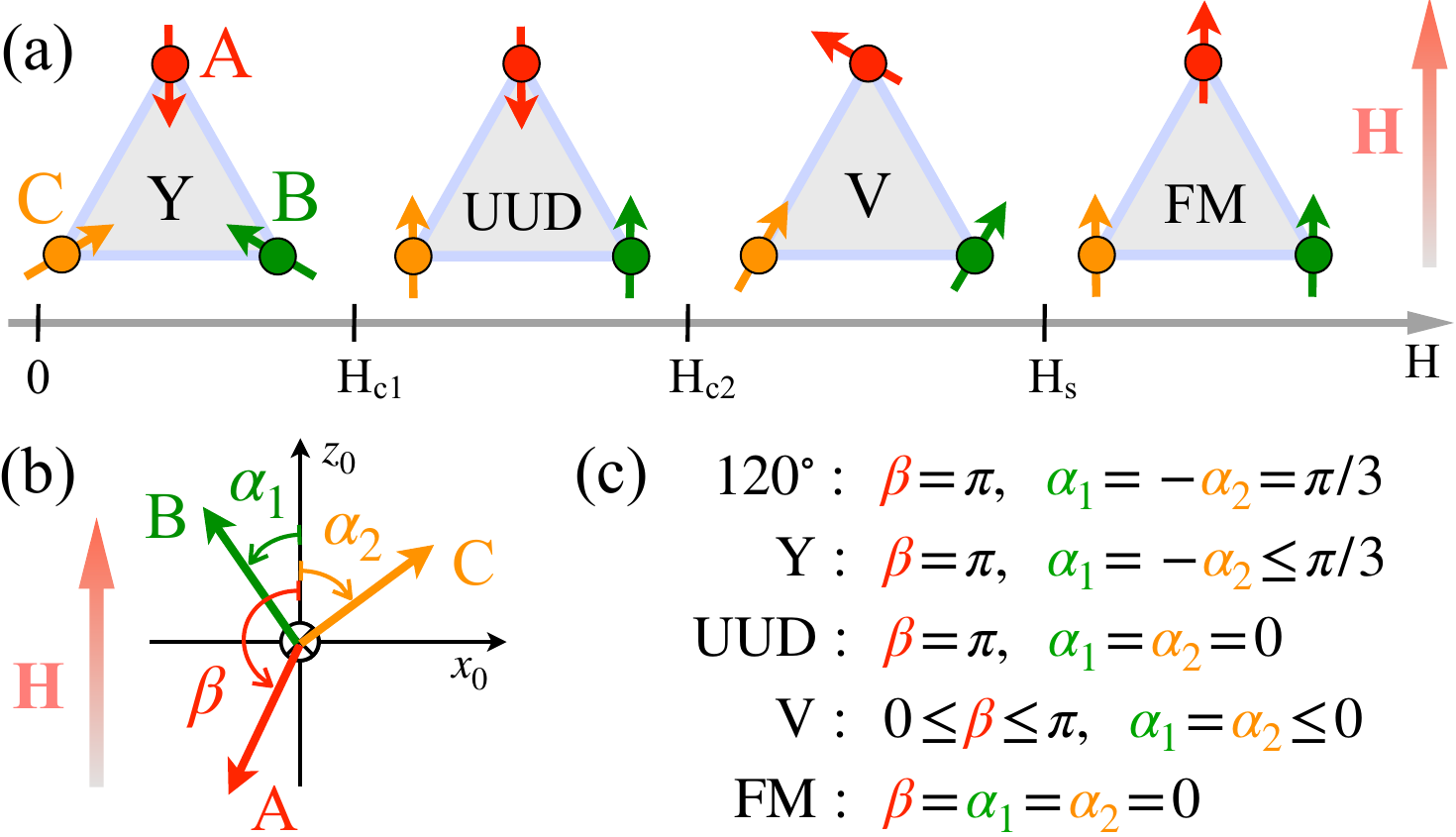}
\caption{(a) The schematics of the evolution of  magnetic order with the field 
from  120${\degree}$ state at $H\!=\!0$ to the Y phase, with the transition to the UUD 
plateau phase at $H_{c1}$, from the UUD phase to V phase
at $H_{c2}$, followed by a transition to the saturated FM phase at the saturation field $H_s$. 
The representative three-sublattice spin structures are shown.
(b) Angles of spins with the laboratory $z_0$ axis (field direction) for an 
arbitrary coplanar three-sublattice structure. Angles $\widetilde{\alpha}\!=\!\{\beta,\alpha_1,\alpha_2\}$
correspond to the $\alpha\!=\!\{{\rm A},{\rm B},{\rm C}\}$ sublattices. 
(c) Sets of $\widetilde{\alpha}$ for all  phases.}
\label{Fig_angles}
\end{figure}

For the model (\ref{H_1}), the field-evolution of the classical ground states 
is known from the earlier works \cite{Chubukov1991,Starykh2015,Sakikabara1} with the 
schematics of the evolution of  magnetic order with the field shown in Fig.~\ref{Fig_angles}(a).
At $H\!=\!0$,  spins assume a 120${\degree}$ configuration that was  confirmed for EuC$_6$ by the 
muon-spin spectroscopy \cite{Lamura2012}. A finite field continuously deforms it into the so-called 
Y-structure followed by a transition to the UUD (plateau) state at $H_{c1}$. The spin
angles and  the  field direction are shown in Fig.~\ref{Fig_angles}(b) and (c). 
The higher field induces a transition from the UUD phase to the 
V phase at $H_{c2}$ and to the fully  polarized FM phase at the saturation field $H_s$. 
It is worth noting that in all ordered phases, spin configurations are coplanar   
and belong to the three-sublattice structure with the same unit cell, see 
 Fig.~\ref{Fig_lattice} and Fig.~\ref{Fig_angles}.

In the earlier studies of EuC$_6$~\cite{Chen,Sakikabara1,Sakikabara2,Sakikabara3}, 
the minimal model (\ref{H_1}) was also augmented by the ring-exchange term
\begin{align}
\label{H_K}
&\mathcal{H}_K = K\sum_{\langle ijk\ell \rangle} \Big( Q_{ij}Q_{k\ell}+Q_{i\ell}Q_{jk}-Q_{ik}Q_{j\ell}\Big),
\end{align}
where  $Q_{ij}\!=\!\left(\mathbf{S}_i  \cdot\mathbf{S}_j\right)$ and spins belong to the elementary 
nearest-neighbor four-site  plaquettes $\langle ijk\ell \rangle$.

After some deliberation, one can write the classical energy of the model (\ref{H_1}) with the ring-exchange 
term (\ref{H_K}) for an arbitrary coplanar three-sublattice structure as
\begin{align}
&\frac{E_{cl}}{NS^2J_1} =  
(1-k)\big(\cos\widetilde{\alpha}_{AB}+\cos\widetilde{\alpha}_{AC}+\cos\widetilde{\alpha}_{BC}\big)
\nonumber\\
\label{E_cl}
&\quad\quad\ \
+3j_2-b\big(\cos^2\widetilde{\alpha}_{AB}+\cos^2\widetilde{\alpha}_{AC}+\cos^2\widetilde{\alpha}_{BC}\big) 
\\
&-h\big(\cos\widetilde{\alpha}_{A}+
\cos\widetilde{\alpha}_{B}+\cos\widetilde{\alpha}_{C}\big)
+2k\big(\cos\widetilde{\alpha}_{AB}\cos\widetilde{\alpha}_{AC}
\nonumber\\
&\phantom{-h\big(\cos\widetilde{\alpha}_{A}+}
\ \ \ \ +\cos\widetilde{\alpha}_{AB}\cos\widetilde{\alpha}_{BC}+
\cos\widetilde{\alpha}_{AC}\cos\widetilde{\alpha}_{BC}\big),
\nonumber
\end{align}
where $N$ is the number of sites in the triangular lattice, the
dimensionless field and exchange parameters are in units of the nearest-neighbor exchange 
$J_1$, $h\!=\!g\mu_BH/3J_1S$, $j_2\!=\!J_2/J_1$,  $b\!=\!BS^2/J_1$, and $k\!=\!KS^2/J_1$,
spin angles with the field direction $\widetilde{\alpha}_\alpha\!=\!\{\beta,\alpha_1,\alpha_2\}$
correspond to the $\alpha\!=\!\{{\rm A},{\rm B},{\rm C}\}$ sublattices according to 
Fig.~\ref{Fig_angles}(b) and (c), and mutual angles of spins  are
$\widetilde{\alpha}_{AB}\!=\!\beta-\alpha_1$, $\widetilde{\alpha}_{AC}\!=\!\beta-\alpha_2$, 
and $\widetilde{\alpha}_{BC}\!=\!\alpha_1-\alpha_2$.

\subsubsection{Tilt angles and critical fields}
\label{Sec_angles}

Energy minimization in (\ref{E_cl}) at a fixed field with respect to spin angles  should produce both the 
equilibrium spin configurations and  critical fields for the transitions between phases. 
Figure~\ref{Fig_angles} shows that in Y and V phases  spin angles depend on 
the field continuously, while spins are (anti)collinear with the field for the full extent of the UUD and FM phases.
In the Y and V phases, the general form of the classical energy in (\ref{E_cl}) simplifies, 
with the energy of the Y phase controlled by one  independent angle and for the V phase by two angles.

For the Y phase, a straightforward algebra  gives an equation  for the angle $\alpha_1$
\begin{equation}
\label{E_clY}
\frac{\partial E^{\rm Y}_{cl}}{\partial \alpha_1} = 0= (1+b)\,x-a_0-6k \,x^2-4b \, x^3\,,
\end{equation}
where $x\!=\!\cos \alpha_1$ and $a_0\!=\!(1+h-3k)/2$. 
Since  cubic equation allows for analytical 
solutions \cite{Wolfram_cubic}, the angles of the spin configuration within the Y phase 
are fully determined by such a solution of (\ref{E_clY}).  In the spin-wave treatment of the 
model (\ref{H_1}) presented below, the equilibrium spin configuration in the Y phase
is obtained from a $k\!=\!0$ version of (\ref{E_clY}).
 
As was first noted in Ref.~\cite{Sakikabara1}, 
the  evolution of $\alpha_1$ with $H$ becomes discontinuous and  transition to the UUD phase turns
first-order at larger values of $B\!>\!0$ and $K\!>\!0$. However, leaving this  detail aside for a moment, one can 
always find a solution for a transition field between the Y and UUD phases by 
{\it assuming}  it to be continuous and putting $\cos \alpha_1\!=\!1$ in (\ref{E_clY}), 
which yields 
\begin{equation}
\label{hc1}
h_{c1}=1-6b-9k\,,
\end{equation}
in agreement with Ref.~\cite{Sakikabara1}. 

The meaning of this critical field is twofold. It is the true critical field for a phase transition at the smaller
values of $B$ and $K$ where it is continuous. 
In what follows, we focus on $K\!=\!0$, ``$B$-only'' model (\ref{H_1}), for which
 continuous transition can be shown to exists up to $b_c\!=\!1/11$. For the values of $b\!>\!b_c$,
the Y phase is stable up to the \emph{higher}  critical field 
\begin{equation}
\label{hc1bc}
\widetilde{h}_{c1}=\sqrt{\frac{4(1 + b)^3}{27b}}-1\,
\end{equation}
at which the angle changes discontinuously. 
However, the critical field in (\ref{hc1}) continues to define the region of 
$\widetilde{h}_{c1}\!>\!h\!>\!h_{c1}$ where the plateau phase is 
(meta)stable, meaning that the spin excitations defined within the UUD phase 
are stable down to $h_{c1}$ in (\ref{hc1}). A detailed consideration of the 
critical fields associated with the first-order transitions is  provided in Appendix~\ref{app:1st}.

Somewhat fortuitously, our choice of parameters for EuC$_6$ discussed below  corresponds 
to $b$ only very slightly larger than $b_c$, so the transitions that we find are very marginally first-order.
Experiments in EuC$_6$ \cite{Chen} have also indicated small hysteresis effects in magnetoresistance
\cite{Chen}, suggesting a correspondence between the two.

For the V phase, energy minimization in (\ref{E_cl}) yields the following equations in the angles,
$\sin\beta\!=\!2\sin\alpha_1$ and
\begin{equation}
\label{E_clV}
h\,\sin\beta= 2\sin\gamma\, \big(1+k+2(k-b)\cos\gamma\big)\,,
\end{equation}
where $\gamma\!=\!\alpha_1+\beta$. For the ``$B$-only'' model (\ref{H_1}) that we focus on below,
one can simplify (\ref{E_clV}) to the equation for $\beta$ in the form $h\!=\!F(\cos\beta, b)$, with  
\begin{equation}
\label{E_clVa}
F(x,b)= \big(x+\sqrt{x^2+3}\big)\big(1-b\big(x\sqrt{x^2+3}-1+x^2\big)\big)\,,
\end{equation}
which can be solved numerically to find $\alpha_1$ and $\beta$ angles of the 
equilibrium spin configuration in the V phase.

An approach to the transitions from the UUD to V and from V to the FM phases 
by assuming their continuity and (anti)collinearity of the spins  in (\ref{E_clV})  yields 
\begin{equation}
\label{hc2hs}
h_{c2}=1+2b-k\,, \ \  \ h_{s}=3\big(1-2b+3k\big)\,,
\end{equation}
also in agreement with Ref.~\cite{Sakikabara1}. While a transition at $h_{c2}$ 
remains continuous for a wide range of parameters, transition to the saturated 
phase for the ``$B$-only'' model (\ref{H_1})  turns first-order at the same 
$b_c\!=\!1/11$ as the Y-to-UUD transition at $h_{c1}$ discussed above, showing a similar phenomenology.
Given that the range of parameters discussed below is only weakly affected by the 
associated discontinuities, we will carry on referring to $h_{c1}$, $h_{c2}$, and $h_{s}$ in 
Eq.~(\ref{hc1}) and Eq.~(\ref{hc2hs}) as to the ``true'' critical fields, see Appendix~\ref{app:1st} for more detail.

\subsubsection{Parameters}
\label{Sec_parameters}

It is useful to consider pure Heisenberg limit of the model (\ref{H_1}) as a reference.
In that case, the dimensionless critical fields $h^0_{c1}\!=\!h^0_{c2}\!=\!1$ and $h^0_{s}\!=\!3$, 
all in units of $3J_1S/g\mu_B$. Thus, as one can see from (\ref{hc1}) and 
(\ref{hc2hs}),   for $B,K\!>\!0$ the biquadratic and ring-exchange terms necessarily open up a 
finite range of fields for the plateau phase. However,  while both terms  drive down $h_{c1}$ from 
its $h^0_{c1}$ value, their effects on $h_{c2}$ and $h_{s}$ are opposite to each other. 
Most importantly, if the additional terms are dominated by the biquadratic one, the critical
fields  $h_{c1}$ and $h_{c2}$ split away from their Heisenberg value in the opposite directions,
with $h_{c1}$ below and $h_{c2}$ above $h^0_{c1}$.  If, however,  the ring-exchange term is the 
leading one, both $h_{c1}$  and $h_{c2}$ shift down from $h^0_{c1}$. 

This observation has a direct impact on the analysis of the phenomenology of EuC$_6$ and 
parameters of the model that follow from it. 
A summary of the  experimental data that is relevant to such an analysis can be found in 
Ref.~\cite{Chen}.  Eu$^{2+}$ spins order antiferromagnetically at $T_N\!\approx\!40$~K, 
with the 120${\degree}$ structure of their zero-field ground state confirmed more recently \cite{Lamura2012}. 
The critical fields of all the transitions discussed above can be inferred directly from the
$T\!\rightarrow\!0$ extrapolations of the associated anomalies in the resistivity data in
Fig.~\ref{Fig_EuC6}(b), which is reproduced from Ref.~\cite{Chen}.
Thus, the experimental value of the saturation field is $H_s^{\rm exp}\!\approx\!21.5$~T, 
while the Y-to-UUD and UUD-to-V transitions are at 
$H_{c1}^{\rm exp}\!\approx\!1.6$~T and $H_{c2}^{\rm exp}\!\approx\!9.0$~T, respectively, see also
Table~\ref{Table1}. 

Given Eq.~(\ref{hc1}) and Eq.~(\ref{hc2hs}), the experimental values of the three critical fields 
are sufficient to uniquely determine three parameters of the model, $J_1$, $B$, and $K$.
In broad strokes, an overall energy scale  dictated by $J_1$ sets an extent of the ordered phases 
that is determined from  the saturation field $H_s$, while the width of the 
plateau between  $H_{c1}$ and $H_{c2}$ and their relation to $H_s/3$ fixes $B$ and $K$. 
The results are listed in the first line of Table~\ref{Table1} where we have also used the 
Lande g-factor g=1.94 \cite{Chen}. 

In agreement with the prior estimates \cite{Chen} and general expectations, the biquadratic and ring-exchange 
terms are much smaller than the leading exchanges, yet they are essential for the existence of the 
unconventional UUD phase. Importantly, the ring-exchange is subleading to the 
biquadratic term with the ratio $B/K\!\approx\!3$. Provided our discussion above, 
the dominance of $B$ over $K$ is clear already from the fact that the UUD-to-V critical field 
$H_{c2}^{\rm exp}$ is substantially larger than $H_{s}^{\rm exp}/3$. 

It is, therefore, rather puzzling to find almost exactly  opposite hierarchy of $B$ and $K$ in 
Refs.~\cite{Chen,Sakikabara1,Sakikabara2,Sakikabara3}, based on the same data for EuC$_6$.
The reason for the difference is the following. With the rest of the phenomenological
constraints being the same, the UUD-to-V critical field in Refs.~\cite{Chen,Sakikabara1} is chosen as 
$\widetilde{H}_{c2}^{\rm exp}\!\approx\!6.4$~T, which is  less than 
$H_{s}^{\rm exp}/3$, hence implying the dominance of $K$ over $B$.
The smaller critical field is inferred from a rather broad magnetization data, which, 
given the second-order nature of the UUD-to-V transition, is strongly affected by the finite-temperature 
effects, see also  Ref.~\cite{Tanaka} on a different material highlighting the same effect. 
It is difficult for us to understand why the lower $\widetilde{H}_{c2}^{\rm exp}$ 
was insisted upon  in the prior works, except for the premeditated importance of the ring-exchange terms.

\begin{table}[t]
\begin{tabular}{ l  c  c  c  c    c  c  c }
\hline \hline
     & \ $J_1$  \ & \ $J_2$  \ & \ $BS^2$  \ &\  $KS^2$ & \ \ \ \ \ \  $H_{c1}$  \ & \ $H_{c2}$  \ &\  $H_{s}$  \ \\ \hline
Exp. \  &\  0.974 \ & -0.783 & 0.086 & 0.029 & \ \ \ \ \ \  1.6 & 9.0 & 21.5    \\ 
Model \ &\ 1.085 \ & -0.728 & 0.1 & 0  & \ \ \ \ \ \  3.91 & 10.35 & 21.39  \\ \hline  \hline
\end{tabular}
\caption{Exchange parameters (K) and critical fields (T), $S\!=\!7/2$. Exp.: experimental 
values of the fields define parameters of the model as described in text. 
Model: chosen parameters of the model (\ref{H_1}) with the resultant critical fields.}
\label{Table1}
\end{table}

The remaining parameter of the model (\ref{H_1}) is the second-neighbor exchange $J_2$,
which is necessary to reconcile the value of the ordering temperature, $T_N$, with that 
of the saturation field, as the two are not fully compatible for the model that contains only 
the nearest-neighbor exchanges. Since the leading mechanism that provides spin couplings in 
EuC$_6$ is believed to be of the RKKY-type \cite{Chen},   the $J_2$-term with $J_2\!<\!0$ is seen as natural. 

Another element of the consideration that is easy to justify is the use of the mean-field 
approximation for the ordering temperature despite the quasi-2D character of EuC$_6$ and 
continuous symmetries of the model (\ref{H_1}). The large spin value $S\!=\!7/2$, aforementioned 
$XXZ$ anisotropy, and the presence of small interplane couplings \cite{Chen} that are ignored 
in our model, all give  strong ground for the use of the mean-field approach \cite{Gingras}
\begin{equation}
\label{T_Nmf}
T_N^{\rm MF}=-\frac{S(S+1)}{3 k_B} \ \lambda_{\rm min} (\mathbf{Q}),
\end{equation}
where $\lambda_{\rm min} (\mathbf{Q})$ is the lowest eigenvalue of the Fourier transform of the 
exchange matrix in (\ref{H_1}) at the ordering vector $\mathbf{Q}$. For the three-sublattice orders, 
$\mathbf{Q}\!= \!(\pm 4\pi/3,0)$ 
and $\lambda_{\rm min} (\mathbf{Q})$ can  also be inferred from the classical energy
in (\ref{E_cl}) as  $\lambda_{\rm min} (\mathbf{Q})\!=\!2E_{cl}^{120\degree}/NS^2$ to yield
\begin{equation}
\label{T_Nmf1}
T_N^{\rm MF}\approx S(S+1) \big(J_1-2J_2\big),
\end{equation}
where  contributions of  small interplane couplings are ignored and 
we have also dropped even smaller and nearly canceling contributions from the $B$ and $K$ terms.
Since $J_1$ is already determined from the critical fields, $T_N^{\rm exp}\!=\!T_N^{\rm MF}$ in 
Eq.~(\ref{T_Nmf1}) gives $J_2$ in the first line of Table~\ref{Table1}.

We note that the experimental constraint on the parameters that is alternative to $T_N^{\rm exp}$ 
could have been the Curie-Weiss temperature, $T_{\rm CW}$. However, the value of $T_{\rm CW}\!=\!+1.3$~K
reported in Ref.~\cite{Sakikabara3} has a ferromagnetic sign,  contradicting to 
all other evidences including $\mu$SR spectroscopy \cite{Lamura2012} that the  $H\!=\!0$ state
of  EuC$_6$ is a 120$\degree$ state. This discrepancy is likely due to the uniform susceptibility 
data taken at a too high value of the field of 1~T \cite{Sakikabara3} that is already close 
to the ferrimagnetic plateau state. In addition, the mean-field value of $T_{\rm CW}$ is proportional to the 
sum of $J_1$ and $J_2$ exchanges, which are of opposite sign and have close values, amplifying 
the errors in the estimates of the individual exchange parameters. Lastly, for the antiferromagnetic state 
it is much more natural to connect to the susceptibility at the corresponding ordering vector, Eq.~(\ref{T_Nmf}), 
 not the uniform one.

Having established the secondary role of the ring-exchange term in EuC$_6$ phenomenology,
we are going to completely ignore such a term  in the model consideration of the scattering of electrons 
by spin excitations presented next. This step is motivated by both  strong
similarity of the effects provided by the ring-exchange to that of the biquadratic terms and 
a considerable cumbersomness of the spin-wave treatment of the ring-exchange in the triangular lattice, 
see Refs.~\cite{Kubo,Yunoki}.

With the number of model parameters reduced, there are more phenomenological constraints 
than there parameters. Fixing one of $H_{c2}$ or $H_{c1}$ to their experimental 
value either  narrows or widens the extent of the plateau by about $4$~T 
compared to the data, with $BS^2$ being 0.06K and 0.17K, respectively. Instead, 
we fix $BS^2$ to an intermediate value of 0.1K, which leads to only a slightly narrower 
plateau and somewhat higher critical fields than in experiment, 
$H^{th}_{c1}\!\approx\!3.9$~T and $H^{th}_{c2}\!\approx\!10.4$~T, 
see Table~\ref{Table1} for a full set of the model parameters. 
This is the set of parameters that will be used henceforth in all calculations of  the magnetoresistivity. 
It corresponds to the dimensionless parameters $j_2\!=\!J_2/J_1\!=\!-0.671$ and  
$b\!=\!BS^2/J_1\!=\!0.0922$.
For the representative pictures of the spin-wave spectra shown in Sec.~\ref{Sec_plots} below we choose a 
close set of $j_2\!=\!-0.8$  and  $b\!=\!0.1$. 

\section{Spin excitations}
\label{Sec_SWT}

In this Section, a general spin-wave approach is formulated for all coplanar three-sublattice  
states in Fig.~\ref{Fig_angles}. In Appendix~\ref{Sec_particular}, we provide  a consideration of the FM,
120${\degree}$,  and  UUD states for which a simplified approach is possible, allowing  to 
obtain  fully analytical results. 

We would like to note that the biquadratic exchange has been widely employed to emulate quantum effects 
in a variety of spin  models, including  Heisenberg and $XXZ$ triangular-lattice models
to stabilize their plateau state. However, we are not aware of the spin-wave theory consideration of the 
model (\ref{H_1}) in the literature, 
with an exception of the early work \cite{Chubukov1991}, which provided a  consideration
of the zone-center, ${\bf k}\!=\!0$, modes.
The possibility of a consistent spin-wave expansion for an arbitrary coplanar three-sublattice structure
presented next was motivated, in part, by a general formalism in Ref.~\cite{Coleta}.

\subsection{General case of a coplanar state}
\label{Sec_general}

For a spin-wave expansion,  the laboratory reference frame  $\{x_0,y_0,z_0\}$ in 
Fig.~\ref{Fig_lattice} and Fig.~\ref{Fig_angles} needs to be rotated to the 
{\it local} reference frame  $\{x,y,z\}$ on each site, so that  the $z$ axis is 
along the direction dictated by a classical spin configuration obtained in Sec.~\ref{Sec_classical}.
For the coplanar states in  Fig.~\ref{Fig_angles}, such a transformation is  a simple rotation 
in the $x_0$--$z_0$  plane, such that $S_\alpha^{y_0}\!=\!S_\alpha^y$ and 
\begin{align}
\label{Sxyz}
&S^{x_0}_\alpha=S^x_\alpha\cos\widetilde{\alpha} -S^z_\alpha\sin\widetilde{\alpha}, \\ 
&S^{z_0}_\alpha=S^z_\alpha\cos\widetilde{\alpha} +S^x_\alpha\sin\widetilde{\alpha}, \nonumber
\end{align}
where $\alpha$  and $\widetilde{\alpha}$ are, respectively, the sublattices and corresponding  spin angles
in Fig.~\ref{Fig_angles}(b).

\subsection{$1/S$-expansion}
\label{Sec_1_S_expansion}

Consider the $1/S$-expansion of each individual term in the model (\ref{H_1}) separately.
For the nearest-neighbor $J_1$ term, it is convenient  to rewrite it first as 
\begin{align}
\label{H_J1}
\mathcal{H}_{J_1} =J_1\sum_{\langle ij \rangle_1}\left(\hat{h}_{ij}^{(e)}+\hat{h}_{ij}^{(o)}\right) ,
\end{align}
where the ``even'' (e) and  ``odd'' (o) parts 
\begin{align}
\label{h_eo}
\hat{h}_{ij}^{(e)}&=S_i^y S_j^y + \cos\widetilde{\alpha}_{ij} \left(S_i^x S_j^x+S_i^z S_j^z\right),\\
\hat{h}_{ij}^{(o)}&= \sin\widetilde{\alpha}_{ij} \left(S_i^z S_j^x-S_i^x S_j^z\right),\nonumber
\end{align}
are separated to distinguish their subsequent contribution of the even and odd powers of 
the bosonic operators to the $1/S$-expansion; here $\widetilde{\alpha}_{ij} \!=\!\widetilde{\alpha}_{i} 
-\widetilde{\alpha}_{j}$ are the angles between neighboring spins.

In the lowest orders, 
the even part yields a contribution to the classical energy and to the harmonic,
${\cal O}(S)$, linear spin-wave theory (LSWT) order of the expansion
 \begin{align}
\label{h_eLSWT}
\hat{h}_{ij}^{(e)}&\Rightarrow S^2 \cos\widetilde{\alpha}_{ij} + \hat{h}_{ij,LSWT}^{(e)}\, ,
\end{align}
while the odd part in (\ref{h_eo}) gives the linear order, ${\cal O}(S^{3/2})$, which must vanish 
upon a summation in (\ref{H_J1}) for the classical energy minimum, followed by the higher-order, 
${\cal O}(S^{1/2})$, anharmonic interactions that can be neglected for the large spin values. 

However, for the biquadratic term of the model (\ref{H_1}) 
\begin{equation}
\label{H_B}
\mathcal{H}_{B} =-B\sum_{\langle ij \rangle_1}\big(\mathbf{S}_i \cdot \mathbf{S}_j\big)^2
=-B\sum_{\langle ij \rangle_1}\left(\hat{h}_{ij}^{(e)}+\hat{h}_{ij}^{(o)}\right)^2 ,
\end{equation}
both even and odd parts play a role in its LSWT order
\begin{equation}
\label{H_B2}
\left(\mathbf{S}_i \cdot \mathbf{S}_j\right)^2\Rightarrow 
2S^2  \cos\widetilde{\alpha}_{ij} \, \hat{h}_{ij,LSWT}^{(e)}
+\Big(\hat{h}_{ij}^{(o)}\Big)^2_{LSWT} \, , \ \
\end{equation}
with their contributions, obtained from the standard Holstein-Primakoff bozonization of spins
in the rotated reference frame,
$S_i^z\!=\! S- a^\dagger_{i}a^{\phantom{\dag}}_{i}$ and 
$S_{i}^-\!=\!a_{i}^\dag\sqrt{2S}$, are 
\begin{align}
\label{h_LSWT}
&\hat{h}_{ij,LSWT}^{(e)}=-\frac{S}{2}\Big[\big(a_i^\dag -a_i\big)\big(a_j^\dag -a_j\big)
 \\ 
&\quad\quad
+\cos\widetilde{\alpha}_{ij}\left(2\big(a_i^\dag a_i+a^\dag_j a_j\big)- 
\big(a_i^\dag +a_i\big)\big(a_j^\dag +a_j\big)\right)\Big],
\nonumber\\
&\Big(\hat{h}_{ij}^{(o)}\Big)^2_{LSWT}= \frac{S^3}{2}
\sin^2\widetilde{\alpha}_{ij}\left(a_i^\dag +a_i - a_j^\dag -a_j\right)^2.\nonumber 
\end{align}

Of the remaining terms in model (\ref{H_1}), 
the Zeeman term is particularly simple, and so is the next-nearest-neighbor $J_2$ term, 
as the former involves only  local energy of bosons while the latter connects spins that belong 
to the same sublattices, giving, in the LSWT order,
\begin{align}
\label{H_H}
\mathcal{H}_{H} =& g\mu_B H\sum_{i}\cos\widetilde{\alpha}_i \, a_i^\dag a_i\, ,\\
\label{H_J2}
\mathcal{H}_{J_2} =& -J_2S\sum_{\langle ij \rangle_2}
\left(a_i^\dag a_i+a^\dag_j a_j-\big(a_i^\dag a_j+a^\dag_j a_i\big)\right) .
\end{align}
We point out, as a side remark, that it is relatively straightforward to modify  
model (\ref{H_1}) and the resultant LSWT Hamiltonian to include the effects of 
the easy-plane anisotropy that is present in EuC$_6$.
However, we did not find significant qualitative changes in the results for some of the key phases 
studied in this work. Given the extra cumbersomeness this anisotropy would introduce 
in the LSWT matrix below, we leave a detailed study of such an extension to a future work.

\subsection{LSWT Hamiltonian}
\label{Sec_LSWT_Hamiltonian}

The LSWT order of the model (\ref{H_1}), explicated in Eqs.~(\ref{H_J1})--(\ref{H_J2}),
is obtained for a general coplanar state. To make further progress, one needs to specify 
spin arrangement for the classical ground state.
In our case, all such states of interest can be represented as the three-sublattice states, 
highlighted in Fig.~\ref{Fig_angles}. Thus, a general approach to all of them can be pursued \cite{Coleta}.

The first step is to switch from the site notation $i$  to the one of the unit cells 
of the three-sublattice structure $\ell$ and sublattice index $\alpha$:
$i\!\rightarrow\!\{\alpha,\ell\}$. 
As a result, the Holstein-Primakoff boson operators 
are split into three species 
$a^{(\dag)}_{\alpha,\ell}\!=\!\{a_\ell^{(\dag)},b_\ell^{(\dag)},c_\ell^{(\dag)}\}$  corresponding to 
$\alpha\!=\!\{{\rm A},{\rm B},{\rm C}\}$ sublattices.
The Fourier transformation for them is
\begin{equation}
\label{FT3}
a^{\phantom{\dag}}_{\alpha,\ell} = \frac{1}{\sqrt{N_c}} \sum_{\bf q}  \,
a^{\phantom{\dag}}_{\alpha,\bf q} \, e^{-i{\bf q}\cdot{\bf r}_{\alpha,\ell}},
\end{equation}
where ${\bf r}_{\alpha,\ell}\! =\! {\bf R}_\ell\! +\! \bm{\rho}_\alpha$ and $N_c\!=\!N/3$ 
is the number of unit cells.
The  sublattice coordinates within the unit cell can be chosen as  
$\bm{\rho}_A\! =\! 0$, $\bm{\rho}_B\! =\! -\bm{\delta}_2$, and
$\bm{\rho}_C \!=\! \bm{\delta}_3$, see Fig.~\ref{Fig_lattice}.

After some algebra, using these boson species and their Fourier transforms, 
the LSWT Hamiltonian for an arbitrary coplanar three-sublattice state reads as
\begin{align}
\label{LSWT_H}
\hat{\cal H}^{(2)} =\frac{3J_1S}{2}\sum_{\bf q}\hat{\bf x}_{\bf q}^\dagger 
\hat{\bf H}_{\bf q}\hat{\bf x}_{\bf q}^{\phantom{\dagger}}\, ,
\end{align}
where $\hat{\bf x}^\dag_{\bf q}\!=\!\left( a^\dag_{\bf q}, b^\dag_{\bf q},c^\dag_{\bf q}, 
a^{\phantom \dag}_{-\bf q},b^{\phantom \dag}_{-\bf q},c^{\phantom \dag}_{-\bf q}\right)$ 
and $\hat{\bf H}_{\bf q}$ is a matrix
\begin{eqnarray}
\label{LSWTmatrix}
\hat{\bf H}_{\bf q}= 
\left( \begin{array}{cc} 
\hat{\bf A}^{\phantom \dagger}_{\bf q} &  \hat{\bf B}^{\phantom \dagger}_{\bf q}\\[0.5ex] 
\hat{\bf B}^\dagger_{\bf q}  & \hat{\bf A}^*_{-\bf q}
\end{array}\right), 
\end{eqnarray}
with the  $3\times 3$ matrices
$\hat{\bf A}^{\phantom \dag}_{\bf q}$ and $\hat{\bf B}^{\phantom \dag}_{\bf q}$  
\begin{eqnarray}
\label{ABany}
\hat{\bf A}^{\phantom \dag}_{\bf q}=
\left( \begin{array}{ccc} 
A_{\bf q} &  D_{\bf q} &  E^*_{\bf q}\\
D^*_{\bf q}  & B_{\bf q} &  F_{\bf q} \\
E_{\bf q}  & F^*_{\bf q} &  C_{\bf q}
\end{array}\right), \ \ \  
\hat{\bf B}^{\phantom \dag}_{\bf q}=
\left( \begin{array}{ccc} 
G &  J_{\bf q} & K^*_{\bf q} \\
J^*_{\bf q}  & H & L_{\bf q}\\
K_{\bf q}  & L^*_{\bf q} &  I
\end{array}\right). \ \ \ \ \ \ 
\end{eqnarray}
The elements of the $\hat{\bf A}^{\phantom \dag}_{\bf q}$ matrix are given by
\begin{align}
A_{\bf q}=&\, h\cos\beta-\cos\widetilde{\alpha}_{AB}-\cos\widetilde{\alpha}_{AC}
-2j_2\big(1-\gamma^{(2)}_{\bf q}\big)
\nonumber\\
&+b\big(1+3\left(\cos 2\widetilde{\alpha}_{AB}+\cos 2\widetilde{\alpha}_{AC}\right)/2\big)  ,\nonumber\\
B_{\bf q}=&\, h\cos\alpha_1-\cos\widetilde{\alpha}_{AB}-\cos\widetilde{\alpha}_{BC}
-2j_2\big(1-\gamma^{(2)}_{\bf q}\big)
\nonumber\\
&+b\big(1+3\left(\cos 2\widetilde{\alpha}_{AB}+\cos 2\widetilde{\alpha}_{BC}\right)/2\big)  ,\nonumber\\
C_{\bf q}=&\, h\cos\alpha_2-\cos\widetilde{\alpha}_{BC}-\cos\widetilde{\alpha}_{AC}
-2j_2\big(1-\gamma^{(2)}_{\bf q}\big)\nonumber\\
\label{ABCD_any}
&+b\big(1+3\left(\cos 2\widetilde{\alpha}_{BC}+\cos 2\widetilde{\alpha}_{AC}\right)/2\big)  , \\
D_{\bf q}=&\, \gamma_{\bf q}\big(1+2b(1-2\cos\widetilde{\alpha}_{AB})\big)\cos^2(\widetilde{\alpha}_{AB}/2),
\nonumber\\
E_{\bf q}=&\, \gamma_{\bf q}\big(1+2b(1-2\cos\widetilde{\alpha}_{AC})\big)\cos^2(\widetilde{\alpha}_{AC}/2),
\nonumber\\
F_{\bf q}=&\, \gamma_{\bf q}\big(1+2b(1-2\cos\widetilde{\alpha}_{BC})\big)\cos^2(\widetilde{\alpha}_{BC}/2),
\nonumber
\end{align}
and of the  $\hat{\bf B}^{\phantom \dag}_{\bf q}$ matrix, respectively,
\begin{align}
G=&  -b\big(\sin^2\widetilde{\alpha}_{AB}+\sin^2\widetilde{\alpha}_{AC}\big) ,\nonumber\\
H=&  -b\big(\sin^2\widetilde{\alpha}_{AB}+\sin^2\widetilde{\alpha}_{BC}\big) ,\nonumber\\
\label{GHI_any}
I=&  -b\big(\sin^2\widetilde{\alpha}_{BC}+\sin^2\widetilde{\alpha}_{AC}\big) ,\\
J_{\bf q}=&  -\gamma_{\bf q}\big(1-2b(1+2\cos\widetilde{\alpha}_{AB})\big)\sin^2(\widetilde{\alpha}_{AB}/2),
\nonumber\\
K_{\bf q}=&  -\gamma_{\bf q}\big(1-2b(1+2\cos\widetilde{\alpha}_{AC})\big)\sin^2(\widetilde{\alpha}_{AC}/2),
\nonumber\\
L_{\bf q}=&  -\gamma_{\bf q}\big(1-2b(1+2\cos\widetilde{\alpha}_{BC})\big)\sin^2(\widetilde{\alpha}_{BC}/2),
\nonumber
\end{align}
where  $h\!=\!g\mu_BH/3J_1S$, $j_2\!=\!J_2/J_1$, and $b\!=\!BS^2/J_1$ as before,
$\widetilde{\alpha}_{AB}\!=\!\beta-\alpha_1$, $\widetilde{\alpha}_{AC}\!=\!\beta-\alpha_2$,
$\widetilde{\alpha}_{BC}\!=\!\alpha_1-\alpha_2$, and 
\begin{align}
\label{gammas}
\gamma_{\bf q}=\frac{1}{3}\sum_{\alpha}  e^{i{\bf q} \cdot\bm{\delta}_\alpha},\ \ \ 
\gamma^{(2)}_{\bf q}=\frac{1}{3}\sum_{\alpha}  \cos {\bf q}{\cdot}\bm{a}_\alpha\, ,
\end{align}
with the first- and second-neighbor translation vectors
$\bm{\delta}_1 \!=\! (1,0)a$,
$\bm{\delta}_2 \!=\! (-1, \sqrt{3})a/2$, 
$\bm{\delta}_3 \!=\!- (1, \sqrt{3})a/2$, and 
$\bm{a}_1 \!=\! (3, -\sqrt{3})a/2$,
$\bm{a}_3 \!=\! (0, \sqrt{3})a$, 
$\bm{a}_2 \!=\! -(3, \sqrt{3})a/2$, respectively, see Fig.~\ref{Fig_lattice}; $a$ is the lattice constant. 

\subsection{Diagonalization}
\label{Sec_Diagonalization}

The eigenvalues of $\hat{\bf g} \hat{\bf H}_{\bf q}$ in (\ref{LSWTmatrix}), 
give magnon eigenenergies $\{\omega_{1{\bf q}},  \omega_{2{\bf q}}, \omega_{3{\bf q}},  
-\omega_{1-{\bf q}},  -\omega_{2-{\bf q}}, -\omega_{3-{\bf q}}\}$ (in units of $3J_1S$). Here $\hat{\bf g}$ is a
diagonal matrix $[1,1,1,-1,-1,-1]$, see Ref.~\cite{Colpa}.
While  magnon energies are crucial for our consideration of the spin-wave scattering 
of electrons that follows, an essential role is also played by the matrix elements, which are related 
to the $U$ and $V$ parameters of the generalized Bogolyubov transformation from the Holstein-Primakoff 
bosons to the ones of the quasiparticle eigenmodes
\begin{align}
\label{Bogo}
a^{\phantom{\dag}}_{\alpha,{\bf q}}=\sum_\gamma\left(U_{\alpha,{\bf q}}^{({\gamma})}\,
A^{\phantom{\dag}}_{\gamma,{\bf q}}+V_{\alpha,{\bf q}}^{({\gamma})}\,
A^{\dag}_{\gamma,-{\bf q}}\right),
\end{align}
with the  quasiparticle operators $A_{\gamma,{\bf q}}\!=\!\left\{A_{\bf q},B_{\bf q},C_{\bf q}\right\}$
and 
\begin{align}
\label{UVnorm}
\sum_\gamma\left(\big|U_{\alpha,{\bf q}}^{({\gamma})}\big|^2-
\big|V_{\alpha,{\bf q}}^{({\gamma})}\big|^2\right)=1\,.
\end{align}
The transformation (\ref{Bogo}) can be written in a matrix form 
\begin{eqnarray}
\label{UVtrans}
\hat{\bf x}_{\bf q}\!=\!\left(\! \begin{array}{c} 
\hat{\bf a}^{\phantom \dagger}_{\bf q} \\[0.5ex] 
\hat{\bf a}^\dagger_{-\bf q}  
\end{array}\!\right) \!=\! 
\left( \!\begin{array}{cc} 
\hat{\bf U}_{\bf q} &  \hat{\bf V}_{\bf q}\\[0.5ex] 
\hat{\bf V}^*_{-\bf q} & \hat{\bf U}^*_{-\bf q}
\end{array}\!\right)\!
\left( \!\begin{array}{c} 
\hat{\bf \cal A}^{\phantom \dagger}_{\bf q} \\[0.5ex] 
\hat{\bf \cal A}^\dagger_{-\bf q}  
\end{array}\!\right)\!= \hat{\bf S}_{\bf q} \cdot\hat{\bf z}_{\bf q}, \ \ \
\end{eqnarray}
where vectors $\hat{\bf a}_{\bf q}\!=\!\big[ a_{\bf q}, b_{\bf q},c_{\bf q}\big]^T$,  
$\hat{\bf a}^\dag_{-\bf q}\!=\!\big[ a^\dag_{-\bf q}, b^\dag_{-\bf q},c^\dag_{-\bf q}\big]^T$ and 
$\hat{\bf \cal A}_{\bf q}\!=\!\big[ A_{\bf q}, B_{\bf q},C_{\bf q}\big]^T$,
$\hat{\bf \cal A}^\dag_{-\bf q}\!=\!\big[ A^\dag_{-\bf q}, B^\dag_{-\bf q},C^\dag_{-\bf q}\big]^T$ are
introduced. It follows that the transformation matrix $\hat{\bf S}_{\bf q}$ diagonalizes  
$\hat{\bf g} \hat{\bf H}_{\bf q}$ in Eq.~(\ref{LSWTmatrix}), see Refs.~\cite{Colpa,Maldonado1993}.
Thus, the $U_{\alpha}^{({\gamma})}$ and $V_{\alpha}^{({\gamma})}$ parameters
can be extracted as the elements of the properly normalized eigenvectors of $\hat{\bf g} \hat{\bf H}_{\bf q}$
from a  diagonalization procedure.

\begin{figure}[t]
\includegraphics[width=\linewidth]{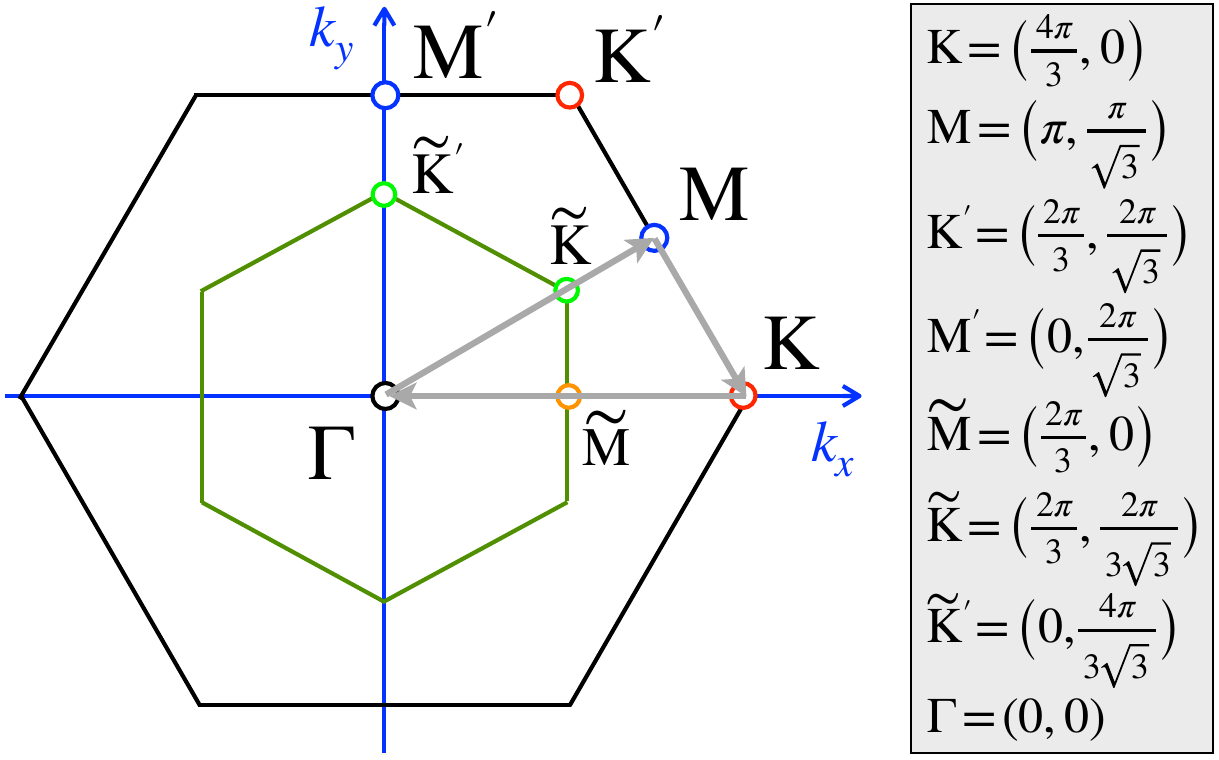}
\vskip -0.2cm
\caption{Full BZ of the triangular lattice (outer hexagon) and magnetic BZ 
of the three-sublattice structures (inner hexagon), high-symmetry points 
in units of inverse lattice spacing $1/a$, and the direction of a representative  K$\Gamma$MK cut.}
\label{Fig_BZs}
\vskip -0.3cm
\end{figure}

With all components of the  $\hat{\bf A}_{\bf q}$ and $\hat{\bf B}_{\bf q}$ matrices 
(\ref{ABany}) given explicitly in (\ref{ABCD_any}) and (\ref{GHI_any}), the $6\!\times\!6$
LSWT Hamiltonian (\ref{LSWTmatrix}) has to be diagonalized numerically. 
We have implemented such a procedure using MATHEMATICA.
In Sec.~\ref{Sec_plots}, we  provide plots of magnon energies throughout  
the Brillouin zone (BZ)  in Fig.~\ref{Fig_BZs}
for the representative field values from all the phases in Fig.~\ref{Fig_angles}(a).

We also point out that although the approach to the multi-flavor boson problem 
discussed here is very general,
there are significant simplifications in our case owing to the high symmetry of the model (\ref{H_1}) 
and the lattice. Specifically, $\hat{\bf A}^*_{-\bf q}\!=\!\hat{\bf A}_{\bf q}$ and 
$\hat{\bf B}^\dag_{\bf q}\!=\!\hat{\bf B}_{\bf q}$ in (\ref{LSWTmatrix}) as 
their off-diagonal matrix elements (\ref{ABCD_any}), (\ref{GHI_any}) are simply proportional to 
the   complex hopping amplitude $\gamma^*_{-\bf q}\!=\!\gamma_{\bf q}$ (\ref{gammas}).
As a result, all eigenenergies of $\hat{\bf g} \hat{\bf H}_{\bf q}$ 
are reciprocal, $\omega_{\gamma -{\bf q}}\!=\!\omega_{\gamma{\bf q}}$,
and $\{\hat{\bf V}^*_{-\bf q},\hat{\bf U}^*_{-\bf q}\}\!=\!\{\hat{\bf V}_{\bf q},\hat{\bf U}_{\bf q}\}$ 
in Eq.~(\ref{UVtrans}).

\subsection{Magnon eigenenergies}
\label{Sec_plots}

\begin{figure*}[t]
\includegraphics[width=\linewidth]{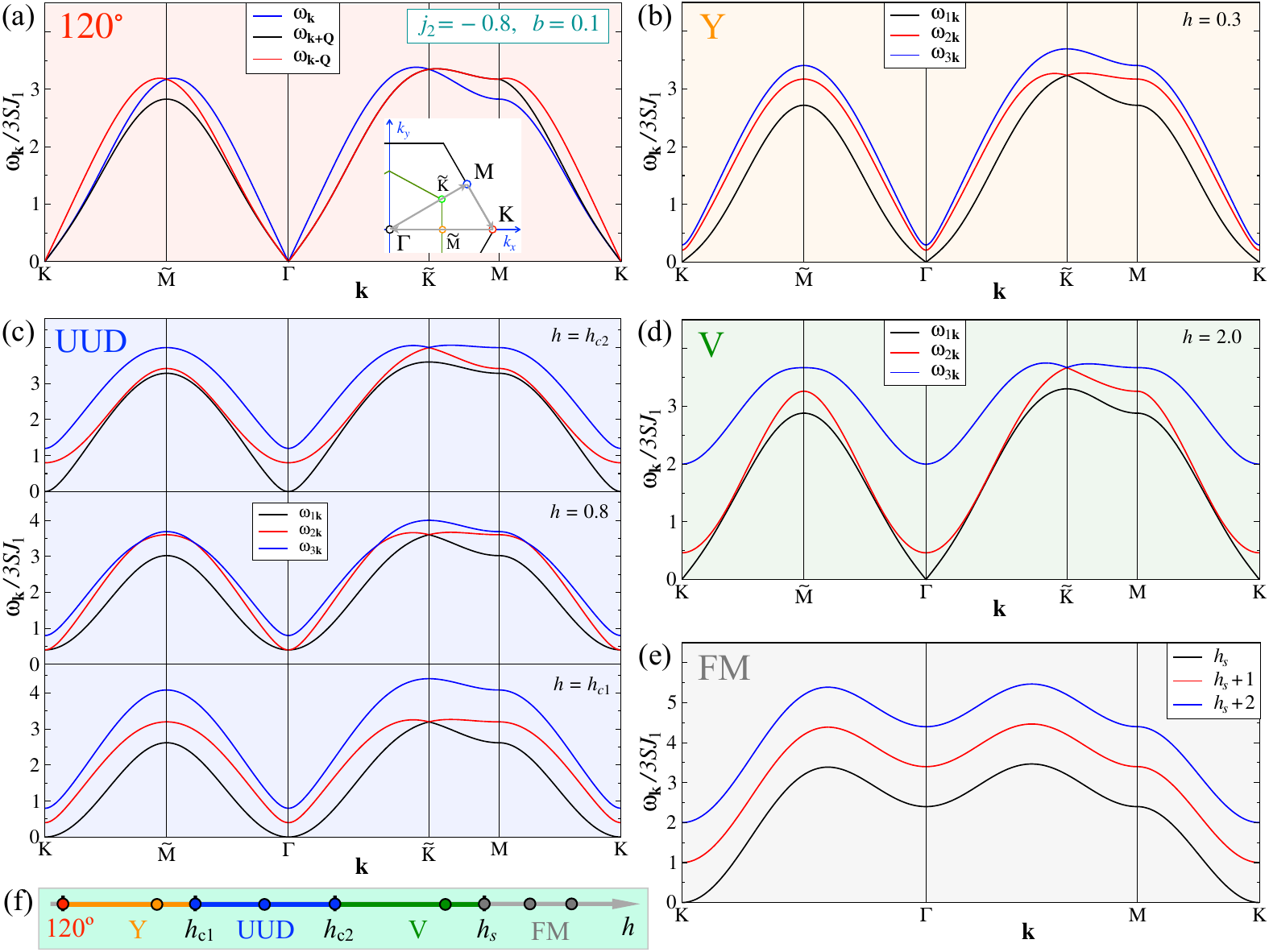}
\vskip -0.2cm
\caption{Magnon energies in units of $3SJ_1$
for several representative field values from the phases sketched in 
Fig.~\ref{Fig_angles}, $j_2\!=\!J_2/J_1\!=\!-0.8$ and $b\!=\!BS^2/J_1\!=\!0.1$.  
(a) 120$\degree$-phase, $h\!=\!0$, (b) Y phase, $h\!=\!0.3$, (c) UUD phase, $h\!=\!h_{c1}\!=\!0.4$,
$h\!=\!0.8$, and $h\!=\!h_{c2}\!=\!1.2$, (d) V phase, $h\!=\!2.0$, (e) FM phase
$h\!=\!h_{s}\!=\!2.4$, $h\!=\!h_{s}+1$, and $h\!=\!h_{s}+2$.
Transitions are at $h_{c1}\!=\!1-6b$, $h_{c2}\!=\!1+2b$, and $h_{s}\!=\!3(1-2b)$.
The 1D phase diagram vs field with representative field values is sketched in (f).}
\label{Fig_wks}
\vskip -0.3cm
\end{figure*}

In Fig.~\ref{Fig_wks}  we provide  plots of  magnon eigenenergies 
for several representative field values from all of the phases sketched in Fig.~\ref{Fig_angles}(a)  
and  for the parameters in model (\ref{H_1}) $j_2\!=\!J_2/J_1\!=\!-0.8$ and $b\!=\!BS^2/J_1\!=\!0.1$.
Energies are in units of $3SJ_1$ and dimensionless field is $h\!=\!g\mu_BH/3J_1S$. 
All plots are along a representative cut K$\Gamma$MK through the full Brillouin zone 
shown in Fig.~\ref{Fig_BZs}.

In zero field, $h\!=\!0$,   magnetic order is the canonical 120$\degree$ phase with the $O(3)$ symmetry
that is  spontaneously broken. Since it is  broken fully  by the choice of the ordering plane and 
by the spin arrangement within the plane, there are three Goldstone modes 
that one can observe in Fig.~\ref{Fig_wks}(a). 
As we discuss in some more detail in Appendix~\ref{Sec_particular} for the 120$\degree$ case, 
the three magnon branches that are defined  within the magnetic BZ can be related to a single 
branch defined within the full BZ using ``rotated'' reference frame for the spin quantization axes.
This allows to represent the full spectrum as the ``original'' branch, labeled by $\omega_{\bf k}$ in 
Fig.~\ref{Fig_wks}(a), and two  modes that are ``shifted''  by the ordering vector $\pm{\bf Q}\!=\!(4\pi/3,0)$.

For a finite in-plane field, the symmetry of the model (\ref{H_1}) 
is lowered to $U(1)$ by the field. Spontaneous breaking of the $U(1)$ symmetry within the 
Y phase in Fig.~\ref{Fig_angles}(a) at $h\!<\!h_{c1}$ results in a single magnon branch with a Goldstone mode
and two gapped branches, as is shown in Fig.~\ref{Fig_wks}(b) for $h\!=\!0.3$.
A characteristic feature of the gapless branch is an upward curvature 
of the dispersion in the long-wavelength limit, 
$\omega_{\bf k}\!\approx\! c |{\bf k}|+r |{\bf k}|^3$ with $r\!>\!0$.  

The UUD phase in Fig.~\ref{Fig_angles}(a) is sandwiched between two critical fields, $h_{c1}\!=\!1-6b\!=\!0.4$
and $h_{c2}\!=\!1+2b\!=\!1.2$. Since the $U(1)$ symmetry is preserved throughout this phase, the spectrum
is, generally, gapped except at the transition points, see Fig.~\ref{Fig_wks}(c) that shows  magnon 
spectra  at both $h_{c1}$ and $h_{c2}$ and at intermediate $h\!=\!0.8$.
The partially polarized, $U(1)$-preserving UUD state is, in a way, similar to the fully polarized 
FM phase, with the spectra for the latter for the fields at and above the saturation field  
$h_s\!=\!3(1-2b)$ shown in Fig.~\ref{Fig_wks}(e). 
Because of the continuous $U(1)$ symmetry of the model (\ref{H_1}),
magnetic field couples to a conserved total magnetization in both UUD and FM cases, which leads to the 
linear dependence of magnon energies in  Fig.~\ref{Fig_wks}(c) and (e)  on the field.  This also  makes the 
transitions at $h_s$ and $h_{c1(2)}$ analogous to  the Bose-Einstein condensation (BEC)  \cite{Batyev,Batista}. 
We note that the absolute  minima of $\omega_{\bf k}$  and the corresponding BEC condensation 
points in the FM case are at the ordering vectors of the three-sublattice order, 
$\pm {\bf Q}\!=\!\pm$K, not at the $\Gamma$ point. To emphasize this feature of the FM phase, the 
magnon energies in Fig.~\ref{Fig_wks}(e) are shown without folding on the magnetic BZ, see also 
Appendix~\ref{Sec_FM}.

The last phase of the model (\ref{H_1}) with a spontaneously broken $U(1)$ symmetry that is 
realized at $h_{c2}\!<\!h\!<\!h_{s}$ is the V phase, see Fig.~\ref{Fig_angles}(a).
Its spectrum is similar to that of the Y phase, having one concave Goldstone and two gapped modes, see 
Fig.~\ref{Fig_wks}(d). It can be seen as interpolating between the  spectra of the UUD phase 
at $h_{c2}$ and that of the FM phase at the saturation field.

\section{Kondo coupling and resistivity}
\label{Sec_Kondo_res}

In this Section we derive the electron-magnon interaction Hamiltonian, originating from the Kondo coupling, 
for a general case of a coplanar spin arrangement   and present the expression for the 
electronic transport relaxation rate due to such a scattering mechanism.

\subsection{Kondo coupling}
\label{Sec_Kondo}

The most reasonable minimal model for the interaction of  conduction electrons with the local spins in the 
magnetic layers of EuC$_6$ is the Kondo coupling
\begin{align}
\label{H_Kondo}
{\cal H}_{int}=J_K\sum_i {\bf s}_i\cdot {\bf S}_i,
\end{align}
where electron spin operators are 
$s^a_i\!=\!\frac12 f^\dag_{i,\alpha}\hat{\sigma}^a_{\alpha\beta}f^{\phantom{\dag}}_{i,\beta}$ with
$\hat{\bm \sigma}_i$ being Pauli matrices. With the external field providing spin quantization axis,
it is natural to split   (\ref{H_Kondo}) into spin-flip and non-spin-flip parts, 
${\cal H}_{int}\!=\!{\cal H}_{int}^{+-}+ {\cal H}_{int}^{zz}$,
\begin{align}
\label{H_Kondo_sf_nsf}
{\cal H}_{int}^{+-}=&\,\frac{J_K}{2}\sum_i \big(f^\dag_{i,\uparrow}f^{\phantom{\dag}}_{i,\downarrow} S_i^{-_0}+
f^\dag_{i,\downarrow}f^{\phantom{\dag}}_{i,\uparrow} S_i^{+_0}\big),\\
{\cal H}_{int}^{zz}=&\,\frac{J_K}{2}\sum_i \big(f^\dag_{i,\uparrow}f^{\phantom{\dag}}_{i,\uparrow} -
f^\dag_{i,\downarrow}f^{\phantom{\dag}}_{i,\downarrow} \big)S_i^{z_0},\nonumber
\end{align}
where $f^{(\dag)}_{i,\uparrow(\downarrow)}$ are  operators of the conduction electrons and 
$\{-_0,+_0,z_0\}$ indices in the operators of local spins refer to the ``laboratory'' reference frame 
$\{x_0,y_0,z_0\}$ associated with the field direction, see Fig.~\ref{Fig_lattice} and  Fig.~\ref{Fig_angles}. 

The Kondo coupling is a standard low-energy approximation, which describes interactions of localized spins 
with electrons at the Fermi surface. In our case, the corresponding electronic states respect the $C_3$ and 
sublattice symmetries of the graphene layer. In Fig.~\ref{Fig_bands}(a), we have demonstrated the folding of the 
two graphene-like bands, originally affiliated with the vicinities of the $K_{\rm gr}$ and $K^\prime_{\rm gr}$ points,
onto the proximity of the $\Gamma$ point in the Brillouin zone of the Eu lattice, see discussion in Sec.~\ref{sec_FS}. 
It is only these two bands that cross the Fermi surface and are encoded in our Fermi-operators $f_{i,\sigma}$ 
in (\ref{H_Kondo}). Therefore, by construction, our Kondo term accounts for the coupling to a linear combination 
of orbitals of the surrounding carbons that respects lattice symmetries mentioned above, but are also projected 
onto the electronic states that belong to these low-energy conduction bands. The couplings to the neighboring 
carbons also involves other electronic states, but they are unrelated to the states near the Fermi surfaces. Thus, 
the Kondo coupling in (\ref{H_Kondo}) provides the most reasonable minimal description of the interaction of 
local spins and conduction electrons. In addition, since we treat the two low-energy bands as independent, 
the coupling to local spins is treated as   diagonal in the band index in  (\ref{H_Kondo}).

For a general coplanar spin configuration in Fig.~\ref{Fig_angles}, the local axes are rotated from the 
laboratory ones (\ref{Sxyz}) to introduce quantized spin excitations for a given spin 
arrangement.   Consider the non-spin-flip part. Here,  according to (\ref{Sxyz}),
$S^{z_0}_i\!=\!S^z_i\cos\widetilde{\alpha} +S^x_i\sin\widetilde{\alpha}$, with
$\widetilde{\alpha}$ being the angle of spin's $z$-axis  on a site $i$ with $z_0$. 
Upon   quantization, $S^z_i$ converts into two-magnon term, while $S^x_i$
yields one-magnon emission/absorption. Similarly to the problem of electron-phonon scattering,
it is the lowest-order coupling that needs to be considered, unless it is forbidden for a 
symmetry reason or its scattering kinematics is suppressed. 
In our case, there are no such constraints and the $S^z$ part is also of the higher order in  
$1/S$ sense. Therefore, we approximate the local spin operators in (\ref{H_Kondo_sf_nsf})
by their single-magnon components 
\begin{align}
S_i^{+_0}\approx &\, 2\,\sqrt{\frac{S}{2}} \, \Big(\cos^2\left(\widetilde{\alpha}/2\right)  \, a^{\phantom{\dag}}_i-
\sin^2\left(\widetilde{\alpha}/2\right) \, a^{\dag}_i\Big),\nonumber\\
\label{single_magnon}
S_i^{z_0}\approx &\, \sqrt{\frac{S}{2}} \, \sin\widetilde{\alpha} \, \Big(a^{\phantom{\dag}}_i + a^{\dag}_i\Big),
\end{align}
where the angle $\widetilde{\alpha}$ depends on the sublattice.

Using Fourier transform (\ref{FT3})  in (\ref{H_Kondo_sf_nsf}) together with (\ref{single_magnon}),
one arrives at
\begin{align}
{\cal H}_{int}^{+-}=&\frac{2\widetilde{J}_K}{\sqrt{3N}}\sum_{{\bf k},{\bf q}} 
\Big[f^\dag_{{\bf k}-{\bf q}\uparrow}f^{\phantom{\dag}}_{{\bf k}\downarrow} 
\sum_\alpha\Big(\cos^2\left(\widetilde{\alpha}/2\right)  a^{\dag}_{\alpha,{\bf q}}\nonumber\\
&\phantom{\frac{2\widetilde{J}_K}{\sqrt{3N}}\sum_{{\bf k},{\bf q}}\Big[f^\dag_{{\bf k}-{\bf q}\uparrow}f}
-\sin^2\left(\widetilde{\alpha}/2\right) a^{\phantom{\dag}}_{\alpha,-{\bf q}} \Big) + {\rm H.c.}\Big],\nonumber\\
\label{H_Kondo_sf_nsf_FT}
{\cal H}_{int}^{zz}=&\,\frac{\widetilde{J}_K}{\sqrt{3N}}\sum_{{\bf k},{\bf q}} 
\Big[f^\dag_{{\bf k}-{\bf q}\uparrow}f^{\phantom{\dag}}_{{\bf k}\uparrow} -
f^\dag_{{\bf k}-{\bf q}\downarrow}f^{\phantom{\dag}}_{{\bf k}\downarrow} \Big]\\
&\phantom{\frac{\widetilde{J}_K}{\sqrt{3N}}\sum_{{\bf k},{\bf q}}\Big[f^\dag_{{\bf k}-{\bf q}\uparrow}}
\times
\sum_\alpha\sin\widetilde{\alpha}\,\Big(a^{\dag}_{\alpha,{\bf q}}+ a^{\phantom{\dag}}_{\alpha,-{\bf q}} \Big) ,
\nonumber
\end{align}
where $\widetilde{J}_K\!=\!\frac12 J_K\sqrt{S/2}$, $N$ is the total number of sites and summation 
in ${\bf k}$ and ${\bf q}$ is over the full Brillouin zone of the triangular lattice, Fig.~\ref{Fig_BZs}.
We note that the single-magnon non-spin-flip terms are nonzero  in the 120$\degree$, Y, and V phases,
in which the angles $\widetilde{\alpha}\!\neq\!\{0,\pi\}$, because in these states the symmetry of the 
Hamiltonian (\ref{H_1})  is broken completely and a spin-flip does not correspond to a particular spin value. 

The last transformation is to the quasiparticle operators given by Eq.~(\ref{Bogo}), which  yields
\begin{align}
{\cal H}_{int}^{+-}=&\frac{\widetilde{J}_K}{\sqrt{3N}}\sum_{{\bf k},{\bf q}} 
\Big[f^\dag_{{\bf k}-{\bf q}\uparrow}f^{\phantom{\dag}}_{{\bf k}\downarrow} 
\sum_\gamma\Big(M^{+-}_{\gamma,{\bf q}}  A^{\dag}_{\gamma,{\bf q}}\nonumber\\
&\phantom{\frac{\widetilde{J}_K}{\sqrt{3N}}\sum_{{\bf k},{\bf q}}
\Big[f^\dag_{{\bf k}-{\bf q}\uparrow}f^{\phantom{\dag}}_{{\bf k}\downarrow}}
+N^{+-}_{\gamma,{\bf q}}  A^{\phantom{\dag}}_{\gamma,-{\bf q}} \Big) + {\rm H.c.}\Big],\nonumber\\
\label{H_Kondo_qp}
{\cal H}_{int}^{zz}=&\,\frac{\widetilde{J}_K}{\sqrt{3N}}\sum_{{\bf k},{\bf q}} 
\Big[f^\dag_{{\bf k}-{\bf q}\uparrow}f^{\phantom{\dag}}_{{\bf k}\uparrow} -
f^\dag_{{\bf k}-{\bf q}\downarrow}f^{\phantom{\dag}}_{{\bf k}\downarrow} \Big]\\
&\phantom{\frac{\widetilde{J}_K}{\sqrt{3N}}\sum_{{\bf k},{\bf q}}\Big[f^\dag_{{\bf k}-{\bf q}\uparrow}}
\times
\sum_\gamma M^{zz}_{\gamma,{\bf q}}\Big(  A^{\dag}_{\gamma,{\bf q}} 
+  A^{\phantom{\dag}}_{\gamma,-{\bf q}} \Big),
\nonumber
\end{align}
with the matrix elements
\begin{align}
M^{+-}_{\gamma,{\bf q}} =&\,  2
\sum_\alpha\Big(\cos^2\left(\widetilde{\alpha}/2\right)  U_{\alpha,-{\bf q}}^{({\gamma})}
-\sin^2\left(\widetilde{\alpha}/2\right) V_{\alpha,-{\bf q}}^{({\gamma})}\Big),
\nonumber\\
N^{+-}_{\gamma,{\bf q}} =&\,  2
\sum_\alpha\Big(\cos^2\left(\widetilde{\alpha}/2\right)  V_{\alpha,-{\bf q}}^{({\gamma})}
-\sin^2\left(\widetilde{\alpha}/2\right) U_{\alpha,-{\bf q}}^{({\gamma})}\Big),\nonumber\\
\label{Kondo_Matrix_elements}
M^{zz}_{\gamma,{\bf q}} =&\,\sum_\alpha\sin\widetilde{\alpha}\,
\Big(U_{\alpha,-{\bf q}}^{({\gamma})}+V_{\alpha,-{\bf q}}^{({\gamma})}\Big).
\end{align}
One can see that while the structure of the non-spin-flip term in (\ref{H_Kondo_qp}) is similar to that 
of the electron-phonon scattering, the spin-flip part is different as the amplitudes of magnon emission 
and absorption by the same $\uparrow(\downarrow)$ electron are, generally, different. 
This is, of course, most obvious in the polarized FM state, in which magnons do have
a definite spin, and, therefore, can be emitted only by electrons with the spin   $\downarrow$ 
and absorbed only by electrons with the spin $\uparrow$.

With the electron-magnon couplings explicated in Eqs.~(\ref{H_Kondo_qp}) and 
(\ref{Kondo_Matrix_elements}), one has a clear path toward a calculation of the electron's relaxation 
rate and, therefore, resistivity as a function of the field and temperature. 

The derivation of the electron-magnon interaction above and the calculation of the relaxation rate  that follows 
can be repeated for individual particular cases of the 120$\degree$, FM, and UUD phases with
an alternative spin-wave formulation considered in Appendix~\ref{Sec_particular}. Each of these considerations
follows the same structure with a varying degree of simplification compared to the 
general case described above. While we do not expose these alternative solutions here as they lead to 
identical outcomes, they do offer an important verification and an analytical insight into the makeup of 
our solution.
 
\begin{figure}[t]
\includegraphics[width=\linewidth]{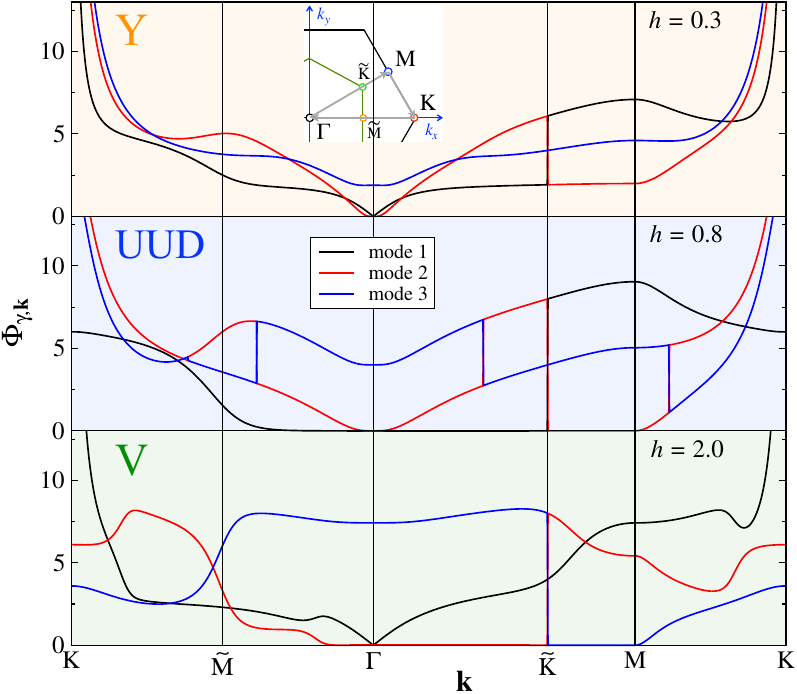}
\vskip -0.2cm
\caption{Combinations $\Phi_{\gamma,{\bf k}}$  (\ref{matrix_elements})  
of the matrix elements  in (\ref{Kondo_Matrix_elements})
for several representative field values along a representative cut K$\Gamma$MK through the full BZ, 
see Fig.~\ref{Fig_BZs}: Y phase, $h\!=\!0.3$,  UUD phase, $h\!=\!0.8$, and V phase, 
$h\!=\!2.0$.}
\label{Fig_matr_el}
\vskip -0.3cm
\end{figure}

We have repeatedly emphasized the importance of the field-induced changes in magnon energies 
and in  electron-magnon matrix elements for our key results that follow next. 
With the representative magnon energies shown in Fig.~\ref{Fig_wks}, we complement them 
with the similar representative plots of the combinations of  matrix elements given in (\ref{matrix_elements}),
which enter the integral expression for the resistivity, see Eq.~(\ref{1_tau_int}) below.
In Fig.~\ref{Fig_matr_el}  we show   combinations $\Phi_{\gamma,{\bf k}}$ from (\ref{matrix_elements}) 
for the three representative field values from different phases,  Y phase at $h\!=\!0.3$,  
UUD phase at $h\!=\!0.8$,  and  V phase at $h\!=\!2.0$. 
The plot is along a representative cut K$\Gamma$MK through the full BZ as in Fig.~\ref{Fig_wks}
and for the same parameter choices in model (\ref{H_1}) as above, 
$j_2\!=\!J_2/J_1\!=\!-0.8$ and $b\!=\!BS^2/J_1\!=\!0.1$.

Since the numeration of   magnon modes in Fig.~\ref{Fig_wks}(b)-(d) is from the lowest to 
highest in energy, the solutions for the matrix elements in Fig.~\ref{Fig_matr_el} switch between  
branches whenever  branches cross. Some of the crossings are at the high symmetry points
and some are not. A general trend that can be observed in Fig.~\ref{Fig_matr_el}  is that some of 
the matrix elements are strongly suppressed around the $\Gamma$ point and are either maximal 
or singular at the $K$ point.

An interesting feature of the matrix elements in the UUD phase  can be noted. There is 
no dependence of $\Phi_{\gamma,{\bf k}}$ on the field, only switching between  branches according
to their numeration. 
That is, while there is a definite reshuffling of the magnon modes vs field that can be seen in 
Fig.~\ref{Fig_wks}(c), the same combination of matrix elements as depicted in the middle panel of  
Fig.~\ref{Fig_matr_el} corresponds to any other point on the magnetization plateau (UUD) phase.
This observation provides an interesting connection between the structure of the quasiparticle states, 
encoded in their wave-functions, and conserved magnetization.
 
\begin{figure*}
\includegraphics[width=\linewidth]{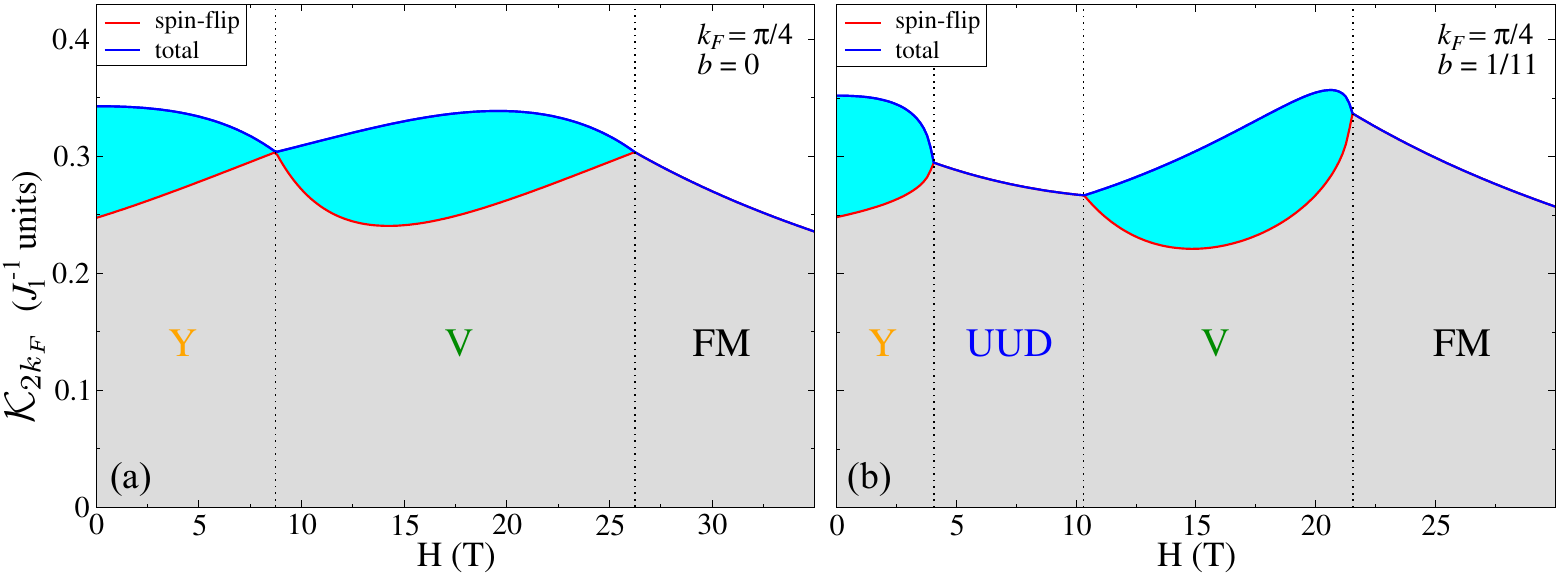}
\vskip -0.2cm
\caption{The kernel ${\cal K}_{2k_F}$, Eq.~(\ref{kernel}), vs $H$ for $k_F\!=\!\pi/4$ and for (a)  $b\!=\!0$, 
and (b)  $b\!=\!1/11$. Contributions of the spin-flip and non-spin-flip channels are represented by shadings,
Y, UUD, V, and FM phases and transitions between them are indicated.}
\label{Fig_kernel}
\vskip -0.3cm
\end{figure*}

\subsection{Resistivity}
\label{Sec_resistivity}

Similarly to  the theory of electron-phonon scattering in the resistivity of metals, 
Fermi energy $E_F$ is by far the most dominant energy scale of the problem, perhaps even 
more so in our case, as the magnon bandwidth, field strengths, and temperature range of interest 
are all  $\alt\!50$~K while $E_F\!\sim\!1$--3~eV \cite{Molodtsov1996}.
Because of that,   magnon-induced scattering of electrons is happening within a thin energy shell around 
the Fermi surface.
For the effectively 2D magnetic excitations,  transport scattering rate can be shown to reduce to a 
1D  integral that is limited by that shell.

With the technical details of the Boltzmann equation approach to the electron-magnon
scatterings in Eq.~(\ref{H_Kondo_qp})  delegated to  Appendix~\ref{App_transport}
and a mild assumption of the circular 2D Fermi surface, we obtain the transport 
relaxation rate for electrons with both spin projections 
\begin{align}
\label{1_tau}
\frac{\hbar}{\tau_{F}}=\frac{\sqrt{3}}{\pi}\,\frac{|\widetilde{J}_K|^2}{E_F}\,(k_F a)^2 \, I_{k_F}(T,H)\,,
\end{align}
where $\widetilde{J}_K\!=\!\frac12 J_K\sqrt{S/2}$, 
$E_F\!=\!\hbar^2k_F^2/2m$ with $m$ being effective electron mass, $k_F$ is a Fermi momentum, 
and the 1D integral given by 
\begin{align}
\label{1_tau_int}
I_{k_F}(T,H)=&\,\int_0^1 \frac{z^2\, dz}{\sqrt{1-z^2}}\\ 
&\,\phantom{\int_0^1}
\times\frac{1}{3}\sum_\gamma \Phi_{\gamma,{\bf q}} \, 
{\rm n}^0_{\gamma,{\bf q}}\big({\rm n}^0_{\gamma,{\bf q}}+1\big)\, 
\frac{\omega_{\gamma,{\bf q}}}{T}\,,\nonumber
\end{align}
with the 2D momentum parametrization along the 1D contour   
${\bf q}\!=\!2k_F(z^2,z\sqrt{1-z^2})$,  Bose distribution function ${\rm n}^0_{\gamma,{\bf q}}$ 
for a magnon with the energy $\omega_{\gamma,{\bf q}}$,  $\gamma$ numerating magnon branches, 
and $\Phi_{\gamma,{\bf q}}$ abbreviating matrix element contribution
\begin{align}
\label{matrix_elements}
\Phi_{\gamma,{\bf q}} =2\big|M^{zz}_{\gamma,{\bf q}}\big|^2+
\big|M^{+-}_{\gamma,{\bf q}}\big|^2+\big|N^{+-}_{\gamma,{\bf q}}\big|^2.
\end{align}
This result in (\ref{1_tau})-(\ref{matrix_elements})  combines the effort of the entire work in a concise 
form. It accumulates the solution of the transport theory that proves the validity of the $1/\tau$-approximation
in our case,  implicitly contains the field-dependence of the magnon spectra and  matrix elements 
(\ref{Kondo_Matrix_elements}) via the spin angles $\widetilde{\alpha}$ and parameters 
of the generalized Bogolyubov transformation (\ref{Bogo}), 
incorporates field-induced transitions between different phases, and  
includes the effect of thermal population of magnetic scatterers on the resistivity. 
As is discussed in the Section~\ref{Sec_results}, contributions of  the thermal distribution of magnons and  
matrix element component  (\ref{matrix_elements}) are both essential
for the resistivity results.
While the general expressions (\ref{1_tau})  and (\ref{1_tau_int}) may not be too intuitive, 
the following consideration will provide essential ingredients for such an intuition. 

\subsubsection{Large-${\bf q}$ insights}
\label{Sec_kernel}

The key elements of the physics packed in Eq.~\eqref{1_tau} can be extracted from the kernel of the integral 
in \eqref{1_tau_int}. We begin by noting that the $z^2$ factor in \eqref{1_tau_int} 
originates from a suppression of the small-angle scattering processes of electrons in the transport relaxation rate.
Thus, the integral is dominated by the large-${\bf q}$  scattering events that correspond to $z \!\rightarrow\! 1$ 
and $q\!\rightarrow\!2k_F$. The  $1/\sqrt{1-z^2}$ factor is due to angular integration in 2D
and also contributes to an enhancement of the large-${\bf q}$ contributions. 

Further intuition, which also lays out expectations for the results presented in the next Section, 
is provided by the remainder of the kernel  in the second line of \eqref{1_tau_int} taken at 
a ``typical'' momentum  ${\bf q}^*\!=\!(2k_F,0)$ and in the high-temperature limit,
approximating Bose-factors as ${\rm n}^0_{\gamma,{\bf q}} \!\approx\! T/{\omega_{\gamma,{\bf q}}}$ 
and omitting an overall prefactor $T/3$, 
\begin{equation}
{\cal K}_{2k_F} = \sum_\gamma \frac{\Phi_{\gamma,{\bf q}^*}}{\omega_{\gamma,{\bf q}^*}} .
\label{kernel}
\end{equation}
Referred to as ``the kernel'' below, ${\cal K}_{2k_F}$ is a sum over the branch index $\gamma$ 
of the ratios of the matrix elements  
(\ref{matrix_elements}) and magnon energies, taken at ${\bf q}^*$.
We also note that the high-temperature approximation is closely relevant to EuC$_6$ phenomenology 
discussed in Sec.~\ref{Sec_results}.
 
The kernel ${\cal K}_{2k_F}$ allows one to analyze contributions of different magnon modes
to the resistivity and also to compare the relative importance of the spin-flip and non-spin-flip channels
in the scattering. While the latter is absent in the collinear UUD and FM phases, it is present 
in the noncollinear Y and V phases, where $\sin\widetilde{\alpha} \!\neq\! 0$, see \eqref{Kondo_Matrix_elements}.
This partitioning of \eqref{kernel} into the channels is done by separating the non-spin-flip matrix element
contribution $|M^{zz}_{\gamma,{\bf q}^*}|^2$ in $\Phi_{\gamma,{\bf q}^*}$ from the rest, 
see \eqref{matrix_elements}.

Figure~\ref{Fig_kernel} shows ${\cal K}_{2k_F}$ \eqref{kernel} vs magnetic field
for a representative $k_F\!=\!\pi/4$ and for two values of the biquadratic parameter $b$ 
in model (\ref{H_1}): (a) Heisenberg, $b\!=\!0$, and (b) $b\!=\!b_c\!=\!1/11$.
The rest of the parameters are from Table~\ref{Table1}, see Sec.~\ref{Sec_parameters}.
In the Heisenberg limit, the UUD phase reduces to a critical point separating Y and V phases.
Contributions of the spin-flip and non-spin-flip scattering channels are shaded by different colors.

The key lesson of Figure~\ref{Fig_kernel} is that the non-spin-flip scattering channel, although 
secondary to the spin-flip one, is responsible for an enhancement of ${\cal K}_{2k_F}$ in the 
noncollinear Y and V  phases relative to the collinear UUD and FM phases where it is not available. 
Accordingly, one should expect the higher transport relaxation rates 
\eqref{1_tau} and higher resistivity in these noncollinear phases. Fig.~\ref{Fig_kernel} also shows that 
the biquadratic interaction enhances non-spin-flip scattering and causes stronger variations of the kernel 
near Y-UUD and V-FM transitions. 
One can anticipate all these trends to persist in the results for the resistivity discussed in Sec.~\ref{Sec_results}.

\subsubsection{$J_K$ estimate and other scatterings}
\label{Sec_J_K_estimate}

Using considerations of the electronic structure of EuC$_6$ provided in Sec.~\ref{sec_FS} and 
assuming two doubly-degenerate bands with cylindrical Fermi surfaces to describe it, the  
3D electronic concentration $n$ is related to the value of the Fermi momentum $k_F$  via
\begin{equation}
\label{n_to_ kF}
n=\frac{k_F^2}{\pi c}\,,
\end{equation}
where $c$ is the interplane distance between Eu layers.
With that and some rearranging, the expression for the resistivity can be cast in the following form
\begin{equation}
\label{rho_R_EF_tau}
\rho=\frac{m}{n e^2 \tau_F}=\frac{c}{4}\, R_K \cdot\frac{\hbar}{E_F\tau_F},
\end{equation}
in which the von Klitzing constant $R_K\!=\!h/e^2\!\approx\!25.8$~k$\Omega$ 
and interplane distance $c$ set the proper units and the relaxation rate of (\ref{1_tau}) is 
made dimensionless by a normalization to the Fermi energy.

One can  use the expression for $\rho$ in Eq.~(\ref{rho_R_EF_tau}) with $\hbar/\tau_F$ from 
(\ref{1_tau}) to estimate the Kondo coupling constant $J_K$ in (\ref{H_Kondo}) that is needed to 
reproduce  experimental values of $\rho$ in EuC$_6$. By taking 
$\rho_{H=0}(24{\rm K})\!\approx\!12.5$~$\mu\Omega$cm from Fig.~\ref{Fig_EuC6}(b) 
and $c\!=\!4.87$~\AA \cite{Sakikabara1}, one obtains $\hbar/E_F\tau_F\!\approx\!0.040$, 
which, if matched to the theory results for 
$k_Fa\!=\!\pi/3$ in Fig.~\ref{Fig_EuC6}(c), yields $J_K/E_F\!\approx\!0.275$. By 
scaling of the value of the Fermi energy in Fig.~\ref{Fig_bands}(a) to what it would be for  $k_Fa\!=\!\pi/3$,
one has $E_F\!\approx\!1.17$~eV and $J_K\!\approx\!0.32$~eV. This estimate is of the same order, 
albeit somewhat larger, than the value 0.15~eV quoted in the early literature \cite{Sugihara1985}. 
However, it seems that most of the discrepancy could be in the factor of two difference in the definition 
of electronic spin in the Kondo coupling (\ref{H_Kondo}), making the remaining difference rather academic.

Empirically, the resistivity of EuC$_6$ changes from about 2~$\mu\Omega$cm 
at  4~K,  to nearly 50~$\mu\Omega$cm at room temperature, see Fig.~\ref{Fig_EuC6}(b) and Ref.~\cite{Chen}.
While the low-temperature value  is in the same range as in the other graphite-intercalated compounds, see 
Ref.~\cite{Dresselhaus_review}, the room-temperature resistivity in EuC$_6$ exceeds that of the 
nearly isostructural non-magnetic GICs such as LiC$_6$ by more than an order of magnitude, 
clearly suggesting that it is the scattering on magnetic degrees of freedom, such as the ones considered in this 
work, that must be dominating the usual phonon and impurity scattering \cite{Chen}.

For the phonon scattering effect in the resistivity of EuC$_6$ for the temperature range relevant to our work, 
the phonon Debye temperatures in various GICs, estimated from the specific heat measurements, range 
from 300~K to 700~K \cite{Dresselhaus_review}, with the phonon spectrum of graphite \cite{graphite_phonons}
suggesting similar value. This implies that the phonon-induced resistivity  for $T\!\alt\!30$~K  should follow the 
strongly non-linear Bloch-Gr\"{u}neisen behavior, inconsistent with experiments. Moreover, as is discussed in 
Sec.~\ref{Sec_results} below, the magnon Debye temperature is about an order of magnitude smaller than that 
of phonons, making phonon contribution to scattering completely negligible in the relevant low-temperature 
regime even for an unphysically large electron-phonon coupling. Thus, this consideration suggest a strongly 
subleading role of  phonons in the resistivity of  EuC$_6$ at {\it all} temperatures. 

As is implied by a comparison of our theory results to experiments in Fig.~\ref{Fig_EuC6} and by the results in the 
Section~\ref{Sec_results} below, one can assume that the ``residual'' resistivity $\rho^{\rm res}$ 
of about 2~$\mu\Omega$cm at 4~K in Fig.~\ref{Fig_EuC6}(b) is mostly associated with impurity scattering. 
It is easy to infer from Eq.~(\ref{rho_R_EF_tau}) that this value corresponds to the mean-free path of 
\begin{equation}
\label{mean_free_path}
\Lambda_{\rm mfp}=\frac{c\, R_K}{2\, \rho^{\rm res}}\, k_F^{-1}\approx 300 \, k_F^{-1},
\end{equation}
which yields $\Lambda_{\rm mfp}\!\approx\!1.3\cdot 10^{-5}$~cm  
for $k_F\!=\!\pi/3a$ and $a\!=\!4.31$~\AA  \cite{Sakikabara1}. This is about a factor $10^2$ smaller than the 
mean-free path in pristine graphite, but is of the same order as in the other GICs \cite{Dresselhaus_review}, 
as the similarity of their residual resistivities has already indicated.

Although many types of defects may play an important role in GICs \cite{Dresselhaus2012}, 
one can model their effect as that of the screened Coulomb centers of charge-$e$ in order to infer an 
overall nominal impurity concentration. This is in accord with the textbook approach \cite{Ziman} to  the 
impurity-scattering in resistivity.
For that, we obtain a modified relation between the mean-free path and impurity concentration  
for the quasi-2D cylindrical Fermi-surface
\begin{equation}
\label{mean_free_path_imp}
\frac{1}{\Lambda_{\rm mfp}\, k_F}=2F(z^2)\, \, \frac{n_{\rm imp}}{n}, 
\end{equation}
with the angular averaging of scattering contained in 
\begin{equation}
\label{integral}
F(z^2)=\int_0^1\frac{x^2\,dx}{\sqrt{1-x^2}\,(1+z^2 x^2)^2}, 
\end{equation}
where $z\!=\!2k_F\lambda$, $\lambda$ is a screening length, and $n$ is the electronic concentration 
from Eq.~(\ref{n_to_ kF}).  For the doubly-degenerate cylindrical Fermi-surfaces of electrons that we use 
here as an approximation, one can find the screening parameter as $z^2 = E_F/(e^2/c)$,
where $c$ is the interplane distance.
Using \eqref{mean_free_path} and \eqref{mean_free_path_imp} 
for $E_F$ corresponding to $k_F\!=\!\pi/3a$ and lattice constants quoted above, we find
that $\rho^{\rm res} = 2$~$\mu\Omega$cm corresponds to $n_{\rm imp}/n\!\approx\!2.3\cdot 10^{-3}$.

Using the same  $k_F$ one can convert impurity concentration per electron to concentration per carbon 
to obtain $n^{\rm per\ C}_{\rm imp}\!\approx\!1.2\cdot 10^{-4}$, which is about 120~ppm. 
In graphite, solubility limits of most impurities 
are very low \cite{imp_concentrations,Kopelevich}, with the major residual impurities that reach the obtained
value often being that of Fe. This observation may provide additional ground for the scenario outlined below
in Sec.~\ref{Sec_discussion} in our discussion of the remaining problems, in which we suggest 
that   sizable variations in the residual resistivity versus field may be related to the magnetic
nature of the impurities and to the scattering due to spin-textures induced by them. 

\section{Results}
\label{Sec_results}

With all   the elements of our approach and  qualitative and quantitative considerations and estimates 
provided above, we can now offer a detailed overview of the results that follow from our theory.
A comparison of the experimental data for the magnetoresistivity in EuC$_6$ vs field with 
our calculations for the model parameters from Table~\ref{Table1} and for 
a representative value of $k_F \!=\! \pi/3$ is given in Figs.~\ref{Fig_EuC6}(b) and (c). 
Given the simplicity of our model and potential additional unaccounted effects discussed in more 
detail in Sec.~\ref{Sec_discussion}, the similarity between experiment and theory is rather astounding. 

This similarity includes high resistivity in the Y phase and its quick roll-down near the Y-UUD transition, 
a gentle downward slope of $\rho$ vs $H$ in the UUD phase, followed by a smooth rise in the V phase. 
The temperature evolution of the $\rho(H)$ curves is also consistent with the data, perhaps with an exception
of the lowest temperatures.
A discrepancy can also be seen in the larger values of $\rho$ in the FM phase and a strong rise toward it
near the V-FM transition in the theory results. This is likely due to a proximity 
to the $H$--$T$ phase transition boundary, where interactions between magnetic excitations, 
neglected in our consideration,  become important.

This successful comparison strongly suggests  the correctness of the advocated mechanism 
of the magnetoresistance in magnetically intercalated graphite as dominated by electron scattering 
on magnetic excitations, which, in turn, allows insights into the nature of such excitations. 

\begin{figure*}
\includegraphics[width=\linewidth]{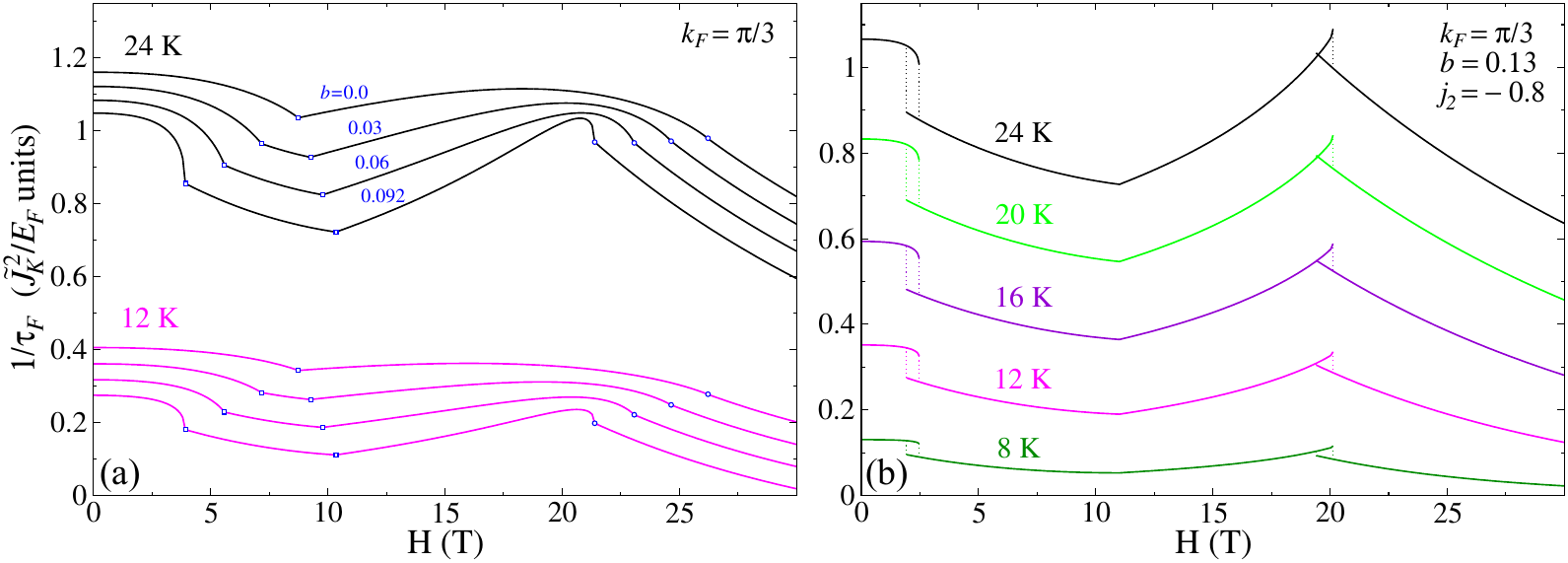}
\vskip -0.2cm
\caption{$1/\tau_F$ vs $H$ from (\ref{1_tau}) for  representative  $k_F \!= \!\pi/3$ and  temperatures, 
exchange parameters are from Table~\ref{Table1}.  
(a) Results for the biquadratic exchange $b\!=\!0$, 0.03, 0.06, and 0.0922 
are displaced down for clarity by the increments of 0.05 in the given units.
 Circles/squares mark transitions between magnetic phases. 
(b) $b\! =\! 0.13$, $j_2\!=\!0.08$, dotted lines indicate discontinuities.}
\label{Fig_rhoBdiff}
\vskip -0.2cm
\end{figure*}

In the following, we present further evidence of the success of our 
theory together with a detailed analysis of the dependence of our results on the key model parameters, 
such as  biquadratic interaction of  spins $b$ and  electron Fermi momentum $k_F$, summarized 
in Figures~\ref{Fig_rhoT}--\ref{Fig_rhoH}. 
This analysis provides implications for the microscopic parameters that should describe EuC$_6$ and also 
offers a glimpse of the prospective new phenomena that can be induced in intercalated magnetic materials and 
similar systems by means of the  chemical-, pressure-, or gate-doping. 

\subsection{$T$-dependence of resistivity}
\label{sec:rho-T}

\begin{figure}[t]
\includegraphics[width=\linewidth]{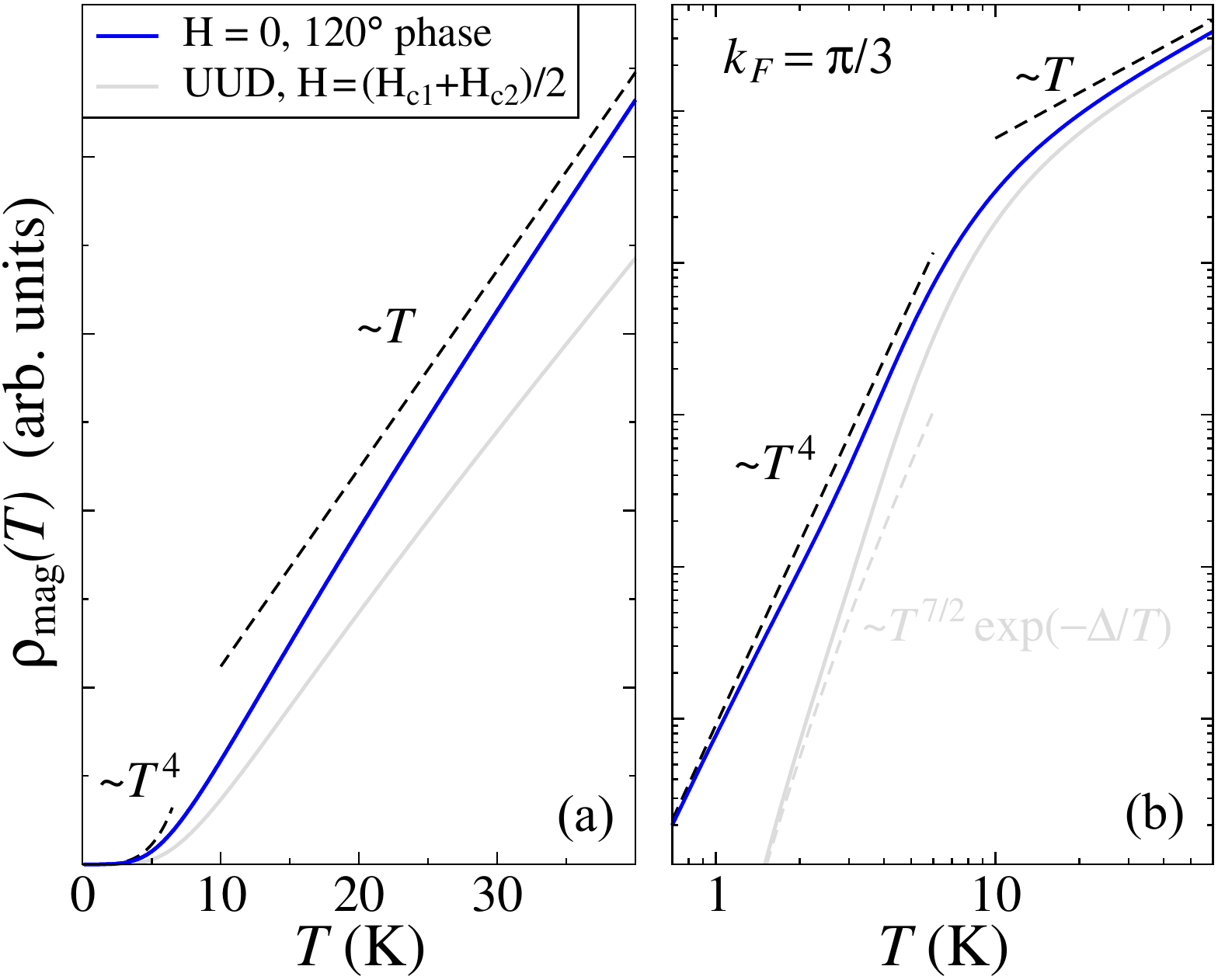}
\vskip -0.2cm
\caption{The $T$-dependence of $\rho_{\rm mag}$ due to magnetic 
scattering (\ref{1_tau}) on the (a) linear and  (b) log-log scale in the 120$\degree$ state, $H\!=\!0$,  
and in the middle of the UUD phase, $H\!=\!(H_{c2}+H_{c1})/2$, 
for the parameters from  Table~\ref{Table1} and $k_F\!=\!\pi/3$. Dashed lines are
Bloch-Gr\"{u}neisen's, $\sim\!T^4$ in 2D, Ohm's, $\sim\!T$, and exponential asymptotics, see text.}
\label{Fig_rhoT}
\vskip -0.3cm
\end{figure}

We complement our results for the field-dependence of $1/\tau_F$ in Fig.~\ref{Fig_EuC6}(c)
by the temperature-dependence of $\rho(H,T)$ at fixed $H$. Our Figure~\ref{Fig_rhoT}(a)
shows the results for  two field values: $H\!=\!0$, 120$\degree$ spin state, 
and for the middle of the UUD phase, $H\!=\!(H_{c2}+H_{c1})/2$. The results are for the same optimal choice of 
parameters to describe  EuC$_6$ from  Table~\ref{Table1} as in Fig.~\ref{Fig_EuC6}(c), and for $k_F\!=\!\pi/3$.

Our Fig.~\ref{Fig_rhoT}(b) shows the same data on the log-log scale in order 
to emphasize two  distinct temperature regimes, the ``low-$T$'' and the ``high-$T$.'' 
The overall energy scale for scattering is set by the magnon bandwidth, which plays the role analogous 
to that of the Debye energy in the electron-phonon resistivity \cite{Ziman}. 
Drawing from this analogy,  a transition between the low- and  high-$T$ regimes can be expected at a fraction 
of the magnetic Debye energy \cite{Ziman,Marder,Ziman_e_ph}, which can be estimated from the magnon spectra 
in Fig.~\ref{Fig_wks} as $W_{\rm m}\!\approx\!10 J_1S$ with some variation between  phases.
Using $J_1\!\approx\!1.1$~K, see Table~\ref{Table1}, and  $S\!=\!7/2$ yields $W_{\rm m}\!\approx\!40$~K. 
Indeed, the transition between the two regimes  can be observed in Fig.~\ref{Fig_rhoT} at  
$W_{\rm m}/5\!\approx\!8$~K. 

This consideration implies that the  majority of experimental results on EuC$_6$ 
in Refs.~\cite{Sakikabara1,Sakikabara2,Sakikabara3}, 
in our Fig.~\ref{Fig_EuC6}(b), which is reproduced   from Ref.~\cite{Chen}, 
and in all our theoretical plots are in that ``high-$T$'' regime, 
$T\!\agt\! 8$~K. The nature of this regime, where resistivity crosses over to a linear-$T$ dependence
as is indicated by the asymptotes in  Fig.~\ref{Fig_rhoT}, is simply an equivalent of the Ohm's law. 
Approximating Bose-factors in \eqref{1_tau_int} by their high-temperature limit,
${\rm n}^0_{\gamma,{\bf q}} \!\approx\! T/{\omega_{\gamma,{\bf q}}}$, 
naturally yields $1/\tau_F\!\propto\!T$. Parenthetically, this also motivates our high-$T$ approximation 
used in the consideration of the kernel in Sec.~\ref{Sec_kernel}.  

Thus, the nearly-linear $T$-dependence of $\rho(T)$ in EuC$_6$ observed in Ref.~\cite{Sugihara1985}
above 8~K  is simply within the onset of the Ohm's regime,  with no need for an artificial backflow scenario 
proposed in that work.
Needless to say, details of the spectra do not matter at high temperatures and the same 
$\rho(T)\! \propto\! T$ dependence should hold for all  field-induced phases, 
as is shown by a comparison of the UUD and 120$\degree$ states in 
Fig.~\ref{Fig_rhoT}. Naturally, in  very high fields, the field-induced gaps $g\mu_B(H\!-\!H_s)\!\gg\!k_B T$ 
will lead to a freeze-out of the magnon scattering. 

The low-$T$ regime is a bit more subtle and depends on the magnetic phase.
The case of the $120^\circ$ state and, by proxy, of the Y and V phases with the Goldstone modes that are
linear at low energies, $\omega_{\bf q}\! \propto\! q$, see Fig.~\ref{Fig_wks}(a), (b) and (d), 
is very much similar to the textbook case of acoustic phonons. For the $120^\circ$ state, one can show 
from Appendix~\ref{Sec_120} that the matrix element contribution \eqref{matrix_elements}  associated 
with the coupling to such a mode is also linear in $q$ in that limit, $\Phi_{\gamma,{\bf q}}\! \propto\! q$. 
Then, a simple power-counting in \eqref{1_tau_int} for $T\!\agt\!\omega_{\bf q}$
using $z\!\propto\!q\!\propto\!T$, yields a 2D analogue of the Bloch-Gr\"{u}neisen asymptotic 
regime $1/\tau_F \!\propto\! T^4$  shown in Fig.~\ref{Fig_rhoT}. 

As opposed to the gapless phases, the UUD and FM phases are gapped away from the transition points,
see Fig.~\ref{Fig_wks}(c) and (e). Thus, one can expect to see an activated behavior of the resistivity, 
$\rho\! \propto \!e^{-\Delta/T}$,  at sufficiently low temperatures. While this regime can be visible 
in Fig.~\ref{Fig_rhoT}(b), in practice its detection requires reaching temperatures $T \ll \Delta$. 
There is also an additional smallness due to a $\propto \!T^{7/2}$ prefactor of the exponent
associated with a suppressed coupling to the lowest mode, $\Phi_{\gamma,{\bf q}}\! \propto\! q^4$. 
An estimate for the gap in the middle of the UUD phase for EuC$_6$ gives 
$\Delta\!\approx\! 1.1S J_1\! \approx\! 4.2$~K, providing a guidance for the future observations.

The locus of magnon momenta ${\bf q}$ that are involved in the scattering 
depends on the value of $k_F$ as we discuss below. However, in the field-polarized FM case, it is, generally, 
away from the energy minimum in Fig.~\ref{Fig_wks}(e),  leading to a larger gap in the exponent, 
which is further increased by the Zeeman energy $g\mu_B(H-H_s)$ away from the saturation field,
accompanied by a more favorable prefactor $\propto \!T^{1/2}$, so the freezing-out should be 
readily observable in the FM phase at higher temperatures.  
 
Lastly, one can naively expect a power-law that is different  from $\propto\! T^4$ 
near the Y-UUD and UUD-V transition points
as both are affiliated with the BEC-like transitions, in which magnon energy is quadratic,
$\omega_{\bf q}\! \propto\! q^2$, see Fig.~\ref{Fig_wks}(c). Although we refrain from discussing it  
in any significant detail, the situation is more complicated as the coupling 
to these BEC modes is different at $H_{c1}$ and $H_{c2}$. In the first case the coupling vanishes, 
maintaining an exponential trend due to higher energy modes, while in the second case it indeed leads to 
a different  power-law $\propto\!T^{7/2}$ due to a suppressed coupling, $\Phi_{\gamma,{\bf q}}\! \propto\! q^4$.

Altogether, the consideration given above presents further evidence of the validity of our theoretical approach, 
providing a physically transparent description of the temperature-dependence of the 
resistivity of EuC$_6$ in the previously accessed temperature regime. It also invites further such
studies in finite fields and especially at  lower temperatures, where resistivity should be a 
sensitive probe of the spin excitation spectra.

\subsection{Magnetoresistivity vs biquadratic-exchange}
\label{sec:rho-heis+q}

The discussion provided below serves two goals. First is to investigate how prevalent are the 
strong anomalies in the magnetoresistivity, $\rho$ vs $H$, in the model of the conduction electrons 
coupled via a coupling (\ref{H_Kondo}) to the spins that are described by the Heisenberg-biquadratic 
model (\ref{H_1}).
Second is to demonstrate that the magnetoresistivity of EuC$_6$ is consistent with the substantial
biquadratic-exchange parameter $b$ in such a model. 

In Figure~\ref{Fig_rhoBdiff}, we present the transport relaxation rate vs field obtained from (\ref{1_tau}) 
for several representative temperatures, Fermi momentum $k_F\!=\!\pi/3$, and 
exchange parameters from Table~\ref{Table1}, except that now we vary the key biquadratic-exchange 
parameter $b\!=\!BS^2/J_1$. The corresponding magnetoresistivity is related to these 
results by a dimensional constant factor, see Eq.~\eqref{rho_R_EF_tau}.

Figure~\ref{Fig_rhoBdiff}(a) shows two sequences of curves, offset for clarity, with the biqudratic exchange
increasing in nearly equal steps from the Heisenberg limit, $b\!=\!0$, to the value  $b\!=\!0.0922$ that we use 
as an optimal choice for EuC$_6$, see Sec.~\ref{Sec_parameters}. Figure~\ref{Fig_rhoBdiff}(b) shows results 
for the  biquadratic exchange $b \!= \!0.13$ that substantially exceeds the ``critical'' value 
$b_c \!= \!1/11$, which corresponds to a change of the  Y-UUD and V-FM transitions to the first-order type
as discussed in detail in Sec.~\ref{Sec_angles} and Appendix~\ref{app:1st}.

The evolution of $1/\tau_F$ with $b$ in Fig.~\ref{Fig_rhoBdiff}(a) features already anticipated trends.
First, the opening of the $1/3$-magnetization plateau (UUD) phase away from the Heisenberg limit, 
see Sec.~\ref{Sec_model}, is clearly visible. Second, the results in Fig.~\ref{Fig_rhoBdiff}(a) are in a close
accord with  the behavior of the kernel, discussed in Sec.~\ref{Sec_kernel} and illustrated in 
Fig.~\ref{Fig_kernel}, providing an explicit confirmation that the transport relaxation rate and magnetoresistivity 
are dominated by the $2k_F$ scattering processes.

The key observation from the results in Fig.~\ref{Fig_rhoBdiff}(a) is that strong roller-coaster like variations 
in magnetoresistivity, such as the ones observed in EuC$_6$, must be associated with the nearly critical 
values of the biquadratic exchange  within our model.
Although some aspects resembling strong variations and indicating clear differences of 
$\rho$ vs $H$ dependence between different phases can already be observed in the pure Heisenberg model, 
see, for example, a kink-like feature at the Y-V boundary in the upper curves in Fig.~\ref{Fig_rhoBdiff}(a),
others are much less pronounced, see a rather small change of slope at the 
the V-FM transition at $H_s$ in the same results.

Our Fig.~\ref{Fig_rhoBdiff}(a) demonstrates that the role of the biquadratic term in model (\ref{H_1})
goes far beyond just establishing the UUD phase boundaries, which are  clearly marked by the 
kinks in $1/\tau_F$. With increasing $b$, the Y-UUD transition becomes steeper upon shifting to 
the lower fields, showing a divergent derivative for $b\!=\!0.0922$ that is related to a similar behavior 
of spin angles  in Fig.~\ref{Fig_cos}. 
Still, the biggest change takes place at the V-FM transition, which too becomes weakly 
first-order, as is elaborated on in  Appendix~\ref{app:1st}. Here, the $1/\tau_F$ field-dependence evolves 
from a nearly featureless one in the Heisenberg limit to a ``shock-wave''-like shape for 
$b\!=\!0.0922$. In contrast, the UUD-V transition remains continuous throughout these transformations, 
although the slope of $1/\tau_F$  at $H_{c2}$ also changes visibly.

Increasing $b$ beyond the critical $b_c$ should lead to a hysteresis in the magnetoresistance. 
Figure~\ref{Fig_rhoBdiff}(b) illustrates the case of $b \!= \!0.13 \!>\! b_c$. 
These results  are obtained by using the local stability of the solutions 
for the magnetic configurations, of their corresponding spin-wave energies, and of electron-magnon matrix 
elements within the overlap regions of the coexisting phases.
For example, from within the Y phase, the Y magnetic configuration persist up to $\widetilde{h}_{c1}$, 
as is described in Appendix~\ref{app:Y-UUD}. From within the UUD phase, the same field region can be 
accessed starting from $h_{c1}\! <\! \widetilde{h}_{c1}$. Therefore, in the overlap region, 
$h_{c1} \!< \!h\! <\! \widetilde{h}_{c1}$, the relaxation rate $1/\tau_F$ can be calculated in two different ways, 
resulting in the sizable discontinuities  in Fig.~\ref{Fig_rhoBdiff}(b), indicated by vertical 
dotted lines marking the overlap intervals of $h_{c1} \!< \!h \!<\! \widetilde{h}_{c1}$ for the Y-UUD and 
$h_{s} \!<\! h \!< \!\widetilde{h}_{s}$ for the V-FM transition. 

We note that the transition regions and discontinuities in Fig.~\ref{Fig_rhoBdiff}(b) are
only illustrative. As is discussed in Sec.~\ref{app:Y-UUD},   the  transition 
between the two overlapping phases should take place at $h_{c}^*$,  
at which the energies of the two phases become equal. At a finite temperature, 
a proper consideration of the first-order transition should include entropic contribution to the 
free energy of the competing phases. 
In addition, one can expect the co-existence region to be affected by  secondary anisotropies 
that are neglected in our minimal model. Nonetheless, we believe that Fig.~\ref{Fig_rhoBdiff}(b)
faithfully represents a qualitative effect of a  strong biquadratic interaction on the 
magnetoresistance across the first-order transition.

Altogether, the results presented in this section  provide an important overview of the characteristic 
evolution of the magnetotransport within the model of electrons coupled to the spin subsystem, which is 
described by the Heisenberg-biquadratic model.  

As is discussed in Sec.~\ref{Sec_parameters}, the microscopic parameters of the spin model  (\ref{H_1}) 
 describing EuC$_6$ are determined entirely from the thermodynamic quantities, such as  
critical fields and transition temperature. Therefore,  for claiming a success
of a theoretical description it is crucial that the resulting set of microscopic parameters yields distinctive features 
that are in accord with a wider phenomenology of the material, especially the one that involves less trivial
quantities such as dynamical response and transport. 
We can claim such a success here, as the parameters chosen to describe EuC$_6$ in Table~\ref{Table1} 
are also the ones that  produce sharp, nearly singular features in magnetoresistivity 
results that follow from our theory and also match closely the observed ones.  

\begin{figure*}
\includegraphics[width=\linewidth]{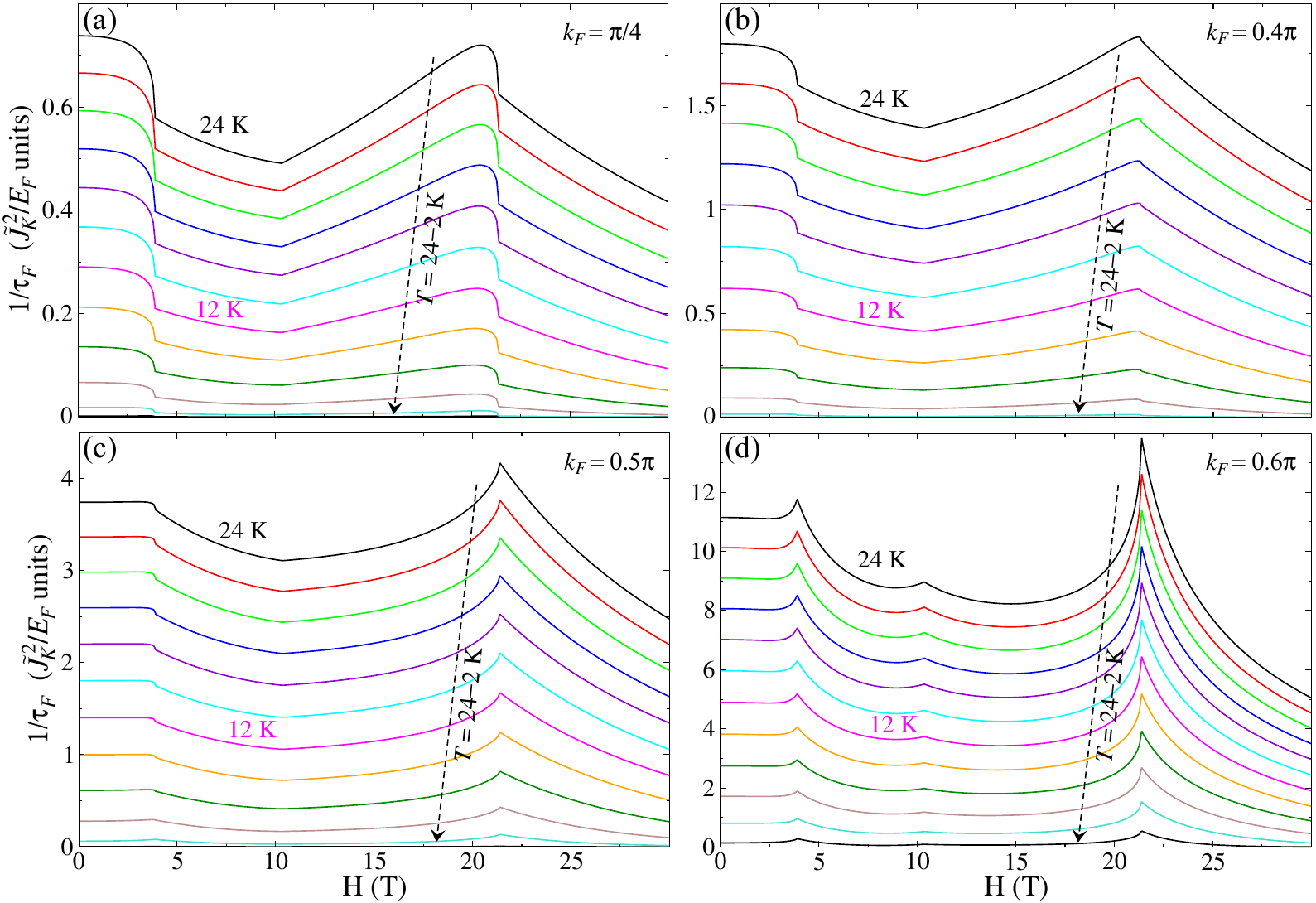}
\vskip -0.2cm
\caption{$1/\tau_F$ from (\ref{1_tau}) vs $H$ for 
(a) $k_F\!=\!\pi/4$, (b) $k_F\!=\!0.4\pi$, (c) $k_F\!=\!0.5\pi$,  and (d) $k_F\!=\!0.6\pi$.
Parameters are from Table~\ref{Table1}.}
\label{Fig_rhoH}
\vskip -0.2cm
\end{figure*}

\subsection{Magnetoresistivity, role of $k_F$}
\label{sec:rho-heis+q:kF}

Two more aspects of our study merit further discussion. 
First, as is mentioned in Sec.~\ref{Sec_phenom}, the electronic band filling fraction in EuC$_6$
and the Fermi momentum $k_F$ parametrizing it are not well-determined. 
While the nominal Eu$^{2+}$ valence naively implies a large Fermi surface,  the electronic structure
and angle-resolved photoemission study \cite{Molodtsov1996} suggested a substantially smaller 
electron fraction in the relevant carbon orbitals and  a smaller $k_F$. 
We would like to weigh in on this subject, with the magnetoresistivity in our model
arguing for a still somewhat smaller Fermi surface, with $k_F\!\alt\!\pi/3$. 

Second, much of the interest in the synthetic materials in general and in the graphite-derived 
systems in particular is due to a significant  flexibility regarding electronic density manipulation.
Then, in addition to varying parameters of the spin model, it is also important to explore the outcomes of 
our theory in a wider range of electronic parameters in order to anticipate potential new effects that can 
be accessible due to such a flexibility. To that end, we discuss some of the larger-$k_F$ results.

Our Figure~\ref{Fig_rhoH} shows the constant-$T$ curves of the transport relaxation rate 
$1/\tau_F$ vs $H$ calculated using (\ref{1_tau}) as in Fig.~\ref{Fig_EuC6}(c) and Fig.~\ref{Fig_rhoBdiff}
for a set of representative temperatures from $T\!=\!24$~K down to $2$~K in 2~K steps. 
Results are for the model parameters  from Table~\ref{Table1} which describe EuC$_6$, 
and for the four different values of the Fermi momentum, $k_F\!=\!\pi/4$, 
$0.4\pi$, $0.5\pi$, and $0.6\pi$. In this case, the field-independent constant factor that relates $1/\tau_F$ to 
magnetoresistivity $\rho(H,T)$, Eq.~\eqref{rho_R_EF_tau}, is different for the four sets 
as they correspond to different electronic concentrations $n$ via (\ref{n_to_ kF}).

Consider $k_F\!=\!\pi/4$  results in  Fig.~\ref{Fig_rhoH}(a) first. 
All of the features in the data are the same as in Fig.~\ref{Fig_EuC6}(c) and as discussed in 
Sec.~\ref{sec:rho-heis+q} for Fig.~\ref{Fig_rhoBdiff}(a), including the steep Y-UUD transition, 
a ``shock-wave'' feature at the V-FM boundary, and  a decline in the FM phase due to the Zeeman-induced 
gap that is depleting magnon population. In agreement with the analysis of $\rho$ vs $T$ in 
Sec.~\ref{sec:rho-T}, the temperature-induced offset of the curves is nearly linear in $T$ except for the lowest 
sets.  

On a closer and more quantitative side, one can argue that in terms of the overall trends in 
magnetoresistivity curves, the $k_F\!=\!\pi/3$ results in Fig.~\ref{Fig_EuC6}(c) provide a somewhat better 
fit to the EuC$_6$ data in Fig.~\ref{Fig_EuC6}(b)  than the $k_F\!=\!\pi/4$ ones. 
Moreover, to match experimental data, the decrease of $k_F$ requires a nearly proportional increase of the 
Kondo coupling constant (\ref{H_Kondo}) relative to the Fermi energy, $J_K/E_F$, 
thus restricting $k_F$ from being too small.

A  surprising trend starts being revealed by the results for the  larger $k_F\!=\!0.4\pi$ 
in Fig.~\ref{Fig_rhoH}(b). Although the features in the constant-$T$ curves are qualitatively similar to  
the $k_F\!=\!\pi/4$ case, 
 changes at the transitions are  less steep and less like the ones in the experimental data 
in Fig.~\ref{Fig_EuC6}(b). They are nearly gone for the Y-UUD boundary in  the $k_F\!=\!0.5\pi$ results 
in Fig.~\ref{Fig_rhoH}(c) and the V-FM transition for this $k_F$ is also marked by the spike-like structures, 
certainly unlike anything observed in EuC$_6$. The $k_F\!=\!0.6\pi$ results in Fig.~\ref{Fig_rhoH}(d) complete 
this unexpected trend, with all the transitions, including the formerly rather featureless UUD-V one, 
showing spikes. 

These qualitative transformations signify a  change in the dominant scattering 
that contributes to the resistivity. Regardless of its nature,  which we discuss below, an immediate 
outcome of this analysis is in the phenomenological restriction on  the size of the 
Fermi surface  in EuC$_6$. As was described in Sec.~\ref{sec_FS}, the trigonally warped Fermi surfaces 
from the band structure in Ref.~\cite{Molodtsov1996} have the extent from 
$k_{F,{\rm min}}\!\approx\!0.45\pi$ to $k_{F,{\rm max}}\!\approx\!0.7\pi$, in a qualitative
agreement with a rigid-band estimate assuming circular Fermi surface and $e/2$ per Eu$^{2+}$ doping
of the carbon bands that gives $k_{F,e/2}\!\approx\!0.43\pi$, see also Fig.~\ref{Fig_bands}. 
However, the magnetoresistivity of EuC$_6$ within our theory suggests a still smaller 
Fermi surface with an optimal  $k_F$ near $\pi/3$.  These results invite more research into 
the band structure and direct measurements of the Fermi surface of EuC$_6$. 

To understand the transformation of the relaxation rates  with $k_F$ in 
Fig.~\ref{Fig_rhoH}, we need to return to the analysis of  $1/\tau_F$ in Eq.~\eqref{1_tau} and 
in Sec.~\ref{Sec_kernel}. Because of the hierarchy $\omega_{\bf q},T\!\ll\!E_F$,  electrons 
participating in a conduction process scatter between momenta that are in a close vicinity of the Fermi surface. 
With an assumption of the circular 2D Fermi surface, 
the magnon momenta that are involved in such a scattering also form a circular locus of points in the 
${\bf q}$ space, see Fig.~\ref{Fig_St}(b) and Fig.~\ref{Fig_tau_alpha} in Appendix~\ref{App_transport}.
These momenta extend from $|{\bf q}|\!=\!0$ to the maximum of $|{\bf q}|\!=\!2k_F$, 
with the small-momentum contribution to  the transport scattering rate in \eqref{1_tau} 
suppressed and large-momentum contribution  enhanced,   
as is discussed in Sec.~\ref{Sec_kernel}.

Then it follows  for the  $k_F\!=\!\pi/3$ case that the typical large-momentum ``$2k_F$''-magnons, 
 responsible for most of the scattering,   are from the set of $|{\bf q}|$ near $2\pi/3$. 
Referring to the Brillouin zones in 
Fig.~\ref{Fig_BZs}, this value corresponds to the proximity of the $\widetilde{M}$ point of the magnetic
Brillouin zone and to the high-energy magnons near the maxima of $\omega_{\gamma,{\bf q}}$, 
see Fig.~\ref{Fig_wks}. 

However, further increase of $k_F$ drives the extent of the ${\bf q}$-contour outside of the first 
magnetic BZ and also brings the  $2k_F$-magnon energy down. Then,  the truly  ``dangerous'' value 
of the Fermi momentum of the circular Fermi surface is $k_F\!=\!2\pi/3$, as it allows magnon momenta to
reach the corners of the full Brillouin zone,  $K$ and $K'$, which correspond to the ordering vector
of all ordered phases, ${\bf Q}\!=\!\pm(4\pi/3,0)$, with  gapless or nearly gapless modes.
Thus, it is the approach of $k_F\!\to\!2\pi/3$, or, rather, $2k_F\! \to \!|{\bf Q}|$, which is responsible for 
the dramatic changes in Fig.~\ref{Fig_rhoH} from (a) to (d).
 
This analysis also shows that at a given $T$, the population of the relevant scatterers for $k_F\!=\!\pi/3$ is 
lower  than that for the larger $k_F$ values, which explains an order-of-magnitude 
enhancement of $1/\tau_F$  from $k_F\!=\!\pi/3$ in Fig.~\ref{Fig_EuC6}(c)  to 
$k_F\!=\!0.6\pi$ in Fig.~\ref{Fig_rhoH}(d) that is only partially accounted for by the $(k_Fa)^2$ factors
in \eqref{1_tau}.

Since the arguments above relies only on $2k_F\!\rightarrow \!|{\bf Q}|$, they suggests a  degree of universality.
Specifically,  we argue that $1/\tau_F$ in all gapless phases should diverge in this limit  as 
$\propto\!|2k_F-Q|^{-1}$, with a field-dependent prefactor, 
leading to an overall increase of the relaxation rates observed in Fig.~\ref{Fig_rhoH}(d). 
In addition, the  $2k_F\! \to \!|{\bf Q}|$ behavior of $1/\tau_F$ should apply 
equally to the pure Heisenberg case, which offers an opportunity for a quantitative analytical 
insight. Using  expressions for the FM and 120$\degree$ phases from Appendix~\ref{Sec_particular}
and high-$T$ limit for the Bose factors in (\ref{1_tau_int}),
neglecting $b$ and $j_2$, and expanding  ${\bf q}$ near ${\bf Q}$, after some  algebra, indeed yields
$I_{k_F} (T,H_s)\!\approx\! (8\pi T/3J_1 S)(2k_F|2k_F-Q|)^{-1}$
for the FM phase at the gapless $H_s$ point. The result is the same for the 120$\degree$ 
phase at $H\!=\!0$, but  smaller by a factor 3/4.

Since $H_s$ is the transition point that exhibits spike-like features in Fig.~\ref{Fig_rhoH}, 
while the 120$\degree$ state is away from the transition, this result confirms our hypothesis 
that the entire set of $1/\tau_F$ is divergent, or nearly divergent in the weakly gapped UUD phase, 
with the spikes being a quantitative effect that is  associated with $\propto\!q^2$ 
Goldstone modes at the transitions compared to $\propto\!q$ modes inside the gapless Y and V phases. 
We can also verify that the factor 3/4 between the $H_s$ and $H\!=\!0$ (120$\degree$ phase) points is 
indeed in a reasonable accord with the results in Fig.~\ref{Fig_rhoH}(d).
The $\propto\!|2k_F-Q|^{-1}$ divergence is also consistent with the difference between   
$k_F\!=\!0.5\pi$ and $k_F\!=\!0.6\pi$ results at $H_s$ in Figs.~\ref{Fig_rhoH}(c) and (d).

This study of the divergence brings in one more important aspect of the problem that has been 
neglected so far. We use a fairly reasonable and certainly simplifying assumption of the cylindrical Fermi surface. 
However, by itself this  assumption does not automatically make $1/\tau_F$   independent of the direction 
of the electron momentum ${\bf k}$, with the angular-dependence originating from the discrete lattice 
symmetry that is still encoded in the spin excitations and electron-magnon matrix elements. 

As may be clear intuitively, the reason for this issue to be important in the context of the  
$\propto\!|2k_F-Q|^{-1}$ divergence is that the ``dangerous'' ${\bf Q}$-vectors correspond to the
discrete points (BZ corners) in the momentum space, see Fig.~\ref{Fig_BZs}, leading to the truly 
divergent $1/\tau_F$  only for these directions. This subject is considered in Appendix~\ref{Sec_1_tau_qp_2}
for a closely related quasiparticle relaxation rate $1/\tau_{\rm qp}$, for which the effect of angular dependence  
can be taken into account without any additional approximations. 

In Appendix~\ref{Sec_1_tau_qp_2}, we demonstrate that the effect of the angle-dependence in 
$1/\tau_{\rm qp}$ is really negligible 
up to $k_F\!\approx\!0.5\pi$  and  even for $k_F\!\agt\!0.6\pi$ it is still very modest, 
confirming the accuracy of our results presented in this work and justifying our 
initial approximation that omitted this effect. 

Given the limitations of the cylindrical Fermi-surface approximation, which should become problematic for 
larger $k_F$, and  possible effects of the Fermi-surface reconstruction at the magnetic 
zone boundaries, it is not entirely clear whether the true divergences will survive, but they may 
still have strong  effects even if avoided. This points to an interesting venue of potential studies of the 
magnetic scattering effects in the large-Fermi-surface EuC$_6$, induced by the chemical-, pressure-, or gate-doping.   

Some of the considered phenomenology is reminiscent of the ``hot spots'' phenomena, much discussed in the 
theory of cuprates \cite{cuprates}, where certain parts of the Fermi surface are suggested to experience strong 
scattering due to the low-energy magnetic excitations with a particular ${\bf Q}$-vector. It is not unthinkable 
that the suggested further studies of the large-Fermi-surface EuC$_6$ may also be able to shed a new light 
on this important problem. 

\subsection{Outlook}
\label{Sec_discussion}

We would like to reiterate that the present study has provided a thorough consideration 
of one of the iconic models in frustrated magnetism, the triangular-lattice Heisenberg model, enriched by the 
biquadratic exchange and coupled to conduction electrons, with the  goal of understanding 
magnetoresistivity  throughout its phase diagram in an external magnetic field. 
The use of this model as a microscopic description of EuC$_6$, with additional approximations for electron bands
and parameters estimated from experimental critical fields and temperature, is clearly a simplification.
Yet, the  evidence of the success of such a description is undeniable, with many, if not most, features
of the magnetoresistivity reproduced, also leading to constraints on the model parameters for both 
localized spins and electron densities discussed above. 

However, the presented  description is not complete. 
Below, we briefly discuss other possible terms in the model that might be missing, their expected effects,
possible sources of the remaining discrepancies of our theory with the experimental data, 
and desirable future studies.

First additional term in the spin model (\ref{H_1}) is the ring-exchange term (\ref{H_K}),
inspired by the early works \cite{He3,Takahashi_77}, see a discussion of it and its secondary role for EuC$_6$ 
in Sec.~\ref{Sec_model}. According to our estimates in Table~\ref{Table1}, the ring-exchange is about
three times smaller than the biquadratic term. Since the symmetry of the model is unaltered by it 
and the field-induced spin-angle dependencies on the parameters that make transitions first-order are 
very similar to the biquadratic-only case \cite{Sakikabara1},  it was reasonable to neglect it.
The only unexplored  outcome of the ring-exchange term is a possible stabilization of an 
additional magnetic state between V and FM phases, with some evidence of such a phase 
in EuC$_6$ suggested by the data, see Fig.~\ref{Fig_EuC6}(b) and Ref.~\cite{Chen}.

Next is the $XXZ$ anisotropy in the $J_1$ and $J_2$ terms, which is necessary to explain  
very different magnetization behavior for the in- and out-of-plane field direction \cite{Chen}.
Given close values of the saturation fields for these directions and nearly isotropic g-factor,
it is expected to be relatively weak, at the level of 10\% \cite{Sakikabara1}, see also Sec.~\ref{Sec_model}.  
However, unlike the ring-exchange, it lowers the symmetry of the model. 
Therefore, for the in-plane field, none of the considered 
phases will have the true Goldstone modes, and gapless excitations will exist only at
the field-induced transitions. This is going to alter the low-$T$ behavior of the resistivity discussed
in Sec~\ref{sec:rho-T}, and also change the dynamical critical exponent at all critical fields from the 
BEC-like ($z\!=\!2$, $\omega_{\bf q}\!\propto\!q^2$) to the relativistic-like 
($z\!=\!1$, $\omega_{\bf q}\!\propto\!q$). However, as most of experimentally-relevant 
theory results is pertinent to the 
``high-$T$'' regime, see  Sec~\ref{sec:rho-T}, the effect is expected to be secondary on them. 
We have performed a limited study of anisotropic $XXZ$  model for some of the phases,
and found no qualitatively significant  differences from the Heisenberg limit results in that regime even for 
strong anisotropy.

Since Eu$^{2+}$ spins are large, the dipole-dipole interactions are not necessarily negligible. However, using
the analysis of Ref.~\cite{Pyrochlores} and the size of the unit cell of EuC$_6$, we estimate this term to be at 
least an order of magnitude smaller than the values of the exchanges. Since the dipolar terms also 
break spin rotational symmetries, one can expect effects similar to that of  the  $XXZ$ anisotropy. 

The spin and electronic degrees of freedom are not purely 2D in EuC$_6$,
with the  3D interplane spin couplings estimated to be of order $0.1J_1$ in Ref.~\cite{Chen}. 
While nominally essential for the finite N\'{e}el temperature, the effect on the spin excitations
can be expected to be minimal. However, even if they are small, both electron and spin 3D dispersions 
can be crucial for the softening of the $2k_F$ singularities discussed in Sec.~\ref{sec:rho-heis+q:kF}.

Perhaps the most significant difference of the outcome of our theory from the magnetoresistance data
in  EuC$_6$ in Figs.~\ref{Fig_EuC6}(b) and (c)  is the larger values of $\rho$ in the FM phase and 
a strong rise toward it near the V-FM transition. 
An obvious reason for the discrepancy is the neglect of the temperature dependence of the phase
transition boundaries in our approach. While it can be dismissed for the Y-UUD and UUD-V boundaries, 
for which the results are in a close accord with the data,
transition at the V-FM boundary has a substantial downward suppression with $T$.
One of the possible approaches, which we leave for future studies, is to include 
 temperature dependence of $H_s$ by accounting for the interactions between magnetic excitations
that are ignored in our consideration.

A less straightforward suggestion is related to the form of the Kondo coupling (\ref{H_Kondo}).
We have two Fermi surfaces, originated from the downfolding of two Dirac graphene bands onto
the Brillouin zone of the Eu-lattice. The coupling to local spins is treated as fully diagonal in the band index in  
(\ref{H_Kondo}). This may, or may not be the full story, with an intriguing prospect 
that the spin arrangement can permit or forbid interband scattering. This is again an
interesting subject for a further investigation.

Another notable difference of our results in Fig.~\ref{Fig_EuC6}(c)  from the data  
in Fig.~\ref{Fig_EuC6}(b) is at the lowest temperatures that are accessed experimentally.
While scattering on magnetic excitations  dies out in our theory,   
magnetoresistivity data retains  sizable differences between  magnetic phases. 
One of the scenarios is due to a  feedback of the spin orders on the electronic 
density of states, producing a different imprint onto the resistivity in different field-induced phases.
Further studies, with a help of electronic structure calculations, can be envisioned here. 

A different scenario for the same effect involves a compelling picture associated with the impurity-induced 
spin-textures \cite{Vojta1,Vojta2}, which generally arise in frustrated spin systems due to
magnetic couplings that are modified in the presence of  impurities. While the impact of such textures 
on the dynamical properties has been recently investigated \cite{Wolfram,park_spin_2021},
their effect on the  resistivity in the field-induced phases is simply unknown. However, 
one can expect the spin-textures to  exist readily in the noncolinear  and be suppressed in the 
collinear phases \cite{Vojta1,Wolfram}, suggesting a  profile that is similar to the one observed in 
the magnetoresistivity. Needless to say, this is a yet another direction for future investigations.

The hybridization between $4f$ electrons of Eu$^{2+}$ ions and conduction electrons  also asks for  
additional insights from the first-principles perspective, as we comment in Sec.~\ref{sec_FS} in explaining 
our choice of the ``minimal'' model description of the Fermi surfaces in EuC$_6$. This subject, 
as well as the role of the valence fluctuations, does require separate studies from both experiment and theory, 
which are well outside the scope of the present work. The success of the proposed effective spin model with 
the Kondo coupling in explaining the ``roller-coaster'' resistivity suggests that both of these open issues, while 
interesting on their own, may only be relevant to the low-energy effective model at the level of phenomenological 
parameters.

Our results can be of a direct relevance to the graphite-derived and related materials of a significant current 
interest that discuss realizations of the conducting electrons that are Kondo-coupled to localized spins $S\!=\!1/2$  
in frustrated lattices, such as transition metal dichalcogenides \cite{Analytis,Musfeldt17,MacDonald_Kondo21}. 
However, a different physical regime dominated by a large Kondo temperature may intervene and prevent the
ordering in a frustrated magnetic system, placing the physics of the $S\!=\!1/2$  case in a new regime. 
The details of how these energy scale compete, whether the present discussion holds, and what new insights are 
required in this case needs further investigation.

We conclude this Section by suggesting several extensions of experimental  work in EuC$_6$. 
As is discussed in Sec.~\ref{sec:rho-T} and above, precise low-temperature measurements would
provide a significant source of information on the field-induced transitions to and from the 
UUD phase that would allow to study intriguing critical behaviors and help determine the effective model
more precisely. As we have expounded on in Sec.~\ref{sec:rho-heis+q:kF}, the tunable-$k_F$ 
experiments can allow to study singular behavior in resistivity that may have significant implications 
to other systems,  and as is briefly mentioned in Appendix~\ref{Sec_1_tau_qp_2}, 
a significant violation of the Wiedemann-Franz law can be expected at the field-induced 
transitions in  EuC$_6$. 

\section{Summary}
\label{Sec_summary}

The main goal of the present study has been to develop a  microscopic theoretical description of the 
highly dramatic evolution of the resistivity in EuC$_6$ with the magnetic field. 
The results and discussions presented in the prior sections provide a strong affirmation that we have 
succeeded in that goal, with our results capturing most of the qualitative and  quantitative features of the 
experimental data. This success is based on a physical picture of the scattering of electrons from the 
graphene-derived bands of the carbon layers  by  spin excitations from the
triangular-lattice Eu-planes.

In the course of this work, we have provided a thorough theoretical investigation of the 
ground states, field-induced phase transitions,  spin excitations, and their couplings to the 
conduction electrons in the paradigmatic two-dimensional triangular-lattice antiferromagnet 
with a biquadratic exchange,  throughout its phase diagram.
Our effort highlights the virtues of the full-scale microscopic approach to the problem,
not only to the spin model, but also to the transport formalism for the spin-flip and non-spin-flip 
channels, allowing  rigorously obtained numerical results to receive comprehensive analytical 
and physical insights and interpretations.

The research advanced in the present study yields predictions of new field-induced and doping-induced 
phenomena in magnetically-intercalated graphite and related systems,  also 
offering an inspiration for bringing together different approaches in the search of new effects 
in the graphite-derived artificial magnetic materials. We anticipate  our effort to be  relevant to 
a broader research in metallic magnets and to provide significant technical guidance for the similar theoretical 
studies. Presently,  our study invites more research into EuC$_6$ electronic, thermodynamic, transport, and 
magnetic properties. 

Lastly, our work has advocated resistivity measurements, combined with a detailed theoretical analysis,  
as a very informative probe of not only  field-induced phase transitions, but also of the unconventional 
spin excitations in magnetic materials. We believe that  synthetic 2D materials may become 
a significant source of potentially novel insights into the nature of exotic spin excitations such as, for example, 
fractionalized spinons in a quantum spin liquid. 

\begin{acknowledgments}

We are indebted to  Roser Valent{\'{\i}}  and Vladislav Borisov for a fruitful discussion regarding 
electronic structure of intercalated graphite and for sharing their unpublished data that provided 
an important first-principles and moral support to our  Fermi-surface  consideration. 
We would like to wholeheartedly thank  KITP for the semi-virtual hospitality during the workshop 
of the pandemic-impacted Fall of 2020 when the bulk of this work was completed. 
An unexpected and enlightening encounter by one of the authors (A.~L.~C.)  with Urobatis Halleri  
during the partial in-person attendance have undoubtedly impacted the reported results.
O.~A.~S. thanks Hassan Allami and Dima Pesin for discussions of experiments on EuC$_6$ and 
initial attempts at the theoretical formulation of the problem. 
A.~L.~C. is grateful to Pavel Maksimov for a substantial 
Mathematica help and to Ilya Krivorotov for an  illuminating discussion 
regarding possible values of the dipole-dipole terms.

The work of A.~L.~C. was supported by the U.S. Department of Energy, 
Office of Science, Basic Energy Sciences under Award No. DE-SC0021221.
The work of O.~A.~S. was supported by the National Science Foundation 
CMMT program under  Grant No. DMR-1928919.
KITP is supported  by the National Science Foundation under  Grant No. NSF PHY-1748958.

\end{acknowledgments}

\appendix

\section{First order transitions}
\label{app:1st}

Here we describe analysis of the first order Y-UUD and V-FM transitions. 
We focus on the Heisenberg-biquadratic model \eqref{H_1} and provide some technical details on the classical 
ground states introduced in Sec.~\ref{Sec_classical}.

\subsection{Y-UUD transition}
\label{app:Y-UUD}

Without the ring-exchange term, $k\!=\!0$, equation on $x\!=\!\cos\alpha_1$ 
in the Y phase, Eq.~\eqref{E_clY}, reduces to 
\begin{equation}
\label{o1}
x^3 - \frac{1+b}{4b} x = -\frac{1+h}{8b}.
\end{equation}
A substitution $x\!=\!q\, \sqrt{(1+b)/3b}$ gives 
\begin{equation}
\label{o2}
q^3 - \frac{3}{4} q= - (1+h)\sqrt{\frac{27 b}{(1+b)^3}}.
\end{equation}
Comparison with the trigonometric identity
\begin{equation}
\label{o3}
\sin^3\phi - \frac{3}{4} \sin\phi = -\frac{1}{4} \sin 3\phi,
\end{equation}
leads to the solution for $x\!=\!\cos\alpha_1$ 
\begin{equation}
\label{o4}
x \!=\! \sqrt{\frac{1+b}{3b}} \sin\!\left(\frac{1}{3}
\arcsin\left[(1+h)\sqrt{\frac{27 b}{4(1+b)^3}}\right]\right).
\end{equation}
It is now easy to check that  the Y-UUD transition for $b \leq b_c \!=\! 1/11$ is continuous and takes place 
at $h_{c1} \!=\! 1- 6b$, see Eq.~\eqref{hc1}, at which $\cos\alpha_1\!=\!1$. For $b\!>\!b_c$, the Y phase remains 
{\em locally} stable up to a critical field  $\widetilde{h}_{c1}\!=\!\sqrt{4(1+b)^3/27b}-1$, see Eq.~\eqref{hc1bc}, 
which is found from the condition that the argument of $\arcsin$ in \eqref{o4} reaches the maximal  value of $1$.

Given that the UUD phase remains {\em locally} stable for all $h\! \geq \!h_{c1}$, 
we observe that field interval $h_{c1} \!\leq\! h\! \leq\!\widetilde{h}_{c1}$ determines the overlap region 
of the Y and UUD phases. 

Within our approach, however, the actual transition between the two phases takes 
place when the classical energies of the two phases become equal. This defines another field, $h_{c1}^*$, 
which can be found as follows. First, with the help of \eqref{o1}, 
the per-site energy of the Y phase  $\widetilde{E}_Y\!=\!E_Y/N S^2 J_1-3j_2$ 
can be simplified  to a quadratic form of $x\!=\!\cos\alpha_1$
\begin{equation}
\widetilde{E}_Y= (1+b)x^2 - 1.5 (1+h) x + h - 1 - b,
\label{o10}
\end{equation}
which, upon equating  with the energy of the UUD phase $\widetilde{E}_{\rm UUD}\! =\! -1 - h - 3b$, yields
\begin{equation}
\label{o5}
x^*  = \frac{3(1+h) - \sqrt{9(1+h)^2 - 32(1+b)(h+b)}}{4(1+b)} .
\end{equation}
For a given $b \!>\! b_c$, the right-hand sides of \eqref{o5} and \eqref{o4} determine  the  
transition field $h_{c1}^*$. The solution is easily obtained numerically using MATHEMATICA.

For our choice of $b\!=\!0.0922$, see  Table~\ref{Table1}, 
which is only  slightly larger than $b_c\!=\!1/11$, the resultant critical fields 
are nearly indistinguishable from each other: 
$\{h_{c1}, h_{c1}^*, \widetilde{h}_{c1}\}\!=\!\{0.4468, 0.446869, 0.446891\}$. 
For the larger values of $b$, the three fields become sufficiently 
different and allow one to study resistivity hysteresis. For example, for $b\!=\!0.13$ 
considered in Sec.~\ref{sec:rho-heis+q} 
we have $\{h_{c1}, h_{c1}^*, \widetilde{h}_{c1}\} \!=\! \{0.22, 0.25556,0.282313\}$. 

\subsection{V-FM transition} 
\label{app:V-FM}

\begin{figure}[t]
\includegraphics[width=\linewidth]{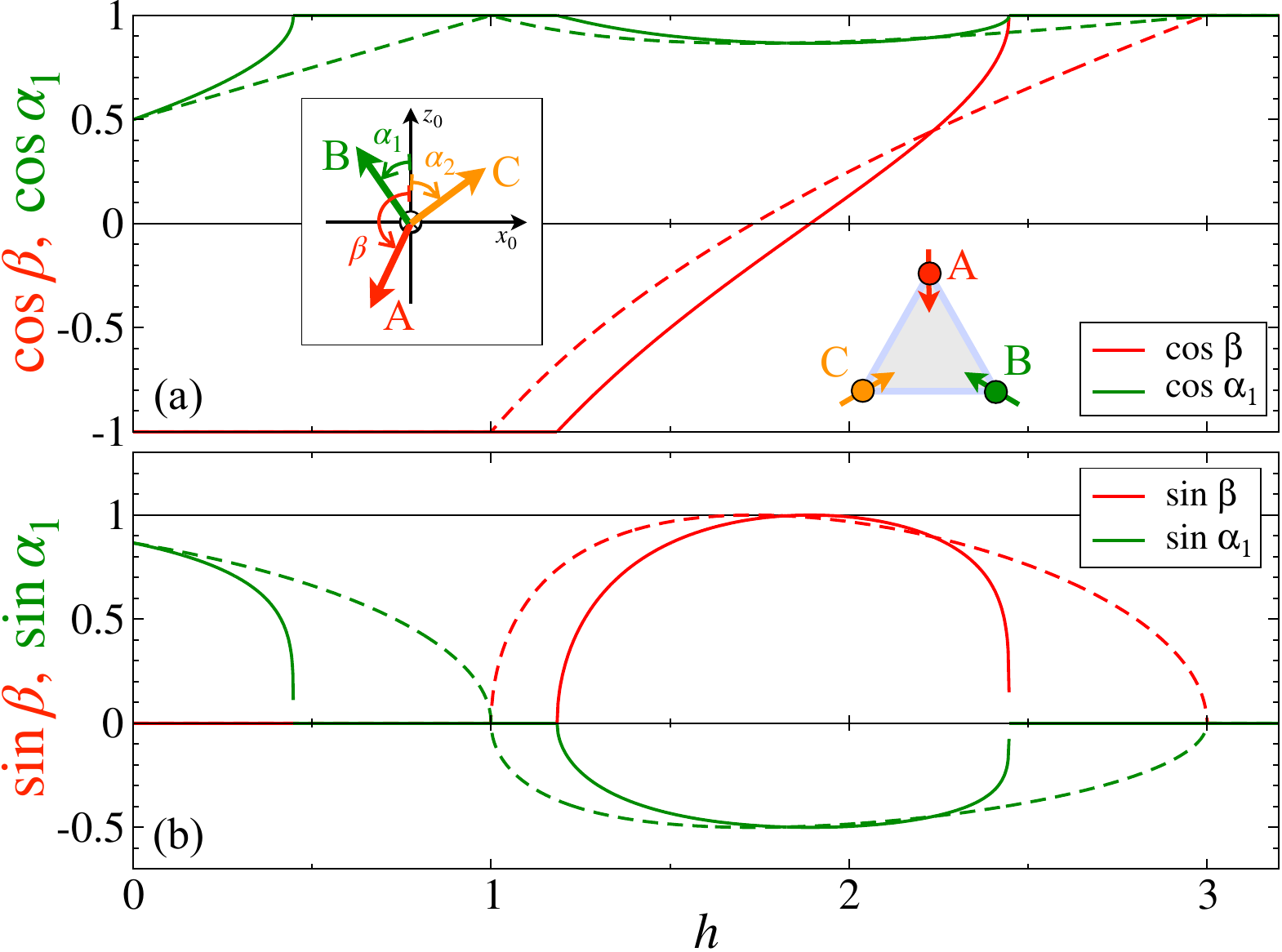}
\vskip -0.2cm
\caption{(a) The cosines and (b) the sines of angles $\alpha_1$ and $\beta$ vs $h$  throughout the 
Y-UUD-V-FM sequence of the phases  in Fig.~\ref{Fig_angles}, $h\!=\!g\mu_BH/3J_1S$. 
Dashed lines are for the pure Heisenberg model, $b\!=\!0$,
solid lines are for $b\!=\!BS^2/J_1\!=\!0.0922$ from Table~\ref{Table1}. Small discontinuities due to 
very weakly first-order transitions can be seen at the Y-UUD and  V-FM transitions, as 
$b\!=\!0.0922\! >\! b_c\! =\! 1/11\! \approx \!0.0909$.}
\label{Fig_cos}
\vskip -0.3cm
\end{figure}

In the V phase, denoting $y\!=\!\cos\beta$, introducing $t\!=\!y+\sqrt{3+y^2}$, and rewriting 
Eq.~\eqref{E_clVa} as a cubic equation for the variable $t$ defined in the interval $1 \!\leq\! t\! \leq\! 3$ yields
\begin{equation}
\label{o6}
t^3 - \frac{2+5b}{b}t = - \frac{2h}{b} .
\end{equation}
This equation is  solved by mapping to the same identity \eqref{o3} as above, with the  result given by  
\begin{equation}
\label{o7}
t = 2 \sqrt{\frac{2+5b}{3b}} \sin\left(\frac{1}{3}\arcsin\left[h \sqrt{\frac{27b}{(2+5b)^3}}\right]\right),
\end{equation}
from which the angles are obtained as 
\begin{equation}
\label{o8}
\cos\beta = \frac{t^2-3}{2t}, \ \ \  \cos\alpha_1 = \frac{t^2+3}{4t}.
\end{equation}
It is easy to check that the V-FM transition is continuous for $b \leq b_c$, with the same  $b_c\!=\!1/11$ as
above, and  the critical field of the transition is given by $h_s \!=\! 3 - 6b$, see Eq.~\eqref{hc2hs}. 
For $b \!>\! b_c$, V phase remains locally stable up to a larger field 
$\widetilde{h}_s \!=\! \sqrt{(2+5b)^3/27b}$, which is found from the 
argument of $\arcsin$ in \eqref{o7} being equal to $1$.

Similarly to the case of the discontinuous Y-UUD transition described above, the actual V-FM transition 
field $h_s^*$ is found by equating energies of the V and FM phases. Some tedious algebra gives the 
energy of the V phase
\begin{equation}
\label{o9}
\widetilde{E}_V = -\frac{b}{8} t^4 + \frac{2+5b}{4} t^2 - h t - \frac{33b}{8} - \frac{3}{2},
\end{equation}
while $\widetilde{E}_{\rm FM}\! =\! 3(1-h-b)$. Solving $\widetilde{E}_V \!=\! \widetilde{E}_{\rm FM}$ 
numerically, with $t$  given by \eqref{o7}, we find that the actual transition field $h_s^*$ satisfies 
$h_s\! <\! h_s^* \!< \!\widetilde{h}_s$. 

For $b\!=\!0.0922$ used in our work, 
we find $\{h_{s}, h_{s}^*, \widetilde{h}_{s}\} \!=\! \{2.4468, 2.44689, 2.44692\}$, 
which are, again, essentially identical. For $b\!=\!0.13$ considered in Sec.~\ref{sec:rho-heis+q}, 
these fields become $\{h_{s}, h_{s}^*, \widetilde{h}_{s}\} \!=\! \{2.22, 2.28231, 2.30258\}$.

The evolution of the cosines and sines of spin angles $\alpha_1$ and $\beta$ with $h$ throughout the 
Y-UUD-V-FM sequence of the phases in Fig.~\ref{Fig_angles} is shown in Fig.~\ref{Fig_cos}
for two representative values of $b$.

\section{Particular cases}
\label{Sec_particular}

With the general spin-wave approach for the coplanar three-sublattice  
states outlined in Sec.~\ref{Sec_general}, it is still immensely useful to have a fully analytical approach  
developed for some of the states. This is for the sake of both explicit analytical results and for an
independent verification of the partially numerical approach of Sec.~\ref{Sec_SWT}.
For the fully polarized FM and 120${\degree}$ states, a single-sublattice formulation of the SWT is 
possible. For the  UUD state, the Hamiltonian 
matrix in (\ref{LSWTmatrix}) can be reduced to $3\!\times\!3$ matrix and solved in a compact form. 

\subsection{Polarized state}
\label{Sec_FM}

In the fully polarized FM state, see Fig.~\ref{Fig_angles}(a), all angles are the same, 
$\widetilde{\alpha}_\alpha\!=\!0$. In the absence of the easy-plane anisotropy,
the off-diagonal $aa$ ($a^\dag a^\dag$) terms cancel out and the LSWT Hamiltonians 
in Eqs.~(\ref{H_J1})--(\ref{H_J2})  reduce to a tight-binding form similar to that of (\ref{H_J2}).
Since there is no distinction between the sublattices in this case, a Fourier transform  
of the Holstein-Primakoff bosons
\begin{equation}
\label{FT_FM}
a^{\phantom{\dag}}_i = \frac{1}{\sqrt{N}} \sum_{\bf q}  \,
\widetilde{a}^{\phantom{\dag}}_{\bf q} \, e^{-i{\bf q}\cdot{\bf r}_i},
\end{equation}
where $N$ is the total number of sites and ${\bf q}$ belongs to the full Brillouin zone of the triangular lattice,
is sufficient to diagonalize the LSWT model. The magnon energy is
\begin{equation}
\label{wq_FM}
\omega_{\bf q} = 3J_1S\Big(h-2(1-2b)\big(1-\overline{\gamma}_{\bf q}\big) 
-2j_2\big(1-\gamma^{(2)}_{\bf q}\big) \Big)\,,
\end{equation}
where  $h\!=\!g\mu_BH/3J_1S$, $j_2\!=\!J_2/J_1$,  $b\!=\!BS^2/J_1$ as before,  
$\gamma^{(2)}_{\bf q}$ is given in (\ref{gammas}), and 
\begin{align}
\label{gamFM}
\overline{\gamma}_{\bf q}=\frac{1}{3}\sum_{\alpha}  \cos {\bf q}{\cdot}\bm{\delta}_{\alpha}\, .
\end{align}
Here, consideration of the Kondo coupling (\ref{H_Kondo})  simplifies substantially 
as the single-magnon spin-conserving terms in the electron-magnon interaction in 
(\ref{H_Kondo_sf_nsf}) are not present, laboratory and local spin axes are the same, and all sublattices 
are equivalent.  Using Fourier transform (\ref{FT_FM})  in (\ref{H_Kondo_sf_nsf}) with (\ref{single_magnon})
and $\widetilde{\alpha}_\alpha\!=\!0$ yields
\begin{align}
\label{H_Kondo_sf_FM}
{\cal H}_{int}^{+-}=\frac{2\widetilde{J}_K}{\sqrt{N}}\sum_{{\bf k},{\bf q}} 
\Big[f^\dag_{{\bf k}-{\bf q}\uparrow}f^{\phantom{\dag}}_{{\bf k}\downarrow} 
 \widetilde{a}^{\dag}_{{\bf q}} + {\rm H.c.}\Big],
\end{align}
where $\widetilde{J}_K\!=\!\frac12 J_K\sqrt{S/2}$ as before.
The spin-flip scattering term is simple, with a matrix element containing no momentum-dependence. 

According to Appendix~\ref{App_sf}, the spin-flip scattering straightforwardly leads 
to the relaxation rate in the form of Eq.~(\ref{1_tau}) with the 1D integral in (\ref{1_tau_int}) taking the form
\begin{align}
\label{1_tau_intFM}
I_{k_F}(T,H)=&\,4\int_0^1 \frac{z^2\, dz}{\sqrt{1-z^2}}\,
{\rm n}^0_{{\bf q}}({\rm n}^0_{{\bf q}}+1)\, \frac{\omega_{{\bf q}}}{T}\,,
\end{align}
with the same momentum parametrization along the 1D contour   
${\bf q}\!=\!2k_F(z^2,z\sqrt{1-z^2})$ and Bose distribution function ${\rm n}^0_{{\bf q}}$ 
with the magnon energy $\omega_{{\bf q}}$ from Eq.~(\ref{wq_FM}). 

Given the simplicity of the polarized FM state, it may be instructive to demonstrate a relation of the 
single-sublattice formalism to the general three-sublattice one described in 
Secs.~\ref{Sec_LSWT_Hamiltonian} and \ref{Sec_Diagonalization}, as the latter is supposed to give an 
identical description. 

For all spins polarized,  $\widetilde{\alpha}_\alpha\!=\!0$,  the off-diagonal term in the Hamiltonian 
matrix (\ref{LSWTmatrix}) $\hat{\bf B}^{\phantom \dagger}_{\bf q}\!\equiv\!0$ and 
\begin{align}
\label{AqFM}
\hat{\bf A}^{\phantom \dagger}_{\bf q}
&=C_{\bf q}\hat{\bf I}+(1-2b)\hat{\bm \Lambda}^{\phantom \dagger}_{\bf q}\,,\\
&C_{\bf q}= h-2(1-2b)-2j_2\big(1-\gamma^{(2)}_{\bf q}\big),\nonumber
\end{align}
where $\hat{\bf I}$ is a $3\times 3$ 
identity matrix  and 
\begin{align}
\label{LambdaFM}
\hat{\bm \Lambda}^{\phantom \dagger}_{\bf q}
= \left( \begin{array}{ccc} 
0 &  \gamma_{\bf q} &  \gamma^*_{\bf q} \\
 \gamma^*_{\bf q}  & 0 &  \gamma_{\bf q}\\
 \gamma_{\bf q}  &  \gamma^*_{\bf q} & 0
\end{array}\right),
\end{align}
with $\gamma_{\bf q}$ from (\ref{gammas}).
Since $[\hat{\bm \Lambda}^{\phantom \dagger}_{\bf q},\hat{\bf I}]\!=\!0$,
the three magnon branches have the energies 
\begin{equation}
\label{wq_FM3}
\omega_{\gamma\bf q} = 3J_1S\big(C_{\bf q}+(1-2b)\lambda_{\gamma\bf q} \big)\,,
\end{equation}
where $\lambda_{\gamma\bf q}$ are the  eigenvalues of   $\hat{\bm \Lambda}^{\phantom \dagger}_{\bf q}$.
A straightforward algebra with (\ref{LambdaFM}) gives $\lambda_{1\bf q}\!=\!2\overline{\gamma}_{\bf q}$
and $\lambda_{2(3)\bf q}\!=\!2\overline{\gamma}_{{\bf q}\pm{\bf Q}}$, with 
$\overline{\gamma}_{\bf q}\!=\!(\gamma_{\bf q}+\gamma^*_{\bf q})/2$ from (\ref{gamFM}) and 
${\bf Q}\!=\!(4\pi/3,0)$. 
Thus, the three magnon branches are the ``original'' single-sublattice result 
 in (\ref{wq_FM}), $\omega_{1\bf q}\!=\!\omega_{\bf q}$, and the other two are 
``shifted'' by the ordering vectors, $\omega_{2(3)\bf q}\!=\!\omega_{{\bf q}\pm{\bf Q}}$. 

In the single-sublattice treatment shown above, the one-magnon
coupling involves $\widetilde{a}^{\phantom{\dag}}_{\bf q}$ ($\widetilde{a}^{\dag}_{\bf q}$) operators
from (\ref{FT_FM}) with a constant matrix element, see Eq.~(\ref{H_Kondo_sf_FM}).
In the three-sublattice approach, matrix elements (\ref{Kondo_Matrix_elements}) of the general form of 
electron-magnon coupling in Eq.~(\ref{H_Kondo_qp}) require the knowledge of the Hamiltonian eigenfunctions. 
In the polarized FM case,  anomalous terms in the transformation to quasiparticles 
(\ref{UVtrans}) are absent, $\hat{\bf V}_{\bf q}\!=\!0$,  and 
all angles are $\widetilde{\alpha}_\alpha\!=\!0$, immediately simplifying  (\ref{Kondo_Matrix_elements})
to just
\begin{equation}
\label{H_Kondo_qpFM}
{\cal H}_{int}^{+-}=\frac{\widetilde{J}_K}{\sqrt{3N}}\sum_{{\bf k},{\bf q},\gamma} 
\Big[M^{+-}_{\gamma,{\bf q}} f^\dag_{{\bf k}-{\bf q}\uparrow}f^{\phantom{\dag}}_{{\bf k}\downarrow} 
  a^{\dag}_{\gamma,{\bf q}}\ + {\rm H.c.}\Big],
\end{equation}
with the matrix elements $M^{+-}_{\gamma,{\bf q}}\!=\!2
\sum_\alpha U_{\alpha,-{\bf q}}^{({\gamma})}$,
where the matrix of vectors $\hat{\bf U}_{\bf q}$ should  diagonalize 
$\hat{\bm \Lambda}^{\phantom \dagger}_{\bf q}$ in (\ref{LambdaFM}). A simple algebra yields 
\begin{eqnarray}
\label{Us_FM}
{\bf U}^{(1)}\!=\!\frac{1}{\sqrt{3}}(1,1,1)^T, \ \
{\bf U}^{(2,3)}\!=\!\frac{1}{\sqrt{3}} (1,e^{\pm i\theta},e^{\mp i\theta})^T, \  \  \ \  \  \ 
\end{eqnarray}
for the ``original''  $\omega_{1\bf q}$ and ``shifted'' $\omega_{2(3)\bf q}$, respectively; here 
$\theta\!=\!4\pi/3$. 

Because of the local nature of the Kondo interaction, the matrix element of the coupling  to an
eigenmode $\gamma$ in (\ref{H_Kondo_qpFM}) is simply proportional to the sum of the components 
of a corresponding vector, $\sum_\alpha U_\alpha^{(\gamma)}$. Thus, as it trivially follows from (\ref{Us_FM}),
the resultant matrix elements of the coupling to the ``shifted'' ${\bf q}\pm{\bf Q}$ modes are identically zero 
and only the ``original'' $\omega_{1\bf q}$-mode  contributes to (\ref{H_Kondo_qpFM}) with   
$M^{+-}_{\gamma,{\bf q}}\!=\!2\sqrt{3}$. 

Needless to say, this renders the coupling Hamiltonians 
in the single-sublattice  and  three-sublattice treatment, (\ref{H_Kondo_sf_FM}) and  (\ref{H_Kondo_qpFM}), 
equivalent.  Their resultant scattering rate is given in (\ref{1_tau}) and (\ref{1_tau_intFM}).

A closely associated problem is the relation between the single-sublattice operators 
$\widetilde{a}^{\phantom{\dag}}_{\bf q}$ ($\widetilde{a}^{\dag}_{\bf q}$) 
to the three-flavor operators  $a^{\phantom{\dag}}_{\alpha\bf q}$ ($a^{\dag}_{\alpha\bf q}$). 
A simple algebra gives 
\begin{align}
a^{\phantom{\dag}}_{\bf q}&=\frac{1}{\sqrt{3}}\left(\widetilde{a}^{\phantom{\dag}}_{\bf q}+
\widetilde{a}^{\phantom{\dag}}_{{\bf q}+{\bf Q}}+
\widetilde{a}^{\phantom{\dag}}_{{\bf q}-{\bf Q}}\right), \nonumber\\
\label{operatorsFBZ}
b^{\phantom{\dag}}_{\bf q}&=\frac{1}{\sqrt{3}}\left(\widetilde{a}^{\phantom{\dag}}_{\bf q}+
e^{i\theta}\,\widetilde{a}^{\phantom{\dag}}_{{\bf q}+{\bf Q}}+
e^{-i\theta}\,\widetilde{a}^{\phantom{\dag}}_{{\bf q}-{\bf Q}}\right), 
\\
c^{\phantom{\dag}}_{\bf q}&=\frac{1}{\sqrt{3}}\left(\widetilde{a}^{\phantom{\dag}}_{\bf q}+
e^{-i\theta}\,\widetilde{a}^{\phantom{\dag}}_{{\bf q}+{\bf Q}}+
e^{i\theta}\,\widetilde{a}^{\phantom{\dag}}_{{\bf q}-{\bf Q}}\right), \nonumber
\end{align}
with the same $\theta\!=\!4\pi/3$. These expressions are general and apply to  
the other cases where both single- and three-sublattice approaches are possible, such as the 
120$^{\degree}$ case discussed next.

\subsection{120$^{\degree}$ state}
\label{Sec_120}

For the 120$^{\degree}$ state, see Fig.~\ref{Fig_angles}(a) for a sketch, the angles 
$\widetilde{\alpha}$ that define spin configuration and transformation to the local reference frame in 
Eq.~(\ref{Sxyz}) can be written as 
\begin{equation}
\label{alpha120}
\widetilde{\alpha}= \widetilde{\alpha}_A-{\bf Q}\cdot{\bf R}_i\,,
\end{equation}
where, according to the choice in Fig.~\ref{Fig_angles}, $\widetilde{\alpha}_A\!=\!\beta\!=\!\pi$,
${\bf R}_i\! =\! {\bf R}_\ell\! +\! \bm{\rho}_\alpha$ with 
$\bm{\rho}_A\! =\! 0$, $\bm{\rho}_B\! =\! -\bm{\delta}_2$, and
$\bm{\rho}_C \!=\! \bm{\delta}_3$, see Fig.~\ref{Fig_lattice},  and ${\bf Q}\!=\!(4\pi/3,0)$ as before.
As a result, all mutual angles are the same up to a sign, $\widetilde{\alpha}_{ij}\!=\!\pm 2\pi/3$, 
making the LSWT Hamiltonian terms in (\ref{h_LSWT}) the same on each bond and rendering 
the division of the lattice in three sublattices unnecessary. The Fourier transform (\ref{FT_FM})
leads to a standard LSWT Hamiltonian \cite{ChubukovJ1J2,Jolicquer,ChZh1,ChZh2,Ivanov}
\begin{align}
\label{H2_120}
\hat{\cal H}^{(2)}=\sum_{{\bf q}} \Big[
A_{\bf q}  \widetilde{a}^\dagger_{{\bf q}}\widetilde{a}^{\phantom{\dag}}_{{\bf q}}   
-\frac{B_{\bf q}}{2} \Big( \widetilde{a}^\dagger_{{\bf q}} \widetilde{a}^{\dagger}_{-{\bf q}}
+{\rm H.c.}\Big)\Big], 
\end{align}
where
\begin{align}
&A_{\bf q}=3J_1S\Big(1+\overline{\gamma}_{{\bf q}}/2-b\big(1-4\overline{\gamma}_{{\bf q}}\big)/2
-2j_2\big(1-\gamma_{{\bf q}}^{(2)}\big)\Big) ,  \nonumber\\
&B_{\bf q}=3J_1S\Big(3\big(\overline{\gamma}_{{\bf q}} +b\big)/2\Big),
\label{ABq120}
\end{align}
with $\overline{\gamma}_{{\bf q}}$ from (\ref{gamFM}) and magnon energy with the parameters of 
the Bogolyubov transformation given by
\begin{eqnarray}
&&\omega_{\bf q}\!=\!\sqrt{A_{\bf q}^2-B_{\bf q}^2},\nonumber\\
&&U^2_{\bf q} +V^2_{\bf q} =\frac{A_{\bf q}}{\omega_{\bf q}},\ \ \ 
2U_{\bf q}V_{\bf q}=\frac{B_{\bf q}}{\omega_{\bf q}}.
\label{UV120}
\end{eqnarray}
Similarly to the FM phase considered above, the relation of the single-sublattice branch $\omega_{\bf q}$ 
to the three-sublattice energies is straightforward, with the three branches given 
by the ``original'' $\omega_{1\bf q}\!=\!\omega_{\bf q}$
and the two ``shifted'' ones, $\omega_{2(3)\bf q}\!=\!\omega_{{\bf q}\pm{\bf Q}}$, see Fig.~\ref{Fig_wks}(a).

The relationship of the   three-flavor operators  $a^{\phantom{\dag}}_{\alpha\bf q}$ ($a^{\dag}_{\alpha\bf q}$)
to the single-sublattice operators $\widetilde{a}^{\phantom{\dag}}_{\bf q}$ ($\widetilde{a}^{\dag}_{\bf q}$) is 
the same as in Eq.~(\ref{operatorsFBZ}) as it is simply a relation between   one- and three-sublattice  
Fourier transforms in (\ref{FT3}) and (\ref{FT_FM}). 
The eigenvectors of the generalized Bogolyubov transformation in the three-sublattice 
case (\ref{Bogo}) are related to the single-sublattice ones in (\ref{UV120}) via a combination of
shifts by $\pm{\bf Q}$ and the phase factors as in (\ref{operatorsFBZ})  in the latter.

The electron-magnon Hamiltonian is obtained from the Kondo coupling (\ref{H_Kondo_sf_nsf}) using 
spin-rotation transformation (\ref{Sxyz}) with the angles (\ref{alpha120}), subsequent spin-operator
bosonization, and Fourier  (\ref{FT_FM}) and Bogolyubov (\ref{UV120}) transforms. Because of the 
non-collinear structure, it contains both  spin-flip and non-spin-flip terms that have the same structure
as the  general form of electron-magnon coupling in (\ref{H_Kondo_qp}), which we will not rewrite here. 

Following the steps outlined in Appendices~\ref{App_non_sf} and \ref{App_sf}, we obtain the relaxation rate 
in the form of Eq.~(\ref{1_tau}) with the 1D integral in (\ref{1_tau_int}) given by
\begin{align}
\label{1_tau_int120}
I_{k_F}(T,H)=&\,\int_0^1 \frac{z^2\, dz}{\sqrt{1-z^2}}\\ 
&\,\phantom{\int_0^1}
\times\sum_\gamma \widetilde{\Phi}_{\gamma,{\bf q}} \, 
{\rm n}^0_{\gamma,{\bf q}}({\rm n}^0_{\gamma,{\bf q}}+1)\, \frac{\omega_{\gamma,{\bf q}}}{T}\,,\nonumber
\end{align}
with ${\bf q}\!=\!2k_F(z^2,z\sqrt{1-z^2})$  and Bose  function ${\rm n}^0_{\gamma,{\bf q}}$  as before,
and with $\gamma$  now numerating  ${\bf q}$ and ${{\bf q}\pm{\bf Q}}$  branches.

The advantage of the form (\ref{1_tau_int120}) over the general expression (\ref{1_tau_int}) 
is in an explicit analytical form of the matrix elements contribution  $\widetilde{\Phi}_{\gamma,{\bf q}}$, 
which are given by 
\begin{equation}
\label{matrix_elements120}
\widetilde{\Phi}_{1,{\bf q}} =2\big(U_{\bf q}-V_{\bf q}\big)^2, \  \ \
\widetilde{\Phi}_{2(3),{\bf q}} =\big(U_{{\bf q}\pm{\bf Q}}+V_{{\bf q}\pm{\bf Q}}\big)^2.
\end{equation}
Using them, an analytic insight into the contributions of ${\bf q}\!\rightarrow\!0$ or 
${\bf q}\!\rightarrow\!{\bf Q}$ regions of integration into the relaxation rate becomes possible.

\subsection{Plateau state}
\label{Sec_UUD}

The up-up-down (UUD), or magnetization plateau state is special as it does not break  rotational 
$U(1)$ symmetry about the field direction, with the spins parallel (up) or antiparallel (down) to the field.
For our choice in Fig.~\ref{Fig_angles}, sublattice $A$ spins are down, 
$\widetilde{\alpha}_A\!=\!\beta\!=\!\pi$, while $B$ and $C$ spins are up, 
$\widetilde{\alpha}_{B(C)}\!=\!0$. With these angles, six out of twelve independent 
matrix elements in the matrices $\hat{\bf A}^{\phantom \dagger}_{\bf q}$ and 
$\hat{\bf B}^{\phantom \dagger}_{\bf q}$ (\ref{ABany})
of the Hamiltonian matrix (\ref{LSWTmatrix}) vanish and the remaining six are complementary,
suggesting a rearrangement in the vector-operator of the Holstein-Primakoff 
bosons that allows to reduce the rank of the Hamiltonian matrix to $3\times 3$,
 \begin{align}
\label{LSWT_H_UUD}
\hat{\cal H}^{(2)} =\frac{3J_1S}{2}\sum_{\bf q}\hat{\bf y}_{\bf q}^\dagger 
\hat{\bf H}_{\bf q}\hat{\bf y}_{\bf q}^{\phantom{\dagger}}\, ,
\end{align}
where $\hat{\bf y}^\dag_{\bf q}\!=\!\left( a^{\phantom \dag}_{-\bf q}, b^\dag_{\bf q},c^\dag_{\bf q}\right)$ 
is the rearranged vector-operator 
with  the  Hamiltonian matrix   
\begin{eqnarray}
\label{LSWTmatrixUUD}
\hat{\bf H}_{\bf q}= 
\left( \begin{array}{ccc} 
\bar{A}_{\bf q} &  \bar{P}_{\bf q} &  \bar{P}^*_{\bf q}\\
\bar{P}^*_{\bf q}  & \bar{C}_{\bf q} &  \bar{H}_{\bf q} \\
\bar{P}_{\bf q}  & \bar{H}^*_{\bf q} &  \bar{C}_{\bf q}
\end{array}\right),
\end{eqnarray}
and matrix elements  given by
\begin{align}
\bar{A}_{\bf q}=&\, 2-h+4b-2j_2\big(1-\gamma^{(2)}_{\bf q}\big) ,\nonumber\\
\bar{C}_{\bf q}=&\, h+4b-2j_2\big(1-\gamma^{(2)}_{\bf q}\big),\nonumber\\
\label{ACH_UUD}
\bar{P}_{\bf q}=&\, -(1+2b)\gamma_{\bf q},\\
\bar{H}_{\bf q}=&\, (1-2b)\gamma_{\bf q},
\nonumber
\end{align}
with all constants as before and $\gamma_{\bf q}$ and $\gamma^{(2)}_{\bf q}$ from (\ref{gammas}).

This approach, first employed in Ref.~\cite{Chubukov1991}, has been  used in a number of more recent 
works \cite{Alicea2009,Starykh2015}. Physically, such a rearrangement takes advantage of the conservation 
of magnetization in the UUD state due to remaining $U(1)$ symmetry for the Heisenberg 
spins, which makes  creation (annihilation) of spin flips in the down-sublattice  equivalent to 
the annihilation (creation) of spin flips in the up-sublattices. 

The standard diagonalization procedure of (\ref{LSWTmatrixUUD}) 
now concerns $\hat{\bar{{\bf g}}} \hat{{\bf H}}_{\bf q}$, where $\hat{\bar{\bf g}}$ is a diagonal matrix $[-1,1,1]$.
The characteristic equation $||\hat{\bar{{\bf g}}} \hat{{\bf H}}_{{\bf q}}-\lambda\hat{{\bf I}}||\!=\!0$ yields
\begin{align}
\label{lambda_UUD}
\lambda^3+b\lambda^2+c\lambda+d=0\,,
\end{align}
which is a cubic equation on $\lambda$ with real coefficients,
\begin{align}
b=&\, \bar{A}_{\bf q}-2\bar{C}_{\bf q} ,\nonumber\\
\label{coeff_UUD}
c=&\, \bar{C}_{\bf q}^2-|\bar{H}_{\bf q}|^2 + 2|\bar{P}_{\bf q}|^2-2\bar{A}_{\bf q}\bar{C}_{\bf q}, \\
d=&\, \bar{A}_{\bf q}\big(\bar{C}_{\bf q}^2-|\bar{H}_{\bf q}|^2\big)
-2\bar{C}_{\bf q}|\bar{P}_{\bf q}|^2+2\Re\big(\bar{P}_{\bf q}^2\bar{H}_{\bf q}\big),
\nonumber
\end{align} 
that can be solved in a number of standard ways \cite{Wolfram_cubic},
with the three roots  $\{\lambda_1,  \lambda_2, \lambda_3\}$ giving   magnon eigenenergies 
$\{-\omega_{1-{\bf q}},  \omega_{2{\bf q}}, \omega_{3{\bf q}}\}$  in units of $3J_1S$. 
At the $\Gamma$ point,  analytic solution of (\ref{lambda_UUD}) simplifies considerably,  
making explicit the linear field-dependence of the magnon energies throughout the UUD phase.
Note that the numeration of the magnon branches in Fig.~\ref{Fig_wks}(c) is from the lowest to 
highest in energy, not necessarily according to the numbering of the solutions above.

The transformation from  Holstein-Primakoff  bosons in (\ref{LSWT_H_UUD}) to the quasiparticles
is similar to the generalized Bogolyubov transformation in (\ref{Bogo}) 
\begin{align}
\label{transUUD}
\bar{a}^{\phantom{\dag}}_{\alpha,{\bf q}}=\sum_\gamma U_{\alpha,{\bf q}}^{({\gamma})}\,
\bar{A}^{\phantom{\dag}}_{\gamma,{\bf q}},
\end{align}
but without the ``anomalous'' $V_{\alpha}^{({\gamma})}$ terms and 
with the ``mixed'' original   $\bar{a}^{\phantom{\dag}}_{\alpha,{\bf q}}\!=\!
\big\{a^{\dag}_{-\bf q},b^{\phantom{\dag}}_{\bf q},c^{\phantom{\dag}}_{\bf q}\big\}$ 
and quasiparticle operators  
$\bar{A}^{\phantom{\dag}}_{\gamma,{\bf q}}\!=\!
\big\{A^{\dag}_{-\bf q},B^{\phantom{\dag}}_{\bf q},C^{\phantom{\dag}}_{\bf q}\big\}$
and normalization that respects the metric   $\hat{\bar{{\bf g}}}$
\begin{align}
\label{UnormUUD}
\sum_\gamma \sigma_\gamma\big|U_{\alpha,{\bf q}}^{({\gamma})}\big|^2=\sigma_\alpha\,,
\end{align}
where $\sigma_\gamma\!\equiv\!\hat{\bar{g}}_{\gamma\gamma}\!=\!\{-1,1,1\}$.

The transformation    (\ref{transUUD}) from  
$\hat{\bf y}_{\bf q}\!=\!\big[ a^{\dag}_{-\bf q},b^{\phantom{\dag}}_{\bf q},c^{\phantom{\dag}}_{\bf q}\big]^T$ 
to 
$\hat{\bf z}_{\bf q}\!=\!\big[ A^{\dag}_{-\bf q},B^{\phantom{\dag}}_{\bf q},C^{\phantom{\dag}}_{\bf q}\big]^T$
vectors, written  in a matrix form is
\begin{eqnarray}
\label{transUUD1}
\hat{\bf y}_{\bf q}= \hat{\bf U}_{\bf q}  \cdot\hat{\bf z}_{\bf q}, \ \ \
\end{eqnarray}
where the transformation matrix $\hat{\bf U}_{\bf q}$
is given by the  normalized eigenvectors of $\hat{\bar{{\bf g}}} \hat{{\bf H}}_{\bf q}$
that can be obtained via a somewhat tedious, but straightforward diagonalization procedure using explicit 
form of $\hat{{\bf H}}_{\bf q}$ in (\ref{LSWTmatrixUUD}),  yielding 
\begin{align}
\label{vectorsUUD}
{\bf U}_{{\bf q}}^{({\gamma})}=\frac{1}{r_{\gamma,{\bf q}}}
\left(
\begin{array}{c} 
\widetilde{R}_{\gamma,{\bf q}}\\
R_{\gamma,{\bf q}} \\
R^*_{\gamma,{\bf q}}
\end{array}\right),
\end{align}
with 
\begin{align}
\widetilde{R}_{\gamma,{\bf q}}=&\, 
\big(\lambda_{\gamma,{\bf q}}-\bar{C}_{\bf q}\big)^2-|\bar{H}_{\bf q}|^2 ,
\nonumber\\
\label{vectorsUUDelements}
R_{\gamma,{\bf q}}=&\, 
\bar{P}_{\bf q}\bar{H}_{\bf q}+\bar{P}_{\bf q}^*\big(\lambda_{\gamma,{\bf q}}-\bar{C}_{\bf q}\big) , \\
r_{\gamma,{\bf q}}^2=&\,\sigma_\gamma \big(2|R_{\gamma,{\bf q}}|^2-\widetilde{R}_{\gamma,{\bf q}}^2\big) ,
\nonumber
\end{align}
where $\lambda_{\gamma,{\bf q}}$ are the eigenvalues obtained from (\ref{lambda_UUD}) 
and (\ref{coeff_UUD}). While somewhat cumbersome, the outlined formalism allows to 
perform calculations without relying on numerical matrix diagonalization procedures.

The derivation of electron-magnon coupling and relaxation rate in the UUD phase
bears a lot of similarity to the polarized FM case in its three-sublattice formulation, see
Sec.~\ref{Sec_FM}. Because of spin collinearity, only spin-flip scattering is present, with 
either emission or absorption 
\begin{equation}
\label{H_Kondo_qpUUD}
{\cal H}_{int}^{+-}=\frac{2\widetilde{J}_K}{\sqrt{3N}}\sum_{{\bf k},{\bf q},\gamma} 
\Big[\widetilde{M}_{\gamma,{\bf q}} f^\dag_{{\bf k}-{\bf q}\uparrow}f^{\phantom{\dag}}_{{\bf k}\downarrow} 
\bar{A}^{\dag}_{\gamma,{\bf q}} \ + {\rm H.c.}\Big],
\end{equation}
where $\bar{A}^{\dag}_{\gamma,{\bf q}}\!=\!
\big\{A^{\phantom{\dag}}_{-\bf q},B^{\dag}_{\bf q},C^{\dag}_{\bf q}\big\}$ as in (\ref{transUUD}) and 
matrix elements $\widetilde{M}_{\gamma,{\bf q}}\!=\!\sum_\alpha U_{\alpha,-{\bf q}}^{({\gamma})}$,
with  vectors ${\bf U}^{({\gamma})}_{\bf q}$  from (\ref{vectorsUUD}).

As in the polarized FM case, the coupling to a mode $\gamma$ in (\ref{H_Kondo_qpUUD}) is given by 
the sum of the components of a corresponding vector, ${\bf U}^{({\gamma})}_{\bf q}$, but
without further simplifications that follow in the FM case.

Lastly, the 1D integral in the relaxation rate (\ref{1_tau}) is identical to Eq.~(\ref{1_tau_int})
\begin{align}
\label{1_tau_intUUD}
I_{k_F}(T,H)=&\,\int_0^1 \frac{z^2\, dz}{\sqrt{1-z^2}}\\ 
&\,\phantom{\int_0^1}
\times\frac13\sum_\gamma \widetilde{\Phi}_{\gamma,{\bf q}} \, 
{\rm n}^0_{\gamma,{\bf q}}({\rm n}^0_{\gamma,{\bf q}}+1)\, \frac{\omega_{\gamma,{\bf q}}}{T}\,,\nonumber
\end{align}
but with a simplified $\widetilde{\Phi}_{\gamma,{\bf q}} \!=\!\big|\widetilde{M}_{\gamma,{\bf q}}\big|^2$.

\section{Transport formalism, $1/\tau$ approximation}
\label{App_transport}

Transport theory, including its Boltzmann version, is an essential chapter in most of the
advanced condensed matter textbooks \cite{Ziman,Marder}. 
Discussion of the electron-phonon scattering, which controls  
resistivity of metals in a wide range of temperatures, is also a necessary part of the story. 
Still, the technical aspects of its theory are usually avoided in this context, opting for some physically motivated
but not rigorous considerations. While some of the required steps may indeed seem sufficiently cumbersome to justify 
such an approach, in the following narrative we would like to dispel this accepted aura from the subject and outline a 
step-by-step derivation of the transport scattering rate under the assumption of the quasi-elastic scattering.
In the process of doing so, we also provide a solution of the problem at hand, that is,
make available a compact expression for the electron-magnon scattering rate in a 2D setting 
relevant to EuC$_6$. 

\subsection{Basics and conventions}
\label{App_1_tau}

For an electric current, the key quantity   to find from the Boltzmann formalism is  
$\delta{\rm f}_{{\bf k},\sigma}\!=\!{\rm f}_{{\bf k},\sigma}\!-{\rm f}^0_{{\bf k},\sigma}$,
a deviation of  the non-equilibrium distribution function from the  equilibrium one, 
where ${\rm f}^0_{{\bf k},\sigma}$ is a Fermi-distribution function. Then, the 
current density is 
\begin{align}
\label{Boltz_j}
{\bf j}=\sum_\sigma\sum_{\bf k}e {\bf v}_{{\bf k}}\delta{\rm f}_{{\bf k},\sigma},
\end{align}
where ${\bf v}_{{\bf k}}\!=\!\partial \varepsilon_{{\bf k}}/\partial {\bf k}$
is the electron velocity. 

The linearized Boltzmann equation (LBE)
\begin{align}
\label{LBE_1}
e \big({\bf E}\cdot {\bf v}_{\bf k}\big)\,\frac{\partial {\rm f}^0_{\bf k}}{\partial \varepsilon_{\bf k}}
={\rm St}_{\bf k}[{\rm f}_{\bf k}],
\end{align}
allows to determine $\delta{\rm f}_{\bf k}$ in the linear-response approximation, 
here ${\bf E}$ is an electric field, ${\rm St}_{\bf k}[{\rm f}_{\bf k}]$ is a collision integral, 
and we ignore electron spin for a moment. 

The crucial  step is the relaxation-time approximation,
\begin{align}
\label{1_tau0}
{\rm St}_{\bf k}[{\rm f}_{\bf k}]\approx -\frac{\delta{\rm f}_{\bf k}}{\tau_{\bf k}},
\end{align}
which  yields the expected linear relation $j^\alpha\!=\!\sigma^{\alpha\beta} E^\beta$, with 
$\sigma^{\alpha\beta}$ being  conductivity tensor. Making further approximation 
of a continuum renders conductivity tensor diagonal, Fermi surface spherical (circular in 2D), and 
naturally suggests the same scattering rate for all $|{\bf k}|\!=\!k_F$. Together with a realization 
that the left-hand side of (\ref{LBE_1}) is singular at $T\!\ll\!E_F$ as 
$-\partial {\rm f}^0_{\bf k}/\partial \varepsilon_{\bf k}\!
\approx\!\delta(\varepsilon_{\bf k}-E_F)$ and using  the relation of the 
density of states at $E_F$ to the electronic density $n$,  leads to the standard
\begin{align}
\label{rho0}
\rho=\sigma^{-1}=\frac{m}{e^2 n}\cdot\frac{1}{\tau_F},
\end{align}
where $1/\tau_F\!=\!1/\tau_{k_F}$ is responsible for all temperature- and field-dependence 
of the resistivity in a metal.

While the achieved progress is not entirely hollow, it is clear that the problem of finding 
resistivity is now converted  to the problem of finding electron transport relaxation rate. The latter
depends on the type of scattering and on the details of microscopic interaction that 
determine the functional form of the collision integral in (\ref{LBE_1}). 
However, the technical problem at hand is more involved, as one needs to {\it rigorously} prove that 
the collision integral for a given type of scattering indeed yields the relaxation-rate 
approximation (\ref{1_tau0}) {\it and} to derive a microscopic expression for $1/\tau_F$ from 
it at the same time.

\subsection{Phonon-like spin-conserving scattering}
\label{App_non_sf}

\begin{figure}[t]
\includegraphics[width=\linewidth]{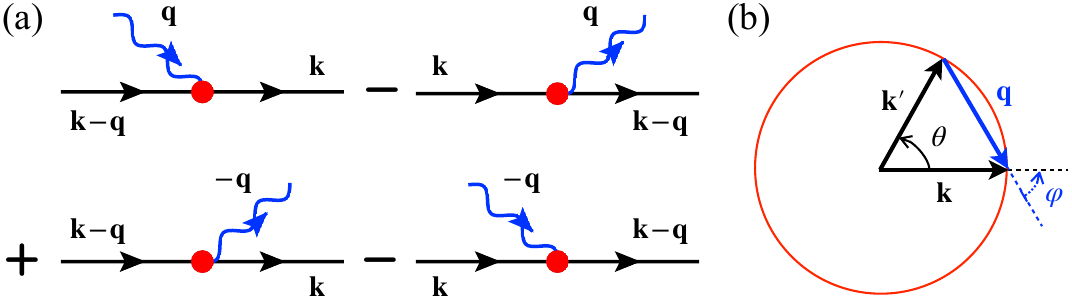}
\vskip -0.2cm
\caption{(a) Schematics of different terms in the collision integral. (b) Momenta of 
electrons and a magnon and their mutual angles.}
\label{Fig_St}
\vskip -0.3cm
\end{figure}

Let us first consider the problem of the non-spin-flip part of electron-magnon scattering 
in (\ref{H_Kondo_qp}). Since  scattering channels for  different 
magnon branches are additive and  electrons of opposite spin do not mix,  interaction 
Hamiltonian can be written in a more general form that is identical to that of electron-phonon coupling
\begin{align}
\label{H_e_ph}
{\cal H}_{int}=\frac{1}{\sqrt{N}}\sum_{{\bf k},{\bf q}} \Big(V_{\bf q}\,
f^\dag_{{\bf k}-{\bf q}}f^{\phantom{\dag}}_{{\bf k}} a^{\dag}_{\bf q} +  {\rm H.c.}\Big),
\end{align}
in which the coupling $V_{\bf q}$ depends only on the bosonic momentum because of the 
adiabatic principle, $E_F\!\gg\!\Theta_D$, where $\Theta_D$ is a phonon (magnon) Debye energy.

Collision integral for the scattering  (\ref{H_e_ph}), schematically 
represented in Fig.~\ref{Fig_St}(a),  has a general form
\begin{align}
\label{St1}
{\rm St}_{\bf k}&[{\rm f}_{\bf k}]=\frac{2\pi}{\hbar N} \sum_{\bf q}\big|V_{\bf q}\big|^2\\
\times&\,\Big\{
\Big({\rm f}_{\bf k'}(1-{\rm f}_{\bf k})\, {\rm n}_{\bf q}-{\rm f}_{\bf k}(1-{\rm f}_{\bf k'})({\rm n}_{\bf q}+1)\Big)\,
\delta_{\varepsilon_{\bf k'},\varepsilon_{\bf k}-\omega_{\bf q}}\nonumber\\
&+\!
\Big({\rm f}_{\bf k'}(1-{\rm f}_{\bf k})({\rm n}_{\bf q}+1)-{\rm f}_{\bf k}(1-{\rm f}_{\bf k'})\, {\rm n}_{\bf q}\Big)\,
\delta_{\varepsilon_{\bf k'},\varepsilon_{\bf k}+\omega_{\bf q}}\Big\},\nonumber
\end{align}
where ${\bf k'}\!=\!{\bf k}-{\bf q}$ and $\delta_{\varepsilon',\varepsilon\pm\omega}\!=\!
\delta(\varepsilon'-\varepsilon\mp\omega)$. Since bosons are in equilibrium, 
${\rm n}_{\bf q}\!=\!{\rm n}^0_{\bf q}$,
linearization of (\ref{St1}) gives
\begin{align}
\label{St_L}
{\rm St}_{\bf k}&[{\rm f}_{\bf k}]=\frac{2\pi}{\hbar N} \sum_{\bf q}\big|V_{\bf q}\big|^2\\
\times&\,\Big\{
\Big(\delta{\rm f}_{\bf k'}\big({\rm n}^0_{\bf q}+{\rm f}^0_{\bf k}\big)-
\delta{\rm f}_{\bf k}\big({\rm n}^0_{\bf q}-{\rm f}^0_{\bf k'}+1\big)\Big)\,
\delta_{\varepsilon_{\bf k'},\varepsilon_{\bf k}-\omega_{\bf q}}\nonumber\\
&+\!
\Big(\delta{\rm f}_{\bf k'}\big({\rm n}^0_{\bf q}-{\rm f}^0_{\bf k}+1\big)-
\delta{\rm f}_{\bf k}\big({\rm n}^0_{\bf q}+{\rm f}^0_{\bf k'}\big)\Big)\,
\delta_{\varepsilon_{\bf k'},\varepsilon_{\bf k}+\omega_{\bf q}}\Big\}.\nonumber
\end{align}

The physically justified shortcut from (\ref{St_L}) to the relaxation-time 
form (\ref{1_tau0})  capitalizes on the adiabatic approximation. 
It advocates an impurity-like, quasi-elastic picture of  the
scattering of the ``fast'' electrons on the ``slow'' bosons \cite{Ziman}, which should  adhere to the relaxation-time 
form by construction. In this case, bosonic thermal population plays the role of the thermally-excited 
``impurity'' concentration.  The resultant transport relaxation rate yields 
qualitatively correct answers  for the resistivity of a metal due to phonon 
scattering in both  low-  and  high-temperature limits, the Bloch-Gr\"{u}neisen and Ohm's laws, respectively. 

Technically, this approach amounts to neglecting $\omega_{\bf q}$ in the energy conservation in (\ref{St_L}),
which is indeed well-justified, and neglecting everything but ${\rm n}^0_{\bf q}$ in the innermost brackets 
of the collision integral in Eq.~(\ref{St_L}). It is the last step that is substantially harder to defend. 
While advanced monographs such as Ref.~\cite{Ziman_e_ph} offer a significantly more refined 
path that we partially follow, there are still additional constructs in it that seem unnecessary.

\subsubsection{Ansatz and solution}
\label{App_ansatz}

A more intelligent way of dealing with the collision integral in (\ref{St_L}) is to reflect on the ultimate form
of $\delta{\rm f}_{\bf k}$ that follows from the LBE (\ref{LBE_1}) and from the 
definition of the current in (\ref{Boltz_j}). This suggests an {\it ansatz}
\begin{align}
\label{df_ans}
\delta{\rm f}_{\bf k}=\frac{e}{m}\,\big({\bf E}\cdot {\bf k}\big)
\bigg(-\frac{\partial {\rm f}^0_{\bf k}}{\partial \varepsilon_{\bf k}}\bigg)\,\widetilde{\chi}_{\bf k}\, ,
\end{align}
which  adheres to the linear-response level of consideration, $\delta{\rm f}_{\bf k}\!\sim\!{\bf E}$,
and respects the reflection {\it anti}-symmetry of $\delta{\rm f}_{-\bf k}\!=\!-\delta{\rm f}_{\bf k}$ that follows
from (\ref{Boltz_j}). It also explicitly separates  the part of the LBE  solution  (\ref{LBE_1}) 
that is strongly peaked near the Fermi energy, $\partial {\rm f}^0_{\bf k}/\partial \varepsilon_{\bf k}$,
from the slowly varying $\widetilde{\chi}_{\bf k}$. By construction, the last quantity, 
$\widetilde{\chi}_{\bf k}$, has the meaning of the relaxation time $\tau_{\bf k}$ 
and is an even function of ${\bf k}$. Last but not the least, in (\ref{df_ans}) we  also use 
continuum-like approximations for electrons, which implies spherical (circular) Fermi surface and
allows us to replace  ${\bf v}_{\bf k}$ with ${\bf k}/m$, a move 
that makes the subsequent steps more straightforward.

We also note that since the dependence of the equilibrium distribution functions on the momentum 
is only via the energy, ${\rm f}^0_{\bf k}\!\equiv\!{\rm f}^0(\varepsilon_{\bf k})$ and
${\rm f}^0_{\bf k'}\!\equiv\!{\rm f}^0(\varepsilon_{\bf k}\pm\omega_{\bf q})$, there is an effective
separation of variables in the ansatz (\ref{df_ans}). This is because 
the singular component of the ansatz, $\partial {\rm f}^0_{\bf k}/\partial \varepsilon_{\bf k}$, 
depends sensitively on the differences $\pm\omega_{\bf q}$ due to scattering  near the Fermi energy, 
but not on the direction of the momentum ${\bf k}$. At the same time, the $({\bf E}\cdot {\bf k})$ 
factor in (\ref{df_ans}) is  solely responsible for the directional dependence of ${\bf k}$, but is largely 
insensitive to the variations in energy due to scattering, as is argued next.

A significant simplification follows from the hierarchy of  energy scales, 
$\varepsilon_{\bf k}\!\gg\!\omega_{\bf k}$. Since conducting electrons
are confined to a proximity of the Fermi energy, the energy conservation 
with an emission or absorption of a boson imply a quasi-elastic nature of the scattering process, 
$\varepsilon_{{\bf k}'}\!\approx\!\varepsilon_{\bf k}$, in a full accord with the qualitative logic outlined above. 
The  implication of this result  is  $|{\bf k}'|\!\approx\!|{\bf k}|$,
valid with the accuracy $O(\Theta_D/E_F)\!\ll\!1$. 
Thus, while one should keep small differences $\pm\omega_{\bf q}$  
in the singular component of $\delta{\rm f}_{{\bf k}({\bf k}')}$, it is only the direction of the 
momentum that can be changing significantly.
This makes  geometry of the scattering
in Fig.~\ref{Fig_St}(b) particularly simple and allows to rewrite $\delta$-functions in (\ref{St_L}) as 
\begin{align}
\label{delta_f}
\delta\left(\varepsilon_{\bf k'}-\varepsilon_{\bf k}\mp\omega_{\bf q}\right)\approx 
\left(\frac{m}{\hbar^2\,kq}\right) \delta\left(\cos\varphi-\frac{q}{2k}\right) ,
\end{align}
where $\varphi$ is the angle of ${\bf k}$ with ${\bf q}$, see Fig.~\ref{Fig_St}(b).

The second important implication of this consideration is for the relation of $\delta{\rm f}_{\bf k'}$
to $\delta{\rm f}_{\bf k}$. Using  (\ref{df_ans}), we write
\begin{align}
\label{df_ans_k'}
\delta{\rm f}_{\bf k'}=\frac{e}{m}\,\big({\bf E}\cdot {\bf k}'\big)
\bigg(-\frac{\partial {\rm f}^0_{\bf k'}}{\partial \varepsilon_{\bf k'}}\bigg)\,\widetilde{\chi}_{\bf k'}\, .
\end{align}
As is discussed already, 
${\rm f}^0_{\bf k'}\!\equiv\!{\rm f}^0(\varepsilon_{\bf k}\pm\omega_{\bf q})$ and $\widetilde{\chi}_{\bf k'}$
is a slowly varying  even function of ${\bf k'}$. Since $|{\bf k}'|\!\approx\!|{\bf k}|$, it is natural to assume 
$\widetilde{\chi}_{\bf k'}\!\approx\!\widetilde{\chi}_{\bf k}$. Consider  Fig.~\ref{Fig_St}(b).
One can break ${\bf k}'$ into components along and perpendicular to ${\bf k}$
\begin{align}
\label{k'_k}
{\bf k'}=\cos\theta \, {\bf k} + \sin\theta\, {\bf k}_\perp\, ,
\end{align}
with the latter component being odd with respect to the mirror reflection, $\theta\!\rightarrow\!-\theta$.
Making a reasonable assumption that $\big|V_{\bf q}\big|^2$ and magnon energies 
$\omega_{\bf q}$ are only weakly dependent on the {\it direction} of ${\bf q}$, it is easy to see that the 
rest of the integrand in (\ref{St_L}) is  even under the mirror reflection. Physically, 
the ${\bf k}_\perp$ component of the contribution of 
a scattered ${\bf k}'$ state to the collision integral is identically canceled by its mirror-pair. 

Thus, up to a factor $\cos\theta$ and a shift of energy in ${\rm f}^0$ from 
$\varepsilon_{\bf k}$ to $\varepsilon_{\bf k'}$, the non-vanishing contribution of $\delta{\rm f}_{\bf k'}$ 
to (\ref{St_L}) can be written as 
\begin{align}
\label{df_ans_k'_final}
\delta{\rm f}_{\bf k'}\Rightarrow
 \bigg(1-\frac{q^2}{2k^2}\bigg) \, \delta{\rm f}_{\bf k} (\varepsilon_{\bf k}\pm\omega_{\bf q})\,,
\end{align}
where we have converted  $\cos\theta$ to $(1-q^2/2k^2)$ using the trigonometry of  Fig.~\ref{Fig_St}(b). 
Combining (\ref{df_ans}), (\ref{delta_f}), and (\ref{df_ans_k'_final}), 
transforms collision integral in (\ref{St_L}) to
\begin{align}
\label{St_L1}
{\rm St}_{\bf k}[{\rm f}_{\bf k}]=\frac{e}{m}\,\big({\bf E}\cdot {\bf k}\big)\,\widetilde{\chi}_{\bf k}\,\,
\frac{2\pi}{\hbar^3 N} \sum_{\bf q}
&\big|V_{\bf q}\big|^2\left(\frac{m}{kq}\right)
\\
\times&\, \delta\left(\cos\varphi-\frac{q}{2k}\right)
\Big\{*\Big\},\nonumber
\end{align}
with 
\begin{align}
\Big\{*\Big\}=&\,\frac{1}{T}\,
\bigg\{\bigg(1-\frac{q^2}{2k^2}\bigg)\,
{\rm f}^0_{\varepsilon-\omega}\big(1-{\rm f}^0_{\varepsilon-\omega}\big)
\big({\rm n}^0_\omega+{\rm f}^0_{\varepsilon}\big)\nonumber\\
\label{St_L2}
&\phantom{\,\frac{1}{T}\,\bigg\{} 
-{\rm f}^0_{\varepsilon}\big(1-{\rm f}^0_{\varepsilon}\big)
\big({\rm n}^0_\omega+1-{\rm f}^0_{\varepsilon-\omega}\big)\\
&\phantom{\,\frac{1}{T}\,} 
+\bigg(1-\frac{q^2}{2k^2}\bigg)\,
{\rm f}^0_{\varepsilon+\omega}\big(1-{\rm f}^0_{\varepsilon+\omega}\big)
\big({\rm n}^0_\omega+1-{\rm f}^0_{\varepsilon}\big)\nonumber\\
&\phantom{\,\frac{1}{T}\,\bigg\{} 
-{\rm f}^0_{\varepsilon}\big(1-{\rm f}^0_{\varepsilon}\big)
\big({\rm n}^0_\omega+{\rm f}^0_{\varepsilon+\omega}\big)
\bigg\},\nonumber
\end{align}
where we used  shorthand notations ${\rm f}^0_{\bf k}\!=\!{\rm f}^0_\varepsilon$ and  
${\rm f}^0_{\bf k'}\!=\!{\rm f}^0_{\varepsilon\pm\omega}$ with
$\varepsilon_{\bf k}\!=\!\varepsilon$ and $\omega_{\bf q}\!=\!\omega$ and an identity
\begin{align}
\label{df0_de}
-\frac{\partial {\rm f}^0_\varepsilon}{\partial \varepsilon}=\frac{1}{T}\, 
{\rm f}^0_\varepsilon\big(1-{\rm f}^0_\varepsilon\big)\,.
\end{align}
For the sake of the subsequent discussion of the spin-flip scattering that involves fewer terms
in the collision integral, we note that the first (last) two lines in (\ref{St_L2}) correspond to 
the first (second) lines in the bracket in (\ref{St_L}) and Fig.~\ref{Fig_St}(a). 

While Eq.~(\ref{St_L2}) seems cumbersome and not intuitive, 
we will demonstrate that it is, in fact, {\it equivalent} to a 
compact and substantially more physical expression (\ref{St_L_final}), which, 
upon an appropriate linearization, will yield both the proof of the $1/\tau$ approximation 
for the collision integral and the sought-after transport relaxation rate (\ref{1_tau_final}). 

We note that the advocated approach of using ansatz of (\ref{df_ans}) and further manipulations with it
are not unfamiliar to the experts, see Refs.~\cite{Ziman_e_ph,Mannari1959,Mills1966}.  
However, they typically resort to an additional integration of both sides of the LBE (\ref{LBE_1}) 
over electronic energy in order to eliminate the singular component that is strongly peaked near the
Fermi energy. Below we demonstrate that such an {\it ad hoc} step is unnecessary.

First reorganization in (\ref{St_L2}) is an ``extraction'' of the bosonic distribution function 
that makes explicit the dependence of the scattering on the boson population
\begin{align}
&{\rm f}^0_{\varepsilon-\omega}\,\big({\rm n}^0_\omega+{\rm f}^0_{\varepsilon}\big)=\,
e^{\omega/T}\,{\rm f}^0_{\varepsilon}\,{\rm n}^0_\omega\,,\nonumber\\
\label{St_reorder1}
&{\rm f}^0_{\varepsilon+\omega}\,\big({\rm n}^0_\omega+1-{\rm f}^0_{\varepsilon}\big)=\,
{\rm f}^0_{\varepsilon}\,{\rm n}^0_\omega\,,
\end{align}
which, together with a ``trade-off of $\omega$,''
\begin{align}
\label{St_reorder2}
e^{\omega/T}\,{\rm f}^0_{\varepsilon}\,\big(1-{\rm f}^0_{\varepsilon-\omega}\big)=\,
{\rm f}^0_{\varepsilon-\omega}\,\big(1-{\rm f}^0_{\varepsilon}\big)\,,
\end{align}
reduces (\ref{St_L2}) to
\begin{equation}
\label{St_L3}
\Big\{*\Big\}\!=\!-\bigg(\frac{q^2}{2k^2}\bigg)\,\frac{{\rm n}^0_\omega}{T}\,
\Big\{{\rm f}^0_{\varepsilon-\omega}\big(1-{\rm f}^0_{\varepsilon}\big)
+ {\rm f}^0_{\varepsilon}\big(1-{\rm f}^0_{\varepsilon+\omega}\big)\Big\}.
\end{equation}
Last  is a ``bosonization'' of the double-fermionic terms
\begin{align}
\label{St_reorder3}
{\rm f}^0_{\varepsilon-\omega}\,\big(1-{\rm f}^0_{\varepsilon}\big)=\,
\big({\rm n}^0_\omega+1\big)\,\big({\rm f}^0_{\varepsilon-\omega}-{\rm f}^0_{\varepsilon}\big)\,,
\end{align}
that leads to the final form
\begin{align}
 \label{St_L_final}
\Big\{*\Big\}\equiv&\,\frac{1}{T}\,\bigg(\frac{q^2}{2k^2}\bigg)\,{\rm n}^0_\omega\big({\rm n}^0_\omega+1\big)
\Big\{{\rm f}^0_{\varepsilon+\omega}-{\rm f}^0_{\varepsilon-\omega}\Big\}\,.
\end{align}
where the naturally-occurring factor $q^2/2k^2\!=\!(1-\cos\theta)$ differentiates
transport scattering rate from the conventional one.
The  bracket in (\ref{St_L_final}) can now  be  expanded 
\begin{align}
 \label{St_expand}
\Big\{{\rm f}^0_{\varepsilon+\omega}-{\rm f}^0_{\varepsilon-\omega}\Big\}\approx 2\,\omega
\bigg(\frac{\partial {\rm f}^0_\varepsilon}{\partial \varepsilon}\bigg)\,,
\end{align}
to yield the singular part of $\delta{\rm f}_{\bf k}$. 
Combining (\ref{St_L_final}) and (\ref{St_expand}), we rewrite collision integral   (\ref{St_L1})   
\begin{align}
\label{St_L_all}
{\rm St}_{\bf k}&[{\rm f}_{\bf k}]=-\bigg[\frac{e}{m}\,\big({\bf E}\cdot {\bf k}\big)\,
\bigg(-\frac{\partial {\rm f}^0_{\bf k}}{\partial \varepsilon_{\bf k}}\bigg)\,
\widetilde{\chi}_{\bf k}\bigg]\,\, \\
\times&\,
\frac{2\pi}{\hbar^3 N} \sum_{\bf q}
\big|V_{\bf q}\big|^2 \bigg(\frac{mq}{k^3}\bigg)\,\delta\left(\cos\varphi-\frac{q}{2k}\right)\,
\frac{\omega_{\bf q}}{T}\,{\rm n}^0_{\bf q}\,\big({\rm n}^0_{\bf q}+1\big),\nonumber
\end{align}
in the long-pursued relaxation-time form (\ref{1_tau0}),
which {\it naturally} replicates $\delta{\rm f}_{\bf k}$ ansatz (\ref{df_ans}) highlighted with the 
square bracket,  also yielding  transport relaxation rate 
\begin{align}
\label{1_tau_final}
\frac{\hbar}{\tau_{\bf k}}=
\frac{\pi}{\varepsilon_{\bf k}N} \sum_{\bf q}\big|V_{\bf q}\big|^2 \bigg(\frac{q}{k}\bigg)
\delta\Big(\cos\varphi \,-&\,\frac{q}{2k}\Big)\\ 
\times&\,\bigg(\frac{\omega_{\bf q}}{T}\bigg)\,{\rm n}^0_{\bf q}\,\big({\rm n}^0_{\bf q}+1\big),
\nonumber
\end{align}
where we have used  $\varepsilon_{\bf k}\!=\!\hbar^2k^2/2m$. 
We also point out that a comparison of the LBE (\ref{LBE_1}) with the collision integral in 
(\ref{St_L_all}) makes explicit the equivalence of the auxiliary function $\widetilde{\chi}_{\bf k}$ from the ansatz 
(\ref{df_ans}) with the transport relaxation time  $\tau_{\bf k}$ in (\ref{1_tau_final}).

One can easily check by  power-counting that for 
the scattering on acoustic phonons ($V_{\bf q}\!\propto\!\sqrt{q}$, 
$\omega_{\bf q}\!\propto\!q$), the relaxation rate   (\ref{1_tau_final}) 
yields Ohm's law, $\rho\!\sim\!T$, for $T\!\gg\!\Theta_D$ and Bloch-Gr\"{u}neisen's law, 
$\rho\!\sim\!T^5$  in 3D, for  $T\!\ll\!\Theta_D$.

We also note that while the provided consideration is for a spherical (circular) Fermi surface,
there is no question in our mind that the results concerning  validity of the 
$1/\tau$-approximation remain correct in general.  The main modification in the relaxation rate 
(\ref{1_tau_final}) should be expected in the constraint on the surface of integration (contour in 2D), 
dictated by the actual shape of the Fermi surface. Thus, one can expect the results 
obtained using this approximation to be quantitatively correct.

\subsubsection{2D case}
\label{App_2D_1_tau}

In the present study, magnons are 2D and the relaxation rate (\ref{1_tau_final}) is given by an integral
along a circular 1D contour in Fig.~\ref{Fig_St}(b) of the radius $k_F$ 
with the magnon momentum varying from 0 to $2k_F$. 
Assuming that $2k_F$ is less than the size of the full BZ, one can make further progress 
and simplify (\ref{1_tau_final}) by an appropriate parametrization. 
Let us first rewrite the summation over the BZ in (\ref{1_tau_final}) as an integral in polar coordinates
\begin{equation}
\label{1Dint_a}
\frac{1}{V_{BZ}}\int  q\,dq \int_0^{2\pi} d\varphi
 \bigg(\frac{q}{k_F}\bigg)
\,\delta\left(\cos\varphi-\frac{q}{2k_F}\right)\widetilde{F}_{\bf q},
\end{equation}
where we have abbreviated part of the integrand as
\begin{equation}
\label{Faux}
\widetilde{F}_{\bf q} = \big|V_{\bf q}\big|^2
\bigg(\frac{\omega_{\bf q}}{T}\bigg)\,{\rm n}^0_{\bf q}\,\big({\rm n}^0_{\bf q}+1\big),
\end{equation}
and $V_{BZ}\!=\!8\pi^2/a^2\sqrt{3}$ is the 2D volume of the triangular-lattice BZ in Fig.~\ref{Fig_BZs}. 

Introducing new variables $z\!=\!\cos\varphi$ and $y\!=\!q/2k_F$ and using the mirror
symmetry of $\widetilde{F}_{\bf q}$ with respect to $\varphi\!\rightarrow\!-\varphi$
simplifies (\ref{1Dint_a}) to
\begin{equation}
\label{1Dint_b}
\frac{\sqrt{3}\, (k_F a)^2}{\pi^2}\int  y^2\,dy \int_0^1 \frac{dz}{\sqrt{1-z^2}}
\,\delta\left(z-y\right)\, 2\widetilde{F}_{\bf q},
\end{equation}
with the momentum ${\bf q}$ belonging to the 1D contour given by the  
parametrization  ${\bf q}\!=\!2k_F(z^2,z\sqrt{1-z^2})$.
Finally,  Eq.~(\ref{1_tau_final}) transforms to 
\begin{align}
\label{1_tau_1Da}
\frac{\hbar}{\tau_{F}}=\frac{\sqrt{3}\, (k_F a)^2}{\pi \,E_F} \int_0^1 \frac{z^2\, dz}{\sqrt{1-z^2}}
\, 2\,\widetilde{F}_{\bf q},
\end{align}
with $\widetilde{F}_{\bf q}$ from (\ref{Faux}), $E_F\!=\!\hbar^2k_F^2/2m$, and we have
explicitly separated a factor of two for the sake of combining this relaxation rate result with the one
due to the spin-flip scattering processes, discussed next.

\begin{figure}[t]
\hskip 1.3cm \includegraphics[width=0.7\linewidth]{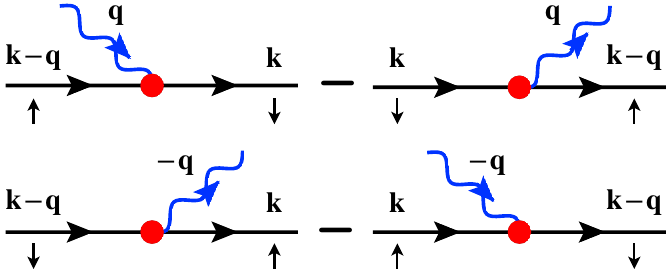}\hfill\
\vskip -0.2cm
\caption{Schematics of the collision integrals in (\ref{St1FM}).}
\label{Fig_StFM}
\vskip -0.3cm
\end{figure}

\subsection{Spin-flip scattering}
\label{App_sf}

The spin-flip component of electron-magnon scattering in (\ref{H_Kondo_qp})
generally contains two  parts, one with emission by spin-down  and one by spin-up electrons,
yielding scattering rates that  have the same structure and are additive.  Therefore,
we consider only one of them and write the interaction Hamiltonian  in a  general form that is 
identical to that of electron-magnon coupling in a ferromagnet
\begin{align}
\label{H_s_f}
{\cal H}_{int}=\frac{1}{\sqrt{N}}\sum_{{\bf k},{\bf q}} \Big(V^\pm_{\bf q}
f^\dag_{{\bf k}-{\bf q}\uparrow}f^{\phantom{\dag}}_{{\bf k}\downarrow} a^{\dag}_{\bf q} +  {\rm H.c.}\Big).
\end{align}
Here, similarly to Eq.~(\ref{H_e_ph}), the coupling $V^\pm_{\bf q}$  depends only on ${\bf q}$ 
because of  its local (Kondo) nature.

Owing to its relevance to the electron-magnon scattering mechanism of the resistivity in ferromagnets, 
the kinetic theory for the model (\ref{H_s_f}) has been the subject of a number of 
works \cite{Mannari1959,Mills1966,Turov1955,Vonsovskii}, which, while yielding correct results, 
have also used unnecessary {\it ad hoc} integrations. 

The principal difference of the problem from the spin-conserving scattering 
is in having two Boltzmann equations, one per spin projection,
with their collision integrals each containing only half of the terms of (\ref{St1}), 
\begin{align}
{\rm St}_{\bf k}&[{\rm f}_{{\bf k}\downarrow}]=\frac{2\pi}{\hbar N} \sum_{\bf q}\big|V^\pm_{\bf q}\big|^2
\delta\left(\varepsilon_{{\bf k'}\uparrow}-\varepsilon_{{\bf k}\downarrow}+\omega_{\bf q}\right)
\nonumber\\
\times&\,\Big\{
\Big({\rm f}_{{\bf k'}\uparrow}(1-{\rm f}_{{\bf k}\downarrow})\, {\rm n}_{\bf q}-
{\rm f}_{{\bf k}\downarrow}(1-{\rm f}_{{\bf k'}\uparrow})({\rm n}_{\bf q}+1)\Big)\Big\},
\nonumber\\
\label{St1FM}
{\rm St}_{\bf k}&[{\rm f}_{{\bf k}\uparrow}]=\frac{2\pi}{\hbar N} \sum_{\bf q}\big|V^\pm_{\bf q}\big|^2
\delta\left(\varepsilon_{{\bf k'}\downarrow}-\varepsilon_{{\bf k}\uparrow}-\omega_{\bf q}\right)\\
\times&\,\Big\{
\Big({\rm f}_{{\bf k'}\downarrow}(1-{\rm f}_{{\bf k}\uparrow})({\rm n}_{\bf q}+1)-
{\rm f}_{{\bf k}\uparrow}(1-{\rm f}_{{\bf k'}\downarrow})\Big)\, {\rm n}_{\bf q}\Big\},
\nonumber
\end{align}
with the energy $\delta$-functions as in the first and second rows of (\ref{St1}), respectively, 
see also Fig.~\ref{Fig_StFM}, ${\bf k'}\!=\!{\bf k}-{\bf q}$ as before.
Linearization in (\ref{St1FM}) gives
\begin{align}
{\rm St}_{\bf k}&[{\rm f}_{{\bf k}\downarrow}]=\frac{2\pi}{\hbar N} \sum_{\bf q}\big|V^\pm_{\bf q}\big|^2
\delta\left(\varepsilon_{{\bf k'}\uparrow}-\varepsilon_{{\bf k}\downarrow}+\omega_{\bf q}\right)
\nonumber\\
\times&\,\Big\{
\Big(\delta{\rm f}_{{\bf k'}\uparrow}({\rm n}^0_{\bf q}+{\rm f}^0_{{\bf k}\downarrow}) -
\delta{\rm f}_{{\bf k}\downarrow}({\rm n}^0_{\bf q}-{\rm f}^0_{{\bf k'}\uparrow}+1)\Big\},
\nonumber\\
\label{St_LFM}
{\rm St}_{\bf k}&[{\rm f}_{{\bf k}\uparrow}]=\frac{2\pi}{\hbar N} \sum_{\bf q}\big|V^\pm_{\bf q}\big|^2
\delta\left(\varepsilon_{{\bf k'}\downarrow}-\varepsilon_{{\bf k}\uparrow}-\omega_{\bf q}\right)\\
\times&\,\Big\{
\Big(\delta{\rm f}_{{\bf k'}\downarrow}({\rm n}^0_{\bf q}-{\rm f}^0_{{\bf k}\uparrow}+1)-
\delta{\rm f}_{{\bf k}\uparrow}({\rm n}^0_{\bf q}+{\rm f}^0_{{\bf k'}\downarrow})\Big)\Big\}.
\nonumber
\end{align}
We follow the narrative of the non-spin-flip consideration, with the  
ansatzes for $\delta{\rm f}_{{\bf k}\sigma}$ as in  (\ref{df_ans})
\begin{align}
\label{df_ansFM}
\delta{\rm f}_{{\bf k}\sigma}=\frac{e}{m}\,\big({\bf E}\cdot {\bf k}\big)
\bigg(-\frac{\partial {\rm f}^0_{{\bf k}\sigma}}{\partial \varepsilon_{{\bf k}\sigma}}\bigg)\,
\widetilde{\chi}_{{\bf k}\sigma}\, ,
\end{align}
and with the same continuum-like approximation for electrons. 
In the present case, it is also important to recall that the 
equilibrium distribution functions depend on the momentum {\it and} spin only via energy, 
${\rm f}^0_{{\bf k}\sigma}\!\equiv\!{\rm f}^0(\varepsilon_{{\bf k}\sigma})$, so that 
${\rm f}^0(\varepsilon_{{\bf k}'\uparrow})\!\equiv\!
{\rm f}^0(\varepsilon_{{\bf k}\downarrow}-\omega_{\bf q})$ 
and ${\rm f}^0(\varepsilon_{{\bf k}'\downarrow})\!\equiv\!
{\rm f}^0(\varepsilon_{{\bf k}\uparrow}+\omega_{\bf q})$.

In the considered case of the field-induced effects in an otherwise spin-compensated system,
the only source of the difference in the Fermi momenta for the  electrons of different spin is  
Zeeman energy. Because $g\mu_BH\!\ll\!E_F$, using the same consideration of 
the quasielastic nature of electron-magnon scattering that is given before Eq.~(\ref{delta_f}), leads to
$\varepsilon_{{\bf k}'}\!\approx\!\varepsilon_{\bf k}$ and $|{\bf k}'|\!\approx\!|{\bf k}|$ in 
all scattering processes regardless of the spin, keeping geometry of the scattering the same as in
Fig.~\ref{Fig_St}(b) and allowing us to rewrite $\delta$-functions in (\ref{St_LFM}) in the form of 
Eq.~(\ref{delta_f}). 

In principle, the same consideration implies $k_{F\uparrow}\!\approx\!k_{F\downarrow}$ 
with the same accuracy of $O(g\mu_BH/E_F)\!\ll\!1$ and suggests that 
$\widetilde{\chi}_{{\bf k}\uparrow}\!\approx\!\widetilde{\chi}_{{\bf k}\downarrow}$, but we 
will arrive to the same conclusion via a different path.

Using the logic leading to (\ref{df_ans_k'_final}), we can rewrite
the non-vanishing contributions of $\delta{\rm f}_{{\bf k}'\sigma}$ to (\ref{St_LFM}) as
\begin{align}
\label{df_ans_k'_finalFM}
\delta{\rm f}_{{\bf k}'\uparrow}\Rightarrow&\,
 \bigg(1-\frac{q^2}{2k^2}\bigg) \, 
 \delta{\rm f}_{{\bf k}\uparrow} \big(\varepsilon_{{\bf k}\downarrow}-\omega_{\bf q}\big)\,,\\
\delta{\rm f}_{{\bf k}'\downarrow}\Rightarrow&\,
 \bigg(1-\frac{q^2}{2k^2}\bigg) \, 
 \delta{\rm f}_{{\bf k}\downarrow} \big(\varepsilon_{{\bf k}\uparrow}+\omega_{\bf q}\big)\,.
 \nonumber
\end{align}
Combining (\ref{df_ansFM}), (\ref{delta_f}), and (\ref{df_ans_k'_finalFM}), 
transforms collision integrals in (\ref{St_LFM}) to
\begin{align}
\label{St_L1FM}
{\rm St}_{\bf k}[{\rm f}_{{\bf k}\sigma}]=\frac{e}{m}\,\big({\bf E}\cdot {\bf k}\big)\,\,
\frac{2\pi}{\hbar^3 N} \sum_{\bf q}
&\big|V^\pm_{\bf q}\big|^2\left(\frac{m}{kq}\right)
\\
\times&\, \delta\left(\cos\varphi-\frac{q}{2k}\right)
\Big\{\sigma\Big\},\nonumber
\end{align}
with 
\begin{align}
\Big\{\downarrow\Big\}=& \frac{1}{T} 
\bigg\{\widetilde{\chi}_{{\bf k}\uparrow}\bigg(1-\frac{q^2}{2k^2}\bigg)
{\rm f}^0_{\varepsilon-\omega}\big(1-{\rm f}^0_{\varepsilon-\omega}\big)
\big({\rm n}^0_\omega+{\rm f}^0_{\varepsilon}\big)\nonumber\\
\label{St_L2FM}
&\phantom{\frac{1}{T}\bigg\{} 
-\widetilde{\chi}_{{\bf k}\downarrow}{\rm f}^0_{\varepsilon}\big(1-{\rm f}^0_{\varepsilon}\big)
\big({\rm n}^0_\omega+1-{\rm f}^0_{\varepsilon-\omega}\big)\bigg\},\  \\
\Big\{\uparrow\Big\}=& \frac{1}{T}
\bigg\{\widetilde{\chi}_{{\bf k}\downarrow}\bigg(1-\frac{q^2}{2k^2}\bigg)
{\rm f}^0_{\varepsilon+\omega}\big(1-{\rm f}^0_{\varepsilon+\omega}\big)
\big({\rm n}^0_\omega+1-{\rm f}^0_{\varepsilon}\big)\nonumber\\
&\phantom{\frac{1}{T}\bigg\{} 
-\widetilde{\chi}_{{\bf k}\uparrow}{\rm f}^0_{\varepsilon}\big(1-{\rm f}^0_{\varepsilon}\big)
\big({\rm n}^0_\omega+{\rm f}^0_{\varepsilon+\omega}\big)
\bigg\},\nonumber
\end{align}
where $\varepsilon\!=\!\varepsilon_{{\bf k}\sigma}$ and $\omega\!=\!\omega_{\bf q}$.

The same sequence of manipulations as in (\ref{St_reorder1}), (\ref{St_reorder2}), and  (\ref{St_reorder3})
reduce Eq.~(\ref{St_L2FM}) identically to
\begin{align}
\Big\{\downarrow\Big\}\equiv\frac{1}{T}\,
{\rm n}^0_\omega\big({\rm n}^0_\omega+1\big)
\bigg[\widetilde{\chi}_{{\bf k}\downarrow}-&\,
\widetilde{\chi}_{{\bf k}\uparrow}\bigg(1-\frac{q^2}{2k^2}\bigg)\bigg]\nonumber\\
\label{St_L_finalFM}
&\times\Big\{{\rm f}^0_{\varepsilon}-{\rm f}^0_{\varepsilon-\omega}\Big\},\\
\Big\{\uparrow\Big\}\equiv\frac{1}{T}\,
{\rm n}^0_\omega\big({\rm n}^0_\omega+1\big)
\bigg[\widetilde{\chi}_{{\bf k}\uparrow}-&\,
\widetilde{\chi}_{{\bf k}\downarrow}\bigg(1-\frac{q^2}{2k^2}\bigg)\bigg]\nonumber\\
&\times\Big\{{\rm f}^0_{\varepsilon+\omega}-{\rm f}^0_{\varepsilon}\Big\},
\nonumber
\end{align}
which, in a retrospect, shows that the two groups of terms in the non-spin-flip consideration 
(\ref{St_L2}) and Fig.~\ref{Fig_St}, associated with two different energy $\delta$-functions, can be 
reduced to a compact form (\ref{St_L_final}) individually. 

Linearization in the last brackets in (\ref{St_L_finalFM}) yields 
\begin{align}
 \label{St_expandFM}
\Big\{\dots\Big\}\approx \,\omega
\bigg(\frac{\partial {\rm f}^0_\varepsilon}{\partial \varepsilon}\bigg)\,,
\end{align}
where we note the factor of two difference with the result of an equivalent step in (\ref{St_expand}).
Bringing together (\ref{St_L_finalFM}) and (\ref{St_expandFM}) transforms collision integrals 
in (\ref{St_L1FM})  to 
\begin{equation}
\label{St_L_allFM}
{\rm St}_{\bf k}[{\rm f}_{{\bf k}\sigma}]\!=\!\bigg[\frac{e}{m}\big({\bf E}\cdot {\bf k}\big)
\bigg(\frac{\partial {\rm f}^0_{{\bf k}\sigma}}{\partial \varepsilon_{{\bf k}\sigma}}\bigg)\bigg]
\Big(A_{\bf k}\widetilde{\chi}_{{\bf k}\sigma}-B_{\bf k}
\widetilde{\chi}_{{\bf k}\bar{\sigma}}\Big),
\end{equation}
where $\bar{\sigma}\!=\!-\sigma$ and $A_{\bf k}$ and $B_{\bf k}$ are auxiliary functions.
Given the symmetry of (\ref{St_L_allFM}) under $\{\uparrow\leftrightarrow\downarrow\}$, one 
can anticipate the following result. The easiest way to proceed is to realize that the content of the 
square bracket in the collision integrals in (\ref{St_L_allFM}) is exactly the left-hand side 
of their corresponding LBEs. Substitution of (\ref{St_L_allFM}) into LBEs with a cancellation
in both sides reduces them to the two identical equations
\begin{align}
\label{St_L_allFMa}
1=&\,A_{\bf k}\widetilde{\chi}_{{\bf k}\sigma}-B_{\bf k}
\widetilde{\chi}_{{\bf k}\bar{\sigma}}\,,
\end{align}
which finally gives $\widetilde{\chi}_{{\bf k}\uparrow}\!=\!\widetilde{\chi}_{{\bf k}\downarrow}$
and provides the relaxation rate for both spin projections in the  form
\begin{align}
\label{1_tau_finalFM}
\frac{\hbar}{\tau_{{\bf k}\sigma}}=\,
\frac{\pi}{2\varepsilon_{\bf k}N} \sum_{\bf q} \big|V^\pm_{\bf q}\big|^2 \bigg(\frac{q}{k}\bigg)
\delta\Big(\cos\varphi\,&-\,\frac{q}{2k}\Big)\\ 
\times&\,\bigg(\frac{\omega_{\bf q}}{T}\bigg)\,{\rm n}^0_{\bf q}\,\big({\rm n}^0_{\bf q}+1\big),
\nonumber
\end{align}
where we note an extra factor $1/2$ compared to (\ref{1_tau_final}).

Further adaptation to the 2D case with spins on  a triangular lattice 
considered in the preceding section gives the same result as in  (\ref{1_tau_1Da})
\begin{align}
\label{1_tau_1Db}
\frac{\hbar}{\tau_{F\sigma}}=\frac{\sqrt{3}\, (k_F a)^2}{\pi \,E_F} \int_0^1 \frac{z^2\, dz}{\sqrt{1-z^2}}
\, \widetilde{F}^\pm_{\bf q},
\end{align}
up to a factor of two  that can be traced to the expansion in
(\ref{St_expand}) and (\ref{St_expandFM}), and, ultimately, to the fact that the spin-flip scattering 
in (\ref{H_s_f}) involves only half the terms of the non-spin-flip one in (\ref{H_e_ph}) for each spin species.
Here $\widetilde{F}^\pm_{\bf q}$ is given by 
\begin{equation}
\label{FauxFM}
\widetilde{F}^\pm_{\bf q} = \big|V^\pm_{\bf q}\big|^2
\bigg(\frac{\omega_{\bf q}}{T}\bigg)\,{\rm n}^0_{\bf q}\,\big({\rm n}^0_{\bf q}+1\big).
\end{equation}

For the full interaction given in (\ref{H_Kondo_qp}), 
collecting all contributions to the scattering rate  
yields the final formula in (\ref{1_tau})-(\ref{1_tau_int}), which takes into account 
three different branches of magnons and all channels of scattering, spin-flip and not.

\subsection{Quasiparticle $1/\tau_{\rm qp}$, angular dependence}
\label{Sec_1_tau_qp}

Our consideration of the transport relaxation rate due to magnon scattering 
would not be complete without  a brief note on the ``regular,'' or quasiparticle, relaxation rate  
and an equally brief remark of the relationship between the two. 
In turn, this analysis also allows us to discuss the angular dependence of the quasiparticle relaxation rate, 
originating from the discrete lattice symmetry that is encoded in the energies of spin excitations and 
matrix elements of the coupling to them, while leaving a substantially more involved study of their 
effect on its transport counterpart to a future work. 

\subsubsection{Quasiparticle vs transport $1/\tau$}
\label{Sec_1_tau_qp_1}

The quasiparticle relaxation rate characterizes a probability of the scattering of an electron 
from the momentum-${\bf k}$ state. It is given by
\begin{align}
\label{1_tau_qp0}
\frac{\hbar}{\tau_{\rm qp, {\bf k}}}=-2\,{\rm Im}\big[\Sigma_{\bf k}(\varepsilon_{\bf k})\big], 
\end{align}
where $\Sigma_{\bf k}(\varepsilon_{\bf k})$ is the on-shell electron self-energy. 

The phonon-like non-spin-flip scattering of the form (\ref{H_e_ph}) 
yields a well-known answer for ${\rm Im}\big[\Sigma_{\bf k}(\varepsilon_{\bf k})\big]$  
via a standard second-order diagrammatic treatment \cite{Mahan}
\begin{align}
\label{1_tau_qp}
\frac{\hbar}{\tau_{\rm qp, {\bf k}}}=\frac{2\pi}{N}
\sum_{\bf q}\big|V_{\bf q}\big|^2
&\,\Big\{
\big({\rm n}^0_{\bf q}-{\rm f}^0_{\bf k'}+1\big)\,
\delta_{\varepsilon_{\bf k'},\varepsilon_{\bf k}-\omega_{\bf q}}\\
&+\!
\big({\rm n}^0_{\bf q}+{\rm f}^0_{\bf k'}\big)\,
\delta_{\varepsilon_{\bf k'},\varepsilon_{\bf k}+\omega_{\bf q}}\Big\},\nonumber
\end{align}
where ${\bf k'}\!=\!{\bf k}-{\bf q}$ and $\delta_{\varepsilon',\varepsilon\pm\omega}\!=\!
\delta(\varepsilon'-\varepsilon\mp\omega)$ as before.

While looking deceitfully different from its transport counterpart in (\ref{St_L1}),
simple manipulations using  (\ref{St_reorder2}) and (\ref{St_reorder3}) bring the 
combinations of the  distribution functions in the first and second terms of (\ref{1_tau_qp}) to 
\begin{align}
 \label{1_tau_qp_relation}
\mp\,{\rm n}^0_\omega\big({\rm n}^0_\omega+1\big)\,
\frac{\big({\rm f}^0_{\varepsilon}-{\rm f}^0_{\varepsilon\mp\omega}\big)}
{{\rm f}^0_\varepsilon\big(1-{\rm f}^0_\varepsilon\big)}\,,
\end{align}
respectively. Recognizing the denominator in (\ref{1_tau_qp_relation}) 
for $(-T)(\partial {\rm f}^0_\varepsilon/\partial \varepsilon)$ and
expanding the numerators in $\omega$ gives
\begin{eqnarray}
\label{1_tau_qp1}
\frac{\hbar}{\tau_{\rm qp, {\bf k}}}\approx \frac{4\pi}{N}
\sum_{\bf q}\big|V_{\bf q}\big|^2
\,\frac{\omega_{\bf q}}{T}\,{\rm n}^0_{\bf q}\,\big({\rm n}^0_{\bf q}+1\big)
\, \delta\big({\varepsilon_{{\bf k}-{\bf q}}-\varepsilon_{\bf k}}\big), \ \ \ \ \ \ \
\end{eqnarray}
where we have also neglected $\omega_{\bf q}$ in the  delta-function.
Further simplification of the latter for the spherical (cylindrical) Fermi surface  (\ref{delta_f})
and parametrization of the electron mass the with energy  $\varepsilon_{\bf k}\!=\!\hbar^2k^2/2m$
used above, brings the quasiparticle relaxation rate in (\ref{1_tau_qp1}) close to an expression of its 
transport version  (\ref{1_tau_final})
\begin{align}
\label{1_tau_qp_final}
\frac{\hbar}{\tau_{{\rm qp},{\bf k}}}=
\frac{\pi}{\varepsilon_{\bf k}N} \sum_{\bf q}\big|V_{\bf q}\big|^2 \bigg(\frac{2k}{q}\bigg)
\delta\Big(\cos\varphi \,&-\,\frac{q}{2k}\Big)\\ 
\times&\,\bigg(\frac{\omega_{\bf q}}{T}\bigg)\,{\rm n}^0_{\bf q}\,\big({\rm n}^0_{\bf q}+1\big).
\nonumber
\end{align}
As is expected, the only difference between the two rates is an extra factor
$q^2/2k^2\!\equiv\!(1-\cos\theta)\!=\!2\cos^2\varphi$  in the integrand of the 
transport relaxation rate that is responsible for the suppression of the small-angle scattering in it;
as before, $\theta$ and $\varphi$ are the angles between ${\bf k}$ and ${\bf k}'$ and   
${\bf k}$ and ${\bf q}$, respectively, see Fig.~\ref{Fig_St}(b). 

Using the same approximations as in Appendix~\ref{App_non_sf} and considering
the case of 2D, the quasiparticle relaxation rate  (\ref{1_tau_qp_final})  at $k\!=\!k_F$ reduces to 
\begin{align}
\label{1_tau_qp_1D}
\frac{\hbar}{\tau_{{\rm qp},F}}=\frac{\sqrt{3}\, (k_F a)^2}{\pi \,E_F} \int_0^1 \frac{dz}{\sqrt{1-z^2}}
\,\widetilde{F}_{\bf q},
\end{align}
with $\widetilde{F}_{\bf q}$ from (\ref{Faux}),   ${\bf q}\!=\!2k_F(z^2,z\sqrt{1-z^2})$, 
and the difference from the transport rate in
(\ref{1_tau_1Da}) given by an extra factor $2z^2$ in the latter. Needless to say, for the spin-flip scattering 
(\ref{H_s_f}), an equivalent of the self-energy in (\ref{1_tau_qp})  contains only half the terms,
resulting in an extra factor 1/2 in (\ref{1_tau_qp_final}) and (\ref{1_tau_qp_1D}), 
identically to the transport case (\ref{1_tau_finalFM}) and (\ref{1_tau_1Db}) 
considered in Appendix~\ref{App_sf}.

While the technical tricks of converting standard  expression 
for the quasiparticle relaxation rate (\ref{1_tau_qp}) into the form that is akin to the transport 
one (\ref{1_tau_qp_final}) may not be well-known, the resultant similarity between them is in a 
full accord with the textbook expectations.
Thus, it is  prudent on our side to investigate the resulting difference between the two rates for our 
problem with the actual electron-magnon couplings and energies using Eqs.~(\ref{1_tau})-(\ref{matrix_elements})
and their quasiparticle analogues. 

\begin{figure}[t]
\includegraphics[width=\linewidth]{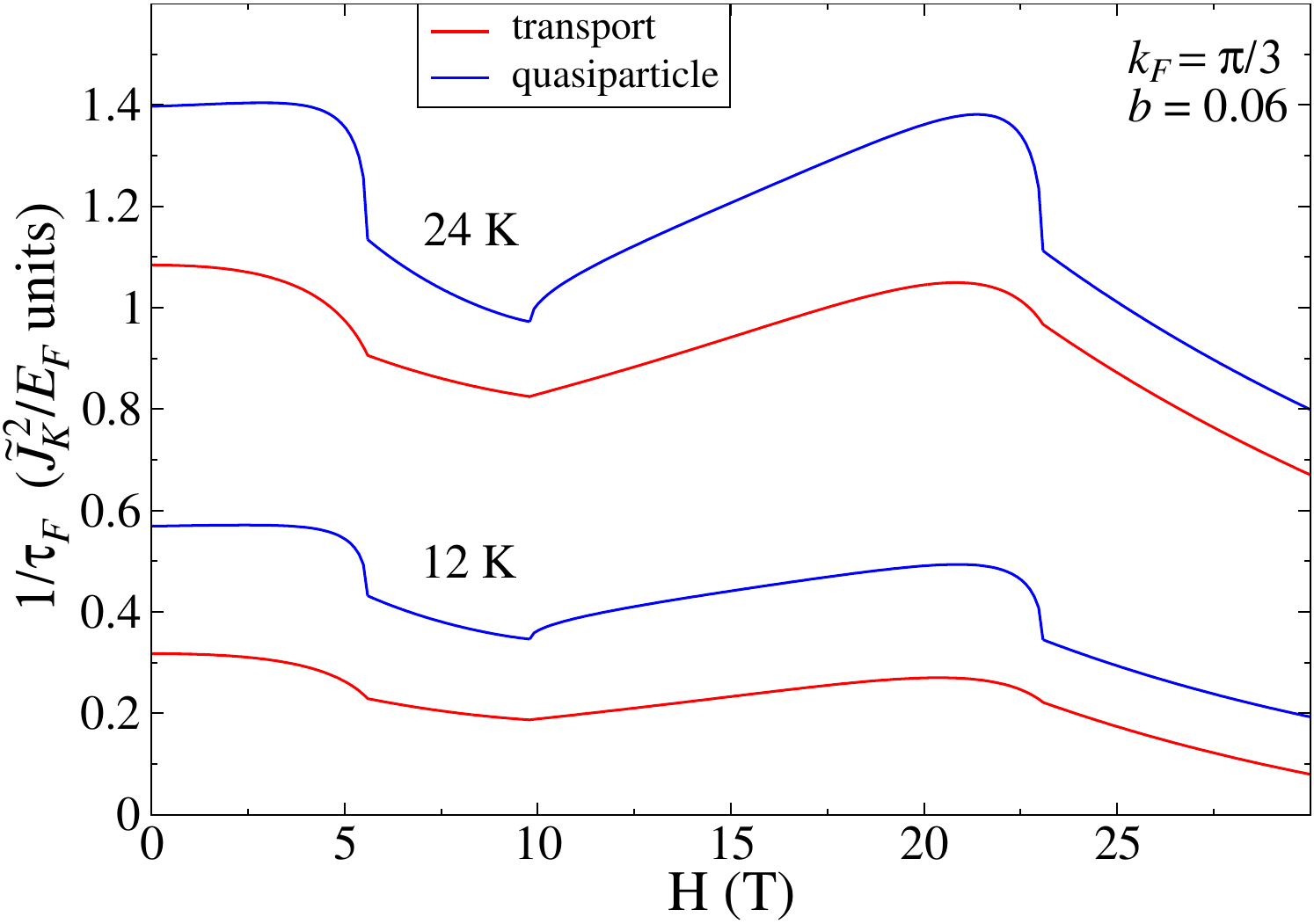}
\vskip -0.2cm
\caption{Transport and quasiparticle relaxation rates from Eqs.~(\ref{1_tau})-(\ref{matrix_elements}) 
and their quasiparticle analogues vs field for representative temperatures, $b\!=\!0.06$, and $k_F\!=\!\pi/3$;
other parameters are from Table~\ref{Table1}. Transport rates are from Fig.~\ref{Fig_rhoBdiff}(a).}
\label{Fig_taus}
\vskip -0.3cm
\end{figure}

The result of such a comparison 
is presented in Fig.~\ref{Fig_taus} for a representative value of $b\!=\!0.06$ and $k_F\!=\!\pi/3$.
The transport rates reproduce the results shown in Fig.~\ref{Fig_rhoBdiff}(a). 
Although the overall field-dependence of the rates is quite similar, the quasiparticle $1/\tau_{\rm qp}$ exhibits 
more pronounced variations from phase to phase,  with an exception of the polarized FM region, 
which shows only a modest overall offset from the transport one. 

The evolution of $1/\tau_{\rm qp}$ with the biquadratic exchange reveals more dramatic differences.
Upon approaching the first-order transition, $b\!\rightarrow\!b_c$, the Y-UUD and V-FM 
phase boundaries in $1/\tau_{\rm qp}$ become sharper, turning into  discontinuities at $b\!=\!b_c$.
Further increase to $b\!>\!b_c$ suggest truly divergent relaxation rates to occur at the (metastable) boundaries
discussed in Appendix~\ref{app:1st}. This is in contrast with the evolution of the transport 
relaxation rate, which remains continuous at these phase boundaries up to  $b\!=\!b_c$ and develops 
finite discontinuities for $b\!>\!b_c$, as is  demonstrated in Figs.~\ref{Fig_rhoBdiff}(a) and (b), respectively.  

Since the difference of the rates is in the suppression of the small-angle scattering in the 
transport case, the small-momentum contribution to the quasiparticle rate 
is an obvious culprit for such a dichotomy. 
Detailed analysis reveals the key role of the non-spin-flip scattering, 
present in the noncollinear Y and V phases, in precipitating singular behavior of $1/\tau_{\rm qp}$. 

In the proximity of the transitions, $h\!\rightarrow\!h_{c1(s)}$, where the 
Goldstone mode softens to $\omega_{\bf q}\!\propto\! q^2$, see Fig.~\ref{Fig_wks},
these processes provide a dangerous  $\propto\!1/\omega_{\bf q}$ term to the small-momentum 
equivalent of the kernel (\ref{kernel}), resulting in
the leading contribution to $1/\tau_{\rm qp}$ that scales as  $\sim\!\sin^2\!\alpha_1/\sqrt{h_c-h}$. 
From Appendix~\ref{app:1st}, $\sin^2\!\alpha_1\!\propto\!(h_c-h)$ for $b\!<\!b_c$, still giving a 
continuous behavior of $1/\tau_{\rm qp}$ vs $h$, as is shown in Fig.~\ref{Fig_taus}. 
However, for $b\!\rightarrow\!b_c$, the  field-dependence of the angle switches to  
$\sin^2\!\alpha_1\!\propto\!\sqrt{h_c-h}$, resulting in a finite contribution  to $1/\tau_{\rm qp}$ at $h_c$ and 
culminating in a discontinuity. Further increase  to $b\!>\!b_c$ leaves the angle finite at the 
transition field $h_c$, see Fig.~\ref{Fig_cos}(b), producing  a true singularity,  
$1/\tau_{\rm qp}\!\propto\!1/\sqrt{h_c-h}$.

This discussion opens up an interesting possibility of the studies of the field-tuned 
singular behavior in the scattering rate due to small-angle scattering processes that 
can lead to a number of observables including catastrophic violation of the Wiedemann-Franz law.

\subsubsection{Angular dependence of $1/\tau_{\rm qp}$}
\label{Sec_1_tau_qp_2}

Another aspect of the quasiparticle relaxation rate makes its consideration worthwhile.
The diagrammatics-derived expression for it in Eq.~(\ref{1_tau_qp}) is exact to the second order  
in the coupling. The approximations leading to (\ref{1_tau_qp1})  involves only a highly-justified 
energy hierarchy, $\omega_{\bf q}\!\ll\!E_F$. 
The next step, resulting in (\ref{1_tau_qp_final}),  is a fairly reasonable assumption of the spherical 
(cylindrical) Fermi surface. However, this last assumption does not automatically make 
$1/\tau_{\rm qp}$ independent of the {\it direction} of the momentum ${\bf k}$ on the Fermi surface.
It is an additional  step of assuming an even mirror-symmetry of the entire kernel $\widetilde{F}_{\bf q}$ 
with respect to a reflection of the momentum ${\bf q}$ about the direction of the momentum ${\bf k}$
which does that, leading to the final expression  (\ref{1_tau_qp_1D}), 
see  the line preceding Eq.~(\ref{1Dint_b}).

Although this assumption is valid for ${\bf k}$ along the high-symmetry directions, dictated by the $C_6$ 
symmetry of the triangular lattice in our case, the angular-dependence 
of the quasiparticle relaxation rate, encoded in the energies of spin excitations and 
matrix elements of the coupling to them, should remain.

A proper consideration of this problem for the transport relaxation rate has an additional
complicating factor because it also requires a consistency in the assumption of the angle-dependence 
in relating  scattered-state distribution function $\delta{\rm f}_{\bf k'}$
to $\delta{\rm f}_{\bf k}$, thus modifying Eq.~(\ref{df_ans_k'_final}). However,
the simplicity of this problem for the quasiparticle relaxation rate makes it particularly appealing
to solve. Thus, we return to the relaxation rate in Eq.~(\ref{1_tau_qp_final}) and advance it one step further 
to include such an angular-dependence properly,  leaving its study  in the transport counterpart to a future work. 

\begin{figure}[t]
\includegraphics[width=\linewidth]{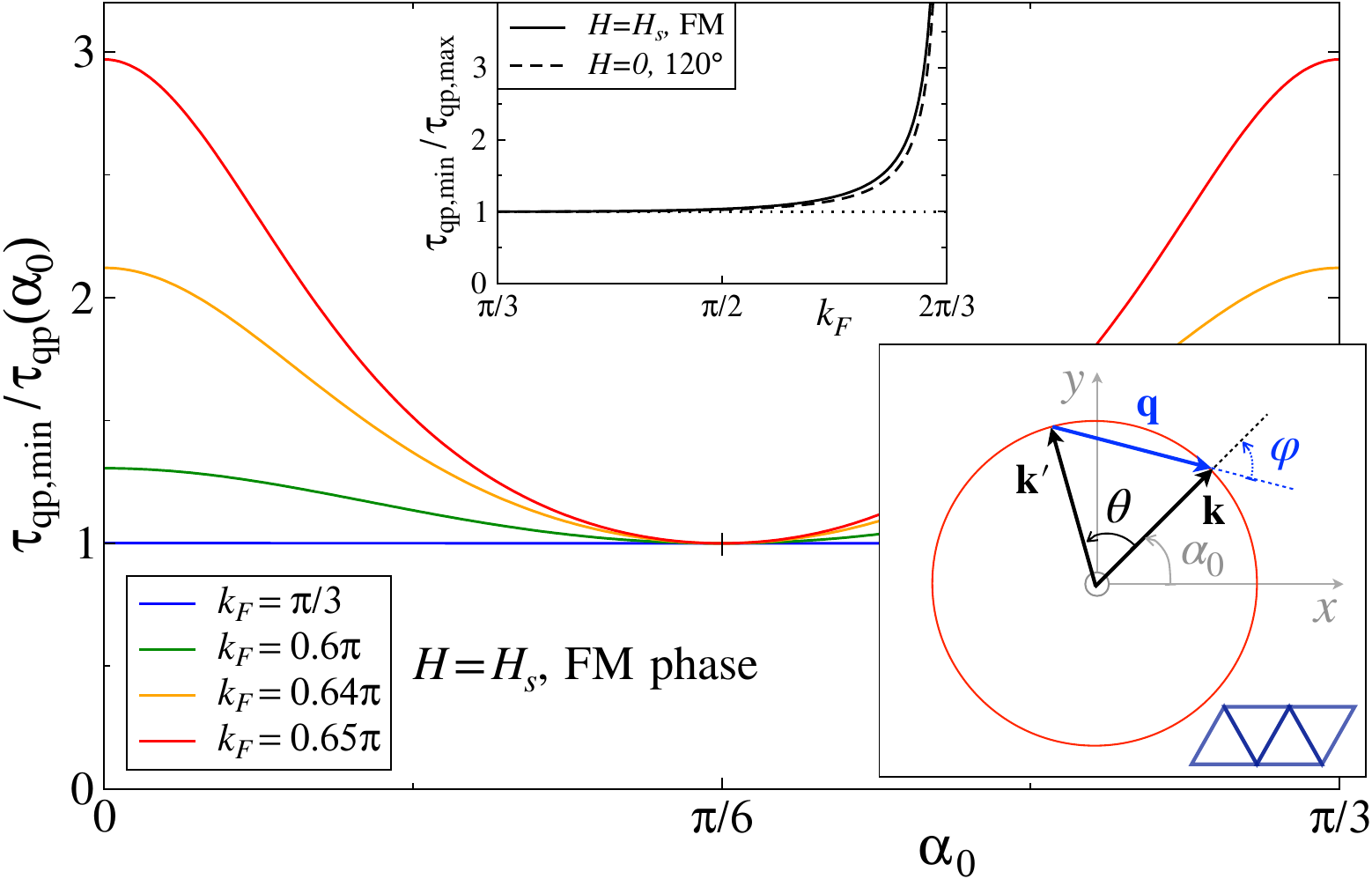}
\vskip -0.2cm
\caption{$1/\tau_{\rm qp}$ vs $\alpha_0$ for $H\!=\!H_s$ normalized to its value at $\alpha_0\!=\!\pi/6$ 
for the parameters in Table~\ref{Table1} and several values of $k_F$. Upper inset: 
$\tau_{\rm qp}(\pi/6)/\tau_{\rm qp}(0)$ for the FM  (solid line) and 
120$\degree$ state (dashed line) vs  $k_F$. Right inset: scattering diagram, angles, 
and laboratory axes, cf. Fig.~\ref{Fig_St}(b).}
\label{Fig_tau_alpha}
\vskip -0.3cm
\end{figure}

Although the direction of the electron momentum ${\bf k}$ is not explicitly present in 
(\ref{1_tau_qp_final}), it is implicitly tied to that of the magnon momentum ${\bf q}$ via the 
mutual angle $\varphi$, see the scattering diagram in the right inset of Fig.~\ref{Fig_tau_alpha}.
This Figure shows that in the reference frame formed by $({\bf \hat{k}}, {\bf \hat{k}}_\perp)$, 
where ${\bf \hat{k}}_\perp$ is the axis perpendicular to the unit vector ${\bf \hat{k}}$,
the magnon momentum ${\bf q}$ is parametrized as ${\bf q}\!=\!q(\cos\varphi,\sin\varphi)$. However, this frame 
 itself is rotated by the angle $\alpha_0$ with respect to the laboratory reference frame, represented here by  
$(x,y)$, with the $x$ axis tied to one of the bond-directions of the triangular lattice.   
Then, straightforwardly,   momentum ${\bf q}$ parametrization in the $(x,y)$ reference frame
is  ${\bf q}\!=\!q\big(\!\cos(\varphi\!+\!\alpha_0),\sin(\varphi\!+\!\alpha_0)\big)$. 

Since the absolute value of the momentum ${\bf q}$ is tied to the angle $\varphi$ via 
$q\!=\!2k_F\cos\varphi$, the integral in (\ref{1_tau_qp_final}) can still be reduced to a 1D form
\begin{align}
\label{1_tau_qp_1Dalpha}
\frac{\hbar}{\tau_{{\rm qp},F}}=\frac{\sqrt{3}\, (k_F a)^2}{\pi \,E_F} \int_{-\pi/2}^{\pi/2} d\varphi\,\frac12\,
\widetilde{F}_{\alpha_0}(\varphi),
\end{align}
where $\widetilde{F}_{\alpha_0}(\varphi)\!\equiv\!\widetilde{F}_{\bf q}$ with 
\begin{align}
\label{Faux1}
&\widetilde{F}_{\bf q} = \big|V_{\bf q}\big|^2
\Big(\frac{\omega_{\bf q}}{T}\Big)\,{\rm n}^0_{\bf q}\,\big({\rm n}^0_{\bf q}+1\big),\ \ \ {\rm and} \\
&{\bf q}=2k_F\cos\varphi\,\big(\!\cos(\varphi\!+\!\alpha_0),\sin(\varphi\!+\!\alpha_0)\big)\,.\nonumber
\end{align}
We note that Eq.~(\ref{1_tau_qp_1Dalpha}) does not have any additional approximations beyond the 
ones already present at the step (\ref{1_tau_qp_final}). Ignoring $\alpha_0$, using symmetry of
$\widetilde{F}_{\bf q}$ to $\varphi\!\rightarrow\!-\varphi$, and introducing $z\!=\!\cos\varphi$
brings (\ref{1_tau_qp_1Dalpha}) back to (\ref{1_tau_qp_1D}).
Obviously, the same considerations can be used to obtain the spin-flip equivalent of (\ref{1_tau_qp_1Dalpha}).

With this result (\ref{1_tau_qp_1Dalpha}), we can study the effect of the angular dependence of 
the quasiparticle relaxation rate. The findings for the polarized FM phase at the saturation 
field $H\!=\!H_s$ and for the 120$\degree$  $H\!=\!0$ state as representative points of our analysis are
summarized in  Fig.~\ref{Fig_tau_alpha}. In  Fig.~\ref{Fig_tau_alpha} 
we used the same model parameters as in the rest of the paper when applied to EuC$_6$, see 
Table~\ref{Table1}, and investigated the angle-dependence of $1/\tau_{\rm qp}$ as a function of $k_F$,
with  very similar results in both cases. 

As is discussed in Sec.~\ref{Sec_results}  for the transport case, the overall rates grow with $k_F$. 
Since we are interested in the angular dependence, but not the absolute values, 
the main panel in Fig.~\ref{Fig_tau_alpha} shows 
the quasiparticle relaxation rate from (\ref{1_tau_qp_1Dalpha}) as a function of $\alpha_0$ 
normalized to its minimum value, which is in the middle of the two principal (bond) directions of the 
triangular lattice, denoted as $\pi/6$. The results are shown for the FM phase and for several values of $k_F$. 
The upper inset shows the ratio of the rates at the maximum ($\alpha_0\!=\!0$) and minimum  
($\alpha_0\!=\!\pi/6$) vs $k_F$ for both FM and 120$\degree$ states.  

One can see that the effect of the angle-dependence in $1/\tau_{\rm qp}$ is really negligible up to 
$k_F$ about $\pi/2$ and is still very modest up to $k_F\!\agt\!0.6\pi$, justifying our initial approximation 
that neglected it and confirming the correctness of our results presented throughout the main body of the
paper.  In a sense, the effect is reminiscent of the accuracy of the sextupole field as nearly circular away 
from the close vicinity of the poles.

However, upon approaching the value of $k_F\!=\!2\pi/3$, the rates for the principal directions diverge
while the rates at $\alpha_0\!=\!\pi/6$ stay finite. As is discussed in Sec.~\ref{Sec_results}, this 
effect is due to a singularity associated with the scattering by the gapless Goldstone magnons
with ${\bf Q}\!=\!\pm(4\pi/3,0)$ and equivalent ordering vectors matching $2k_F$ for the momenta ${\bf k}$ 
in the principal directions. In that sense, the $\alpha_0\!=\!0$ and $\alpha_0\!=\!\pi/6$ directions
of the electron momentum ${\bf k}$ for $k_F\!\rightarrow\!2\pi/3$ become close analogues of the 
``hot'' and ``cold'' spots for the scattering, familiar from the cuprate superconductors \cite{cuprates}.
As is mentioned above, while the same behavior is expected to occur for the transport scattering rates,
the derivation of it becomes more complicated as one needs to account for the angular dependence in
relating non-equilibrium distribution functions for different directions self-consistently.

\bibliography{refs-EuC6}

\end{document}